\documentclass[reprint,floatfix,amsmath,amssymb,aps,longbibliography]{revtex4-2}
\usepackage{algorithm}
\usepackage{algpseudocode}
\usepackage{dsfont} 
\usepackage{float} 
\usepackage[colorlinks=true,linkcolor=black,citecolor=black,filecolor=black,urlcolor=black,breaklinks=true]{hyperref}
\usepackage[capitalize]{cleveref}
\usepackage{physics}
\usepackage{enumerate}
\usepackage{tikz} 
\usepackage{mathtools}
\usepackage{xargs}   
\usepackage{xcolor}  
\usepackage{amsthm}
\usepackage{lmodern}
\newtheorem{theorem}{Theorem}
\newtheorem{corollary}{Corollary}[theorem]
\newtheorem{observation}[corollary]{Observation}
\newtheorem{conjecture}[corollary]{Conjecture}
\newcommand{\robs}[1]{Observation~{\ref{#1}}}
\newcommand{\rconj}[1]{Conjecture~{\ref{#1}}}

\usepackage{thmtools}
\declaretheoremstyle[notefont=\bfseries,notebraces={}{},%
    headpunct={},postheadspace=1em,bodyfont=\itshape]{mystyle}
\declaretheorem[style=mystyle,numbered=no,name=Theorem]{thm-hand}

\newcommand{\m}{\mathrm}
\newcommand{\diag}{\m{diag}}
\newcommand{\I}{\m{i}}
\newcommand{\Ito}{It\^o\ }
\newcommand{\calp}{\mathcal{P}} 
\newcommand{\calq}{\mathcal{Q}} 
 
\newcommand{\calh}{\mathcal{H}} 
\newcommand{\cald}{\mathcal{D}} 
\newcommand{\calr}{\mathcal{R}} 
 
\newcommand{\calx}{\mathcal{X}} 
\newcommand{\calv}{\mathcal{V}} 
 
\newcommand{\calu}{\mathcal{U}} 
\newcommand{\cals}{\mathcal{S}} 
\newcommand{\cala}{\mathcal{A}} 
\newcommand{\calw}{\mathcal{W}} 
\newcommand{\sfl}{\textbf{\textsf{L}}}
\newcommand{\sft}{\textbf{\textsf{T}}}
\newcommand{\sfp}{\textbf{\textsf{P}}}
\newcommand{\sfu}{\textbf{\textsf{U}}}

\DeclarePairedDelimiterX\inproduct[2]{\langle\negthinspace\negthinspace\langle}{\rangle\negthinspace\negthinspace\rangle}{#1 \delimsize\vert #2}

\DeclarePairedDelimiter\lcom{\text{[\negthinspace[}}{\text{]\negthinspace]}}

\DeclarePairedDelimiterX\superdyad[2]{\lvert}{\rvert}{#1 \delimsize\rangle\negthinspace\negthinspace\rangle \negthinspace\langle\negthinspace\negthinspace\langle #2}

\makeatletter
\def\namedlabel#1#2{\begingroup
    #2%
    \def\@currentlabel{#2}%
    \phantomsection\label{#1}\endgroup
}
\makeatother

\begin{document}

\title{Asymptotic Fate of Continuously Monitored Quantum Systems}
\author{Finn Schmolke}
\affiliation{Institute for Theoretical Physics I, University of Stuttgart, D-70550 Stuttgart, Germany}

\begin{abstract}
  A quantum trajectory is the natural response of a quantum system subject to external perturbations due to continuous indirect measurement.
  We completely characterize the asymptotic behavior of continuously monitored quantum systems in finite dimensions and show that generically, spontaneous irreversible localization transitions on the level of individual realizations occur, where the evolution becomes effectively constrained to one of the irreducible components of the total Hilbert space.
  More generally, localization can be either complete, where the strongest possible confinement is achieved, or incomplete, where localization terminates prematurely.
  The full description contains, but is not restricted to, asymptotic purification, the abelian structure of symmetries, and classical noise.
  On the trajectory level, symmetries and conserved quantities are no longer respected and localization transitions occur concurrently with violations of ergodicity.
  As a result, a generalized update rule emerges that effectively projects the system onto one of several possible time evolutions.
  The update comes equipped with a generalized Born rule that assigns probabilities to these irreversible events.
  Spontaneous transitions thus occur probabilistically and can deviate considerably from the behavior of the ensemble.
  In particular, time and ensemble average no longer commute which gives rise to global violations of ergodicity, while on a local level, ergodicity is restored.
  We show that these violations are captured by the mean fidelity between the time and ensemble averaged states, resulting in a participation ratio which depends solely on the effective distribution of the initial state over the substructures of the Hilbert space.
  We explicitly illustrate the asymptotic behavior of continuously monitored systems in a series of examples demonstrating, among other phenomena, stabilization of many-body scar states and generation of Bell states from local measurement.
  Finally, we present two algorithms, one based on simultaneous block diagonalization and one based on quantum trajectories to identify all the minimal orthogonal subspaces of the Lindblad equation and all the extremal stationary states which are supported on them.
\end{abstract}

\maketitle 
\section{Introduction}
Understanding quantum measurement is of fundamental importance to quantum science and technology.
In standard quantum theory, a projective measurement produces one of the eigenvalues $l$ of a Hermitian observable $L$, while at the same time representing a drastic intervention on the evolution of the system that collapses the wave function.
Due to the newly acquired information, the measurement process is associated with an update rule that maps the state onto one of the eigenspaces of $L$, corresponding to the obtained measurement outcome.
Performing sequences of measurements makes the system concede more information at the expense of inducing more complex dynamics.
The response to these external perturbations gives rise to a quantum trajectory, which, in a broad sense, describes the evolution of the state undergoing unitary self-evolution interrupted by generalized measurements at random times.
The backaction intrinsic to the measurement process allows quantum trajectories to probe higher order statistics of the density operator and thus expose features that are hidden to the evolution of the ensemble average.
In fact, a new class of quantum phase transitions has recently been discovered in the dynamics of entanglement that is revealed upon perturbing the system with randomly distributed measurements, but entirely masked in the average dynamics \cite{Cao2019,Li2018,Szyniszewski2019,Skinner2019,Gullans2020,Lang2020,Choi2020,Szyniszewski2020,Buchhold2021,Alberton2021,Ippoliti2021,Minato2022}.
These classes of quantum trajectories are typically discrete-time processes that require external control to apply unitary or projective gates in quantum-circuit models.
Here, we focus on a different but related class that arises due to the continuous monitoring of a quantum system, thereby representing an autonomous process.

The most general instantaneous update rule is given by positive operator valued measures that describe the effective map on a reduced state when a projective measurement is performed on a larger composite system, constituting an indirect observation of the reduced system \cite{Nielsen2010,Gardiner2004,Wiseman2009,Barchielli2009,Jacobs2014,Albarelli2024,Jordan2024}.
Indirect measurements may be performed continuously by permanently monitoring the environment to continually gain a little information, while at the same time causing only little disturbance \cite{Gardiner2004,Wiseman2009,Barchielli2009,Jacobs2014,Albarelli2024,Jordan2024}.
Such a setup is, for instance, naturally realized by a trapped, laser-driven atom in a cavity, where absorption and emission processes, due to the interaction with the electromagnetic field, leave discrete imprints on the detected photocurrent.
The reduced system still evolves in time, experiencing random fluctuations that make the state trace a stochastic path in Hilbert space, giving rise to a continuous-time quantum trajectory.
In general, a competition is realized between the measurement and the coherent dynamics.
Mutual compatibility, $[H,L] = 0$, gradually projects the system onto one of the common eigenspaces of $H$ and $L$, effectively performing a prolonged projective measurement associated with a continuous wave function collapse \cite{Jordan2013}.
Nontrivial quantum features arise when the different processes are at least partially incompatible, with $[H,L_j] \neq 0$ and or $[L_i,L_j] \neq 0$ for some $i,j$ \cite{Cao2019,Li2018,Szyniszewski2019,Skinner2019,Gullans2020,Lang2020,Choi2020,Szyniszewski2020,Buchhold2021,Alberton2021,Ippoliti2021,Minato2022}.

Quantum trajectories have been originally conceived to facilitate numerical simulations of large open systems, since an ensemble of pure states can be evolved in place of the more memory intensive density matrix.
Nowadays, modern trajectory theory complements and successively extends the instantaneous update rules and accurately captures the behavior of quantum systems under continuous monitoring that can since be routinely observed in experiments \cite{Sauter1986,Nagourney1986,Bergquist1986,Vijay2011,Murch2013,Rist2013,Sun2014,Weber2014,Hacohen_Gourgy2016,Minev2019}.
Quantum trajectories have helped deepen our understanding of quantum measurement theory and led to a new, more detailed view on quantum dynamics \cite{Sauter1986,Nagourney1986,Bergquist1986,Vijay2011,Vijay2012,Murch2013,Rist2013,Sun2014,Weber2014,Hacohen_Gourgy2016,Minev2019}.
Indeed, the minimal invasiveness of weak measurements allows resolving the statistics of simultaneous non-commuting measurements \cite{Hacohen_Gourgy2016} and offers a unique perspective on the nature of the wave function collapse \cite{Murch2013,Jordan2013,Weber2014,Minev2019}.
Moreover, applying feedback control enables steering trajectories along desired paths in Hilbert space and perform error correction to counteract decoherence \cite{Vijay2012,Weber2014}.

Mathematically, quantum systems subject to continuous monitoring are described by stochastic master equations and correspond to an unraveling of a Lindblad equation, which is reproduced upon taking the average over the ensemble of realizations \cite{Ueda1989,Molmer1993,Carmichael1993,Plenio1998,Gardiner2004,Breuer2007,Wiseman2009,Barchielli2009,Jacobs2014,Daley2014,Landi2024,Albarelli2024,Jordan2024}.
Whereas quantum systems undergoing Lindbladian evolution are guaranteed to posses at least one stationary state \cite{Evans1977}, conventional wisdom dictates that a quantum trajectory will exhibit indefinite noisy dynamics due to the incessant random perturbations caused by the continuous measurement.
The asymptotic properties of Lindbladians have been under extensive study since their inception \cite{Gorini1976,Lindblad1976,Davies1970,Frigerio1977,Frigerio1978,Lindblad1999,Breuer2007,Baumgartner2008_1,Baumgartner2008_2,Baumgartner2012,Albert2014,Albert2016,Carbone2016,Nigro2019,Zhang2020,Zhang2024,Hironobu2024}, resulting in a complete characterization of the entire state space structure and all the stationary states \cite{Lindblad1999,Baumgartner2008_1,Baumgartner2008_2,Carbone2016}.
At the same time, little is known about the asymptotics of quantum trajectories.

In this article, we provide a complete description of the long-time behavior of the quantum jump and diffusive homodyne unraveling for arbitrary finite-dimensional quantum systems.
We derive the necessary and sufficient conditions for quantum trajectories to admit invariant states, wich are in a sense the eigenstates of the generalized measurement process, and investigate their attractiveness on the level of single realizations.
More generally, we show that a generic continuously monitored quantum system undergoes spontaneous irreversible localization transitions in time, confining and purifying the system to varying degrees.
In \cref{sec:proof-incomplete-minimal,sec:incomplete-composite} we proof our main theorems on asymptotic localization of arbitrary quantum trajectories.
We will gradually unfold and analyze in detail the implications of this general result to eventually assemble a complete picture of the asymptotic behavior of continuously monitored systems.

A fundamental assumption in the study of classical and quantum stochastic processes is the ergodic hypothesis which states that the long-time average of a single trajectory coincides with the ensemble average.
Combining localization properties with previously known results on ergodicity, we specify when this assumption is actually justified and reveal that quantum trajectories generally break ergodicity even in the absence of symmetries.
In addition, we provide a more refined description and show that ergodicity is locally restored when purification and localization take place unimpeded.
In special situations, the system may moreover spontaneously escape the influence of the detector, whereupon the backaction disappears and the measurement effectively acts as like an environment inducing classical noise on the system, which always results in ergodic dynamics.

Measurements compatible with the system Hamiltonian can be considered as an effective dragged out projective measurement, associated with a gradual collapse of the wave function.
Here, we make this argument rigorous and show that this notion extends to arbitrary finite-dimensional quantum systems, where an effective update rule emerges supplied with a generalized Born rule.
We investigate the general behavior in several contexts and demonstrate localization transitions in systems ranging from a single qubit to complex many-body systems.

\subsection{Preliminaries}
Our starting point is the quantum master equation of Lindblad type 
\begin{align}
    \dot{\rho} 
    = \sfl(\rho)
    = -\I[H,\rho] + \sum_k \left(L_k\rho L_k^\dagger - \frac{1}{2}\left\{L_k^\dagger L_k,\rho\right\}\right).
    \label{eq:me}
\end{align}
It describes the general evolution of a system coupled to an environment given certain physically motivated approximations \cite{Breuer2007}.
Energy and information may be exchanged with its surrounding in an incoherent way, inducing decay, dephasing and general decoherence \cite{Baumgartner2008_2,Zurek2003}.
The Lindblad jump operators $L_k$ account for the dissipative action of the environment on the system.
Given a generator $\sfl$, the decomposition into unitary and dissipative parts is not unique; some actions can be ascribed to either Hamiltonian $H$ or the jump operators $L_k$.
There exist attempts to favor special choices of the $L_k$, but we make here no restriction.
If not stated otherwise, we denote by $\rho$ the density matrix corresponding to the solution of the Lindblad equation.
The master equation \eqref{eq:me} may be formally integrated to evolve an initial state
\begin{align}
  \rho(t) = \sft^t(\rho(0)) = \rho(0) + \int_0^t \dd{s} \sfl(\rho(s)).
\end{align}
The Lindblad equation is guaranteed to posses at least one stationary state that will be assumed in the long-time limit.
If multiple stationary state exists, any convex combination is also stationary.
Of special importance are the linearly independent---extremal---states, corresponding to the vertices of the convex hull, all of which are mutually orthogonal.
We will elaborate more on the properties of the state space of the Lindblad equation in \cref{sec:space-structure}.

The Lindblad equation \eqref{eq:me} can be unraveled into different ensembles of quantum trajectories, corresponding to different physical measurement schemes or realizations of an abstract process \cite{Ueda1989,Molmer1993,Carmichael1993,Plenio1998,Gardiner2004,Breuer2007,Wiseman2009,Barchielli2009,Jacobs2014,Daley2014,Landi2024,Albarelli2024,Jordan2024,Jordan2024}.
By construction, all unravelings share the property that the ensemble average reproduces the Lindblad equation \cref{eq:me}.
Here, we study the two most prominent unravelings, quantum jump detection and diffusive, homodyne detection.
Quantum jumps resolve the spontaneous sudden transitions between states of different energy under photon emission or absorption and have been experimentally accessible for a long-time \cite{Sauter1986,Nagourney1986,Bergquist1986,Vijay2011,Sun2014,Minev2019}.
The evolution of the system conditioned on the sequence of measurement outcomes follows a quantum jump trajectory, that is described by the stochastic master equation \cite{Ueda1989,Carmichael1993,Dalibard1992,Molmer1993,Plenio1998,Breuer2007,Jacobs2006,Wiseman2009,Jacobs2014,Daley2014}
\begin{align}
    \dd{\rho}_J
    =& \sfl_J(\rho_J)\\
    =& -\I[H,\rho_J] \dd{t} + \sum_k\left(\langle L_k^\dagger L_k\rangle \rho_J - \frac{1}{2} \{L_k^\dagger L_k,\rho_J\}\right) \dd{t} \notag \\
    &+ \left[\frac{L_k\rho_J L_k^\dagger}{\langle L_k^\dagger L_k\rangle}-\rho_J\right] \dd{N}_k.
    \label{eq:jump-sme}
\end{align}
The independent stochastic Poissonian increments $\dd{N_j}\dd{N_k} = \delta_{jk}\dd{N_j}$ are equal to either zero (no detection) or one (detection). 
The probability to make an observation is $\mathbb{E}[\dd{N_k}(t)] = \langle L_k^\dagger L_k\rangle \dd{t}$, where $\mathbb{E}[\bullet]$ denotes the classical average over the ensemble of trajectories, and $\langle \bullet \rangle \equiv \tr[\rho \bullet]$ is the usual quantum mechanical average with respect to $\rho$.

Under certain conditions one may take the limit to many but weak measurements by introducing a strong local oscillator according to $L_k \to L_k + \alpha_k\mathds{1}$ with complex amplitude $\alpha_k$ and implementing a homodyne detection scheme to arrive at the diffusion approximation of the quantum jump trajectory \cref{eq:jump-sme} \cite{Wiseman2009,Pellegrini2009,Landi2024}.
While the Lindblad equation \cref{eq:me} is invariant under the joint transformation 
\begin{align}
  &L_k \to L_k + \alpha_k \mathds{1} \notag\\
  &H \to H + \frac{1}{2\I}\sum_k (\alpha^\ast_k L_k - \alpha_k L^\dagger_k) + \beta \mathds{1},
  \label{eq:local-oscillator}
\end{align}
with $\alpha_k \in \mathbb{C}$ and $\beta \in \mathbb{R}$ \cite{Breuer2007,Wiseman2001}, the quantum jump unraveling \cref{eq:jump-sme} is generally not.
Introducing the local oscillator fundamentally changes the measurement setup and the two unravelings really describe separate physical processes.
A rigorous proof on the existence of the diffusive limit of quantum jumps has been established in Ref.~\cite{Pellegrini2009}.
When this procedure can be carried through successfully, it yields a diffusive quantum trajectory described by the \Ito stochastic master equation \cite{Wiseman2009,Barchielli2009}
\begin{align}
    \dd{\rho_\xi} 
    =& \sfl_\xi(\rho_\xi)\\
    =& -\I[H,\rho_\xi] \dd{t} + \sum_k \left(L_k\rho_\xi L_k^\dagger - \frac{1}{2}\left\{L_k^\dagger L_k,\rho_\xi\right\}\right) \dd{t} \notag \\
    &+ \left[L_k\rho_\xi+\rho_\xi L_k^\dagger - \langle L_k+L_k^\dagger\rangle \rho_\xi\right]\dd{W_k}.
    \label{eq:homodyne-sme}
\end{align}
Here, $\dd{W_k(t)}$ denotes mutually independent Wiener processes with zero mean, $\mathbb{E}[\dd{W_k}] = 0$, and unit variance, $\mathbb{E}[\dd{W_j(t)}\dd{W_k(t^\prime)}] = \delta_{jk}\delta(t-t^\prime) \dd{t}$.
The experimental realization of diffusive quantum trajectories became available with the advent of superconducting quantum circuits and their observation thus has a more recent history \cite{Murch2013,Rist2013,Weber2014,Hacohen_Gourgy2016}.

Although different unravelings generally yield different distributions of the density matrix, some properties are shared.
We will use the index $c=(J,\xi)$ to indicate general conditional evolution for relations that apply to both unravelings simultaneously.
Most notably, the Lindblad equation is reproduced on average, that is
\begin{align}
  \rho(t) 
  = \mathbb{E}[\rho_\m{c}(t)] 
  = \lim_{N \to \infty} \frac{1}{N}\sum_{c=1}^N \rho_\m{c}(t),
\end{align}
where $\mathbb{E}[\bullet]$ denotes the average over the ensemble of quantum trajectories.
Both \cref{eq:homodyne-sme,eq:jump-sme} may be formally integrated to yield a completely positive and trace preserving map that evolves an initial state $\rho(0)$ along a quantum trajectory.
We denote by
\begin{align}
  \rho_\m{c}(t) 
  = \sft^t_\m{c}(\rho(0)) 
  = \rho(0) + \int_0^t \dd{s}\sfl_\m{c}(\rho_\m{c}(s)),
  \label{eq:non-linear-map}
\end{align}
the formal solution of the stochastic differential equations \cref{eq:homodyne-sme,eq:jump-sme}.

\begin{figure}[t]
  \centering 
  \begin{tikzpicture}
    \node (a) [label={[label distance=-0.1cm]125: \textbf{(b)}}] at (4.2,0) {\includegraphics{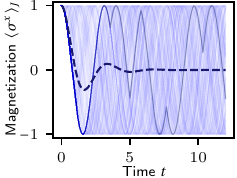}};	
    \node (b) [label={[label distance=-0.1cm]125: \textbf{(a)}}] at (0,0) {\includegraphics{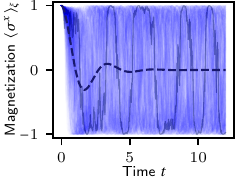}};
    \draw (0.2,1.7) node {\textsf{Diffusion}};
    \draw (4.4,1.7) node {\textsf{Jumps}};
  \end{tikzpicture}
  \caption{Continuous monitoring of a qubit in the absence of steady state degeneracies. 
  (a) Diffusive unraveling \cref{eq:homodyne-sme} and 
  (b) quantum jump trajectories \cref{eq:jump-sme} corresponding to the same Lindblad equation \eqref{eq:me} (dashed line).
  Overlapping trajectories give rise to darker regions rendering the color shading proportional to the probability density. A typical realization is highlighted in gray.}
  \label{fig:introduction}
\end{figure}

\emph{Introductory example: Continuous measurement of a qubit.}---To illustrate the typical behavior of quantum trajectories with competing unitary and measurement dynamics in the absence of symmetries and steady state degeneracies, we consider a single continuously monitored qubit with Hamiltonian $H = \omega \sigma^x$, where $\omega$ is the level spacing.
The magnetization in $z$-direction is continuously monitored with jump operator $L = \sqrt{\Gamma}\sigma^z$, where $\Gamma$ controls the measurement strength.
In \cref{fig:introduction} we plot the evolution of the $x$-polarization for a subensemble of $250$ quantum trajectories alongside the corresponding average evolution due to the Lindblad equation.
Individual trajectories roam around randomly to indefinitely explore the whole Hilbert space, without ever reaching stationarity.
The main objective of this work is to discuss the most general case in arbitrary finite dimensions that may feature any number of symmetries and steady state degeneracies and apply the results to relevant physical scenarios.

\section{Overview}
\label{sec:main}
Before describing our analysis we first provide an overview over the main results.
We study the long-time behavior of continuously monitored quantum systems in finite dimensions.
We focus on the diffusive, homodyne and the quantum jump unraveling because they are the most prominent and can be realized experimentally.
The properties of diffusive quantum trajectories are always studied first and the subsequent analysis of quantum jumps is then provided separately, where similarities and differences are highlighted.

Our full characterization starts with the concept of invariant states along quantum trajectories.
The Lindblad equation describes the dissipative dynamics of an open quantum system and is guaranteed to posses at least one stationary state that will be assumed in the limit of infinite time as the system either equilibrates or settles into a non-equilibrium.
Continuous measurement represents a different type of open system description, where the competition between coherent dynamics and measurement is generally expected to never come to rest.
Invariant states of quantum trajectories are in a sense a generalization of stationary states to stochastic master equations.
We provide the necessary and sufficient conditions for quantum trajectories to admit these invariant states:
The non-linear stochastic map has certain exceptional, regular points in Hilbert space that are unstable, where any tiny deviation will lead to stochastic dynamics.
It turns out that the only deterministically stable invariant states are pure.
By choosing appropriate measurement operators, we show that trajectories can stabilize and single out individual many-body scar states of a chaotic spin-1 Hamiltonian (\cref{sec:scar states}).

Invariant states are not yet the whole wisdom.
It has recently been realized that individual quantum jump trajectories do not always respect the symmetries and conserved quantities of the Lindblad equation \cite{Munoz2019,Tindall2023}.
Here, we show in full generality that quantum trajectories are susceptible to the Hilbert space structure imposed by the Lindblad equation.
To prepare for the general analysis, we therefore first establish the corresponding structure theorems derived by Baumgartner and Narnhofer in Refs.
\cite{Baumgartner2008_2,Baumgartner2012}, that characterize the complete decomposition of the Hilbert space into its irreducible components, the minimal orthogonal subspaces.
We proof that, generically, individual realizations select minimal orthogonal subspaces and successively localize into them.
The long-time average then reproduces the unique (extremal) stationary state of the Lindblad equation supported on the corresponding minimal subspace.
There are however still non-generic cases where subspaces are in a certain sense indistinguishable from the perspective of the quantum trajectory and, as a consequence, the system is unable to localize further.
In general, we differentiate between complete localization where further confinement in Hilbert space is impossible and incomplete localization where still smaller substructures exist but localization stops prematurely.
We show that, in the infinite-time limit, any quantum trajectory exhibits one of three characteristic behaviors:
\begin{enumerate}[(a)]
  \item Partial localization occurs; trajectories do not purify and evolve independently, there is no measurement backaction just classical noise.
  \item Partial localization occurs; evolutions are unitarily equivalent in minimal orthogonal subspaces and asymptotic purification occurs in each of the subspaces individually but not globally.
  \item Full localization occurs, the trajectory does not purify but is subject to classical noise, or asymptotic purification and full localization occur concurrently; the quantum system enters completely a minimal orthogonal subspace.
\end{enumerate}

These spontaneous localization transitions in Hilbert space occur probabilistically and can lead to considerable deviations from the evolution of the mean, since the system will be confined to only a small fraction of the available state space.
Knowledge of the ensemble average is hence not sufficient to provide information about the behavior of individual trajectories.
To quantify how strongly they deviate from the Lindblad equation, we compute the mean fidelity between the time and ensemble averaged trajectories.
Given an initial state, we compare the infinite-time average of individual trajectories with the time averaged asymptotic state of the Lindblad equation.
Generically, in the absence of special symmetries between subspaces, the mean fidelity reduces to a participation ratio that depends only on the effective distribution of the initial state over the Hilbert space and is in particular independent of the microscopic structure and the subspace dimensions.
In essence, the measurement backaction inherent to quantum measurement is responsible for purification, localization and breaking of ergodicity.
Indeed, we show that classical noise always induces ergodic evolution, where the trajectory explores the entire available state space.
However, even amid the strongest deviations from the ensemble average, ergodicity can be locally restored to hold in each of the irreducible subspaces individually.
Taking the time average of a single trajectory will then always result in one of the extremal stationary states of the Lindblad equation.

Having established the full characterization, we provide a general physical interpretation of the long-time behavior of quantum trajectories.
Any initial state is assigned to an effective, asymptotic mapping that projects the system onto an orthogonal subspace, while the generator of the time evolution is itself projected onto a new, valid quantum trajectory with a new, restricted Hamiltonian and measurement process.
This mapping gives rise to a generalized measurement update rule supplied with a generalized Born rule.
A situation similar to standard quantum measurement arises that is generalized in multiple ways (\cref{sec:Born}).
The standard measurement postulate can be recovered as a special case which is illustrated with the introductory example of a monitored qubit in \cref{sec:single-qubit}.
We show that quantum trajectories induce parity selection in a non-linear Kerr resonator (\cref{sec:Kerr}), boost coherent oscillations of non-interacting qubits (\cref{sec:two-qubits}), and create Bell states from local measurement (\cref{sec:ring}).

\section{Outline}
The paper is organized as follows.
In \cref{sec:invariant-states}, we introduce the concept of steady states along quantum trajectories and derive the necessary and sufficient conditions on the Hamiltonian and the Lindblad jump operators to admit them.
Before a more general analysis can be performed, it is necessary to establish the structuring of the Hilbert space.
In \cref{sec:space-structure}, we provide the structure theorems for a complete decomposition of the Hilbert space induced by the Lindblad equation.
In \cref{sec:transition}, this knowledge is used to show that quantum trajectories spontaneously undergo irreversible localization transitions in Hilbert space.
In \cref{sec:incomplete}, we then investigate the infinite-time behavior of quantum trajectories in full generality and provide the necessary and sufficient conditions for complete and incomplete localization.
Equipped with these general results, a physical interpretation is possible.
In \cref{sec:erg-theorem}, we quantify the ergodic properties of quantum trajectories and show that continuous measurement gives rise to an effective generalized update rule in \cref{sec:Born}.
\Cref{sec:examples} provides several examples and special cases, including the limits of strong measurement and classical noise.
In \cref{sec:sbd} we finally present an algorithm based on simultaneous block diagonalization and propose a trajectory-based method to resolve the full Hilbert space structure and identify all extremal stationary states of the Lindblad equation.

\section{Invariant States}
\label{sec:invariant-states}
Our aim is to characterize the asymptotics of continuously monitored quantum systems.
The focus lies on quantum trajectories that can be unraveled from a corresponding Lindblad equation.
For Lindbladian evolution, the asymptotic behavior has been completely classified and is uniquely determined by the stable invariant manifold containing the stationary states.
For quantum systems under continuous measurement the behavior is richer.
As a first step towards full characterization, we study the equivalent invariant manifold of states on the trajectory level.

The structure of this manifold enjoys a straightforward derivation that directly follows from the stochastic master equations themselves without resorting to the heavier machinery that is required later to derive more general statements.
It turns out that any arbitrary state can be made invariant for some quantum trajectory, however the only stable states are pure states.
It is thus impossible to deterministically stabilize mixed states along quantum trajectories.

A state $\rho$ is invariant under the non-linear stochastic map, \cref{eq:non-linear-map}, if
\begin{align}
    \sft_\m{c}^t(\rho) = \rho.
\end{align}
The state $\rho$ as defined above is both stationary and invariant under the map $\sft_\m{c}^t(\bullet)$. However, there are cases where $\rho$ may still carry time dependence undergoing residual unitary oscillations such that 
\begin{align}
  \sft_\m{c}^\tau(\rho(t)) = \rho(t+\tau)
\end{align}
with an effective Hamiltonian $\tilde H$ according to
\begin{align}
  \rho(t+\tau) = e^{-i\tilde H \tau}\rho(t) e^{\I\tilde H \tau}.
\end{align}
An invariant state needs to stay invariant at all times and for all realizations of the stochastic process.
Clearly, if such states exist, the stochastic and dissipative terms in \cref{eq:homodyne-sme,eq:jump-sme} need to vanish separately, that is, every such state necessarily has to fulfill 
\begin{align}
  \sfl(\rho_\xi) &= 0, \\
  L_k\rho_\xi + \rho_\xi L^\dagger_k &= \langle L_k+L^\dagger_k\rangle \rho_\xi,\ \forall k,
  \label{eq:inv-diff}
\end{align}
for the diffusive trajectory and
\begin{align}
  \sfl(\rho_J) &= 0, \\
  L_k\rho_J L^\dagger_k &= \langle L^\dagger_k L_k\rangle \rho_J,\ \forall k,
  \label{eq:inv-jump}
\end{align}
for the quantum jump detection model.
Since both unravelings average to the Lindblad equation, an asymptotic state of quantum state diffusion or quantum jump trajectory is also an asymptotic state of the Lindblad equation but the converse is not true.
The invariant states thus represent the special subset of the stationary states of the Lindbladian that satisfy the additional constraints \cref{eq:inv-jump,eq:inv-diff} respectively.

Assume $\rho$ is invariant.
Without loss of generality we can always express the state in its eigenbasis.
If the state is not full rank, then we group together the nonzero eigenvalues in a common block.
In such a basis the state, the Hamiltonian and the Lindblad jump operators assume the following block structure
\begin{align}
  \rho = 
  \begin{bmatrix}
      D & 0 \\ 0 & 0
  \end{bmatrix}
  \
  H = 
  \begin{bmatrix}
    H_{11} & H_{12} \\ H_{21} & H_{22}, 
\end{bmatrix},
\
  L_k = 
  \begin{bmatrix}
      L_{k,11} & L_{k,12} \\ L_{k,21} & L_{k,22}
  \end{bmatrix}.
  \label{eq:block form}
\end{align}
Note that $\rho$ has full rank on its support, thus, by construction, $D$ is always full rank.
We can now state the necessary and sufficient conditions for the homodyne unraveling to admit invariant states.
\begin{theorem}
  \label{th:diff-inv}
  A state $\rho$ is invariant under the diffusive unraveling, \cref{eq:homodyne-sme}, if and only if the Hamiltonian commutes with the state, $[H,\rho] = 0$, and it holds that
  \begin{align}
    L_k &=
    \begin{bmatrix}
      A_k& L_{k,12} \\
      0 & L_{k,22}
    \end{bmatrix}, \ \forall k,\\
    H_{12} &= -\I\frac{1}{2} \sum_k A_k^\dagger L_{k,12},
  \end{align} 
  with $A_k = S_k + \mathds{1}a_k$, where $S_k$ is a anti-Hermitian operator that commutes with the state, $[S_k,\rho] = 0, \forall k$
  and $a_k \in \mathbb{C}$, is an arbitrary complex number.
  The block matrices $H_{22}$ and $L_{k,22}$ are arbitrary.
\end{theorem}
The proof is given in \cref{sec:invariant-diff}.
For quantum jumps, we analogously obtain the following necessary and sufficient conditions.
\begin{theorem}
  \label{th:jump-inv}
A state $\rho$ is invariant under the quantum jump unraveling, \cref{eq:jump-sme}, if and only if the Hamiltonian commutes with the state, $[H,\rho] = 0$, and it holds that
\begin{align}
  L_k &= 
  \begin{bmatrix}
    C_k & L_{k,12} \\
    0 & L_{k,22}
  \end{bmatrix}, \ \forall k,\\
  H_{12} &= -\I\frac{1}{2}\sum_k C_k^\dagger L_{k,12},
\end{align}
with $C_k = c_k U_k$, where $U_k$ is a unitary operator that commutes with the state, $[U_k,\rho] = 0, \forall k$
and $c_k \in \mathbb{C}$, is an arbitrary complex number.
The block matrices $H_{22}$ and $L_{k,22}$ are arbitrary.
\end{theorem}
The proof is given in \cref{sec:invariant-jumps}.
The triangular form of the jump operators together with the constraint on the off-diagonal block of the Hamiltonian conspire together to empty out the lower right block, such that the coherent and measurement dynamics eventually drive any initial state into the upper left block (see also \cref{sec:no-trajs-in-D}).
Already at this point it becomes clear that although \cref{eq:homodyne-sme,eq:jump-sme} produce the same mean value, they have different invariance properties.
While the overall form of the operators $L_k$ and $H$ is shared, in order to obtain noise-free invariant states in the long-time limit, the relevant asymptotic dynamics with $L_{k,11}$ needs to have a specific structure adapted to the respective measurement processes.
To admit invariant states, homodyne detection requires anti-Hermitian, quantum jumps unitary jump operators.
This is in fact not too surprising as the diffusive limit of quantum jumps in this case may be understood in terms of the approximation $U \approx \mathds{1}+S$, where $S$ is anti-Hermitian with, $\|S\|\ll 1$, small.

\subsection{Attractiveness}
Clearly, if one manages to prepare the quantum system in an invariant state, it will remain invariant throughout.
However, the question of attractiveness is not immediately obvious.
In contrast to the non-stochastic evolution of the Lindblad equation, the existence of invariant states alone does not yet guarantee that a stationary regime is reached for any initial state.
Fixed points of the Lindbladian may become unstable or vanish altogether along individual trajectories.
The latter case is generic for unique full rank stationary states, cf. \cref{fig:introduction}.
On the other hand, a decoherence-free subspace is both attractive and stable even on the trajectory level but only in a probabilistic sense.
We will elaborate on this case in \cref{sec:dfs}.

For both unravelings, the constructive proofs provide a recipe to generate quantum trajectories with desired invariant states.
One needs to specify the state and then construct $H$ and the set $\{L_k\}$ according to \cref{th:diff-inv,th:jump-inv}.
Since unitary and anti-Hermitian operators are both normal and commute with the invariant state $\rho$, there must exist a basis in which $(S_k,U_k)$ and $\rho$ are simultaneously diagonalizable.
This makes it possible to stabilize arbitrary target states along quantum trajectories, since for any quantum state $\rho$, there always exists a diffusive quantum trajectory and a quantum jump trajectory which has $\rho$ among its invariant states.
For pure states this statement can even be further strengthened.
\begin{observation}
  For any pure state $\ket{\psi}$, there always exists a diffusive or a quantum jump trajectory for which $\ket{\psi}$ is the unique attractive fixed point.
\end{observation}
Any invariant state that is not pure cannot deterministically be stabilized.
Invariant mixed states are generally not attractive.
Any small perturbation, $\rho + \epsilon \rho^{(1)}$, that breaks invariance and commutation with $H$ and $L_{k,11}$ will lead to noisy dynamics (discussed in \cref{sec:classical-noise}) and commutation will not be restored.
We provide more details on invariant states of quantum trajectories in \cref{sec:invariant-appendix} assuming specific structures of the jump operators.

\section{Structuring of the Hilbert space}
\label{sec:space-structure}
The asymptotic behavior of continuously monitored quantum systems goes beyond the invariant states discussed in \cref{sec:invariant-states}.
Quantum trajectories are susceptible to the structure of the Hilbert space imposed by $\sfl$ and are able to identify the irreducible components.
In this section we present the known results on the full characterization of the state space structure of finite-dimensional Lindblad equations, which was derived by Baumgartner and Narnhofer in Refs.~\cite{Baumgartner2008_2,Baumgartner2012}.
It provides the foundation for the ensuing analysis on quantum trajectories.

Given a master equation in Lindblad form, there is a unique partition of the Hilbert space into two orthogonal subspaces
\begin{align}
  \calh = P_\cald \calh + P_\calr \calh = \cald \oplus \calr
  \label{eq:uniqe-split},
\end{align}
where $\cald$ is the maximal decaying subspace with the defining property that it gets completely emptied out in the course of time \cite[Theorem 2]{Baumgartner2012}
\begin{align}
  \cald = \left\{\ket{\psi} \in \calh \big\vert \forall \rho, \lim_{t \to \infty}\bra{\psi}\sft^t(\rho) \ket{\psi} = 0\right\}.
  \label{eq:decaying-subspace}
\end{align}
The probability flows out of $\cald$ into its complement, the asymptotic subspace $\calr$, which is collecting and has itself no outflow.
It is defined by
\begin{align}
  \calr = \left\{\ket{\psi} \in \calh \big\vert \exists \rho \text{ s.t. } \lim_{t \to \infty}\bra{\psi}\sft^t(\rho) \ket{\psi} > 0\right\}.
  \label{eq:asymptotic-subspace}
\end{align}
The asymptotic subspace accommodates all the stationary states of the Lindblad equation.
Equivalently, $P_\cald$ and $P_\calr$ are the projectors onto the largest subspaces $\cald$ and $\calr$ such that for every density matrix $\rho$ it holds that \cite[Theorem 2]{Baumgartner2012}
\begin{align}
  \lim_{t \to \infty} \tr[P_\cald \sft^t(\rho)] &= 0,\\
  \lim_{t \to \infty} \tr[P_\calr \sft^t(\rho)] &= 1.\
\end{align}
The decomposition, \cref{eq:uniqe-split}, is the coarsest possible division into orthogonal subspaces.
There is, however, a more fine grained structure below this macroscopic level.
The decaying subspace, $\cald$, is itself composed of orthogonal subspaces.
It gets drained out monotonously and the outflow takes place in cascades, in which the probability flows down successively between basins (orthogonal subspaces) to ever lower levels until the lowest level, $\calr$, the asymptotic state space is reached, eventually catching the entire flow (see also \cref{fig:sbd-strucutre}).
The decaying subspace affects the transient dynamics and its presence can have considerable impact on the asymptotic behavior of quantum trajectories.
The most relevant structure is nevertheless the finest decomposition of the asymptotic state space because the asymptotic dynamics takes place exclusively inside of $\calr$ (see \cref{sec:no-trajs-in-D}).
The asymptotic subspace can be decomposed into a direct sum of its atomic building blocks, i.e. into subspaces that are not divisible into smaller sets of subspaces, according to \cite[Theorem 7]{Baumgartner2012}
\begin{align}
  \mathcal{R} = \bigoplus_{k=1}^K \calu_k \oplus \bigoplus_{l=1}^M \calx_l.
  \label{eq:decomposition}
\end{align}
There are two types of orthogonal subspaces.
The first part of the decomposition is unique and each subspace $\calu_k$ is itself indecomposable and thus of minimal dimension.
The second part contains only degenerate subspaces where each component, $\calx_l$, is isomorphic to a tensor product structure
\begin{align}
  \calx_l = \bigoplus_{\alpha=1}^{m(l)} \calv_{l,\alpha} \simeq \mathbb{C}^{m(l)} \otimes \calv_l, \quad \calv_l \simeq \calv_{l,\alpha} \ \forall \alpha.
  \label{eq:degenerate-subspaces}
\end{align}
Every subspace $\calv_{l,\alpha}$ is minimal but there are $m(l)$ isomorphic ones, where $m(l)$ denotes the degree of degeneracy of the subspace $\calx_l$.
The decomposition of the degenerate subspaces is unique up to unitary transformation \cite{Baumgartner2012}.
Finding the full structure into minimal orthogonal subspaces is not always obvious because symmetries might be abstract or hidden, not immediately to be inferred from physical reasoning.
\Cref{sec:sbd} is entirely devoted to this task, where we present algorithms to identify the complete decomposition.

The structure of the asymptotic state space has direct implications for the stationary states of the Lindblad equation.
We will elaborate on the consequences for quantum trajectories below (cf. \cref{sec:transition,sec:incomplete}).
In the absence of a decaying subspace, all the probability resides already inside of $\calr$ from the start and every minimal subspace comes equipped with a conserved projector
\begin{align}
  \dot{P}_{\calu_k} &= \sfl^\dagger(P_{\calu_k}) = 0,\\
  \dot{P}_{\calv_{l,\alpha}} &= \sfl^\dagger(P_{\calv_{l,\alpha}}) = 0.
\end{align}
If the decaying subspace is not the empty set, projectors on minimal subspaces are no longer conserved.
Once $\cald$ has been completely drained and all the probability is located within $\calr$, conservation is restored again, now for the infinite-time projectors
\begin{align}
  \dot{P}^\infty_{\calu_k} &= 0, &P^\infty_{\calu_k} &\equiv \lim_{t \to \infty} \left(\sft^t\right)^\dagger(P_{\calv_k}),
  \label{eq:infinite-time-projectors1}\\
  \dot{P}^\infty_{\calv_{l,\alpha}} &= 0, &P^\infty_{\calv_{l,\alpha}} &\equiv \lim_{t \to \infty} \left(\sft^t\right)^\dagger(P_{\calv_{l,\alpha}}).
  \label{eq:infinite-time-projectors2}
\end{align} 

The subspaces $\calu_k$ and $\calx_l$ are related to the presence of abelian and non-abelian symmetries respectively.
Their impact on continuous monitoring and the build-up of quantum coherences is discussed in \cref{sec:incomplete}.

Orthogonal subspaces correspond to mutually orthogonal blocks on the space of bounded linear operators $\mathfrak{B}(\calh)$.
Each minimal subspace, $\calu_k$, supports a unique stationary state $\rho_k$.
The freedom made possible by the degenerate subspaces, $\calx_l$, gives rise to a noiseless subsystem \cite{Knill2000}, with decoherence-free evolution on $\mathbb{C}^{m(l)}$ and a unique steady state on $\calv_l$.
Given the full decomposition of $\calr$, the asymptotic states of the Lindblad equation are thus of the form \cite[Theorem 2]{Baumgartner2008_1}
\begin{align}
  \lim_{t \to \infty} \left\|\rho(t) - \bigoplus_{k=1}^K \lambda_k \rho_k \oplus \bigoplus_{l=1}^M \mu_l e^{-i \tilde H_l t} \sigma_l e^{\I \tilde H_l t} \otimes \tau_l \right\| = 0,
  \label{eq:asym-states}
\end{align}
where $\rho(t)$ is the solution of the Lindblad equation \eqref{eq:me} and $\rho_k$ and $\tau_l$ are extremal stationary states supported on $\calu_k$ and $\calv_l$, respectively.
Evolution of the noiseless subsystem on $\mathbb{C}^{m(l)}$ can at most be unitary with an effective Hamiltonian $\tilde H_l$ and the states $\sigma_l$ remain preserved.
The coefficients $\lambda_k = \tr[\rho(0)P^\infty_{\calu_k}]$ and $\mu_l = \tr[\rho(0)P^\infty_{\calv_l}]$ depend nontrivially on the initial state if the decaying subspace is not the empty set (cf. \cref{eq:infinite-time-projectors1,eq:infinite-time-projectors2}).
They satisfy $\sum_{k=1}^K \lambda_k + \sum_{l=1}^M \mu_l = 1$.

Whenever it is not necessary to explicitly distinguish between the two types of collecting subspaces, we denote by $\calq_j$ a general minimal orthogonal subspace such that the decomposition can be expressed as
\begin{align}
\calr = \bigoplus_{j=1}^N \calq_j = \bigoplus_{k=1}^K \calu_k \oplus \bigoplus_{l=1}^M \bigoplus_{\alpha=1}^{m(l)} \calv_{l,\alpha},
\label{eq:cala}
\end{align}
and $N=K+\sum_{l=1}^M m(l)$.
We accordingly define the asymptotic effective weights
\begin{align}
  \lim_{t\to \infty} \tr[\rho(t) P_{\calq_j}]
  = \tr[\rho(0) P^\infty_{\calq_j}]
  \equiv |\calq^\infty_j|^2,
  \label{eq:effective-weights}
\end{align}
comprising the coefficients with $|\calq^\infty_j|^2 \in \{\lambda_k,\mu_l\}$.

\section{Measurement-induced complete localization transition}
\label{sec:transition}
In this section we show that individual quantum trajectories undergo irreversible localization transitions in Hilbert space, where the quantum system gets spontaneously trapped in one of the irreducible subspaces and subsequent evolution takes place exclusively in this restricted region.
For the sake of clarity, we present calculations for the case where there is no decay, $\cald = \emptyset$, and it therefore holds that $\calh = \calr$.
As shown in \cref{sec:decaying-subspace}, the general case then requires only a slight modification.

Consider an otherwise closed, quantum system subject to continuous measurement that evolves along the diffusive \Ito stochastic differential equation \eqref{eq:homodyne-sme}.
Then the Hilbert space may always be bipartitioned into
\begin{align}
  \calr = P_\calq \calr + P_\calp \calr = \calq \oplus \calp,
\end{align}
where $\calq$ is a minimal subspace with projector $P_\calq$, supporting a unique stationary state of the Lindblad equation and $\calp$ is the orthogonal complement, with projector $P_\calp$ (see \cref{sec:space-structure}).
The probability for the system to be found inside the minimal subspace $\calq$ at time $t$ is then given by the overlap
\begin{align}
  |\calq(t)|^2 \equiv \tr[\rho_\xi P_\calq].
\end{align}
We analyze quantum jumps in \cref{sec:complete-jumps}.
Since total probability is conserved, it holds that $|\calq(t)|^2 + |\calp(t)|^2 = 1$, where $|\calp(t)|^2 \equiv \tr[\rho_\xi P_\calp]$ is the overlap with the orthogonal complement.
The evolution equation for the overlap then follows by \Ito calculus as the stochastic differential
\begin{align}
  \dd{\left(|\calq(t)|^2 \right)}&= \dd{\left(\tr[\rho_\xi(t)P_\calq]\right)}\notag\\ 
  =& \sum_k \bigg(|\calp(t)|^2\tr[\rho_\xi(t)(L_k+L_k^\dagger)P_\calq]\notag \\
  -&|\calq(t)|^2 \tr[\rho_\xi(t)(L_k+L_k^\dagger)P_\calp]\bigg) \dd{W_k(t)}.
  \label{eq:calq}
\end{align}
The overlap $|\calq(t)|^2$ undergoes a free, state-dependent Brownian motion that depends on the microscopic details of the involved subspaces.
The diffusion process may be equivalently described by the probability density $P(|\calq|^2,t)$ that evolves according to the corresponding Fokker--Planck equation
\begin{align}
  \pdv{t} P(|\calq|^2,t) 
  = \pdv[2]{(|\calq|^2)} D(|\calq|^2) P(|\calq|^2,t),
  \label{eq:fpe}
\end{align}
with state-dependent diffusion coefficient 
\begin{align}
  D(|\calq|^2) 
  &= \frac{1}{2}\bigg[\sum_k\bigg(|\calp(t)|^2\tr[\rho_\xi(t)(L_k+L_k^\dagger)P_\calq] \\
  -&|\calq(t)|^2 \tr[\rho_\xi(t)(L_k+L_k^\dagger)P_\calp]\bigg)\bigg]^2.
\end{align}
Our aim is to derive the steady state probability distribution for $|\calq(t)|^2$.
This may be done by solving for the stationary solutions $P^\m{s}(|\calq|^2)$ of the Fokker--Planck equation.
It is, however, more convenient to approach the problem from the trajectory description given in \cref{eq:calq}.
We first note that by \Ito rules the ensemble average of the overlap in each of the subspaces does not change in time 
\begin{align}
  \mathbb{E}\left[\dv{t} \left(|\calq(t)|^2\right)\right] = \mathbb{E}\left[\dv{t} \left(|\calp(t)|^2\right)\right] = 0,
\end{align}
so there is no drift, which is consistent with the fact that projectors on orthogonal subspaces are conserved quantities along Lindbladian dynamics
\begin{align}
  \sfl^\dagger(P_\calq) = \sfl^\dagger(P_\calp) = 0.
\end{align}
On the ensemble level, the overlap with each subspace thus remains equal to its initial value
\begin{align}
  \mathbb{E}\left[|\calq(t)|^2\right] = |\calq(0)|^2, \qquad \mathbb{E}\left[|\calp(t)|^2\right] = |\calp(0)|^2.
  \label{eq:constraint}
\end{align}
Second, we may appreciate that the diffusion process \cref{eq:calq} admits two stable fixed points at the boundaries.
If the system happens to find itself entirely within the minimal subspace ($|\calq(t)|^2 = 1$), the complement will be completely empty ($|\calp(t)|^2 = 0$) and the system remains trapped in $\calq$ with probability one.
Conversely, once the trajectory loses support on the minimal subspace ($|\calq(t)|^2 = 0$), the system will be contained entirely within the complement ($|\calp(t)|^2 = 1$) and the evolution stops, $\dd{\left(|\calq(t)|^2 \right)} = 0$.
The two fixed points are furthermore attractive.
Indeed, the \Ito process \cref{eq:calq} is a local bounded martingale \cite{Oksendal2003,Kuo2005,Roldan2023} and, in the absence of any additional fixed points, the martingale convergence theorem guarantees that every trajectory will eventually converge to one of the boundaries.
Further fixed points can only lie inside the interval $|\calq|^2 \in (0,1)$ and are discussed in detail in \cref{sec:incomplete}.
These must satisfy
\begin{align}
  & |\calp(t)|^2\tr[\rho_\xi(t)(L_k+L_k^\dagger)P_\calq] \notag\\
  &= |\calq(t)|^2 \tr[\rho_\xi(t)(L_k+L_k^\dagger)P_\calp], \ \forall t > T, \ \forall k.
  \label{eq:incomplete-diff}
\end{align}
For generic operators $H$ and $L_k$, the terms on both sides of \cref{eq:incomplete-diff} are independent and the two fixed points $|\calq(t)|^2 \in \{0,1\}$ are the only two stationary solutions of \cref{eq:calq}.
In this case, there is a unique solution of the Fokker--Planck equation compatible with the constraint \cref{eq:constraint}; it is
\begin{align}
  P^\m{s}(|\calq|^2)
  = |\calq(0)|^2\delta(|\calq|^2-1) + (1-|\calq(0)|^2)\delta(|\calq|^2),
  \label{eq:steady-state}
\end{align}
where $\delta(x)$ denotes the Dirac-delta function.
Trajectories asymptotically accumulate at the boundaries with probability given by the initial overlap with the corresponding subspace.

In the presence of multiple minimal subspaces and in the absence of stationary points other than $|\calq|^2\in\{0,1\}$, this procedure may be carried through iteratively until the lowest level of the decomposition is reached.
Indeed, define the first level of the decomposition by $\calq_1 \equiv \calq$ and $\calp_1 \equiv \calp$.
A trajectory localizing in $\calq_1$ has reached an atomic structure and cannot experience more confinement.
Within the complement, further partitions, $\calp_1 = \calq_2 + \calp_2$, may be possible where, again, $\calq_2$ is minimal and $\calp_2$ is the new complement.
The line of arguments following \cref{eq:calq} applies here again and, after $n-1$ steps, results in 
\begin{align}
  \calp_{n-1} = \calq_n + \calp_n,
\end{align}
with $\calq_n$ and $\calp_n$ the $n$th minimal subspace and complement respectively.
This procedure is repeated until the the lowest level of the decomposition is reached, such that eventually, every trajectory is located inside a minimal orthogonal subspace.
The joint steady state probability distribution thus becomes
\begin{align}
  P^\m{s}(|\calq_1|^2,|\calq_2|^2,\ldots)
  = \sum_j |\calq_j(0)|^2\delta(|\calq_j|^2-1).
  \label{eq:steady-state-full-diff}
\end{align}

\subsection{Quantum jumps}
\label{sec:complete-jumps}
For quantum jump trajectories we can follow the same steps as for the diffusive case.
To keep the description concise, we just provide the main results.
The probability $|\calq(t)|^2 \equiv \tr[\rho_J P_\calq]$ to be in the subspace $\calq$ evolves in time according to the stochastic differential equation
\begin{align}
  \dd{\left(|\calq(t)|^2 \right)}
  =& \sum_k \left(|\calq(t)|^2 \langle L_k^\dagger L_k\rangle - \tr[\rho_J (L_k^\dagger L_k) P_\calq]\right)\dd{t} \notag\\
    &+ \left(\frac{\tr[\rho_J (L_k^\dagger L_k) P_\calq]}{\langle L_k^\dagger L_k\rangle}-|\calq(t)|^2\right) \dd{N_k},
  \label{eq:calq_jump}
\end{align}
where $\langle L_k^\dagger L_k\rangle = \tr[\rho_J (L_k^\dagger L_k)]$.
The above equation is a state-dependent Poisson process with an additional drift term.
Note that the drift and jump terms always have opposite signs.
Deterministic inflow of probability can thus only be followed by an outgoing jump that decreases again the overlap $|\calq(t)|^2$, and vice versa.
The ongoing competition between drift and jumps stops once the trajectory hits upon a stable invariant fixed point.
By inspection, two fixed points located at the boundaries are readily identified.
Once the system is entirely inside the minimal subspace ($|\calq(t)|^2 = 1$), the complement will be empty ($|\calp(t)|^2 = 0$), whereupon $\langle L_k^\dagger L_k\rangle = \tr[\rho_J (L_k^\dagger L_k)P_\calq]$ and both the drift and the jump term in \cref{eq:calq_jump} go to zero and the stochastic process terminates.
Similarly, if the system reaches the opposite boundary and finds itself entirely inside of the complement ($|\calp(t)|^2 = 1$), the minimal subspace is left vacant ($|\calq(t)|^2 = 0$), where it holds that $\tr[\rho_J (L_k^\dagger L_k) P_\calq] = 0$ and the time evolution stops.

Computing the ensemble average in \cref{eq:calq_jump}, we again find with, $\mathbb{E}[\dd{N}] = \langle L_k^\dagger L_k\rangle \dd{t}$, the relations 
\begin{align}
  \mathbb{E}\left[\dv{t} \left(|\calq(t)|^2\right)\right] = \mathbb{E}\left[\dv{t} \left(|\calp(t)|^2\right)\right] = 0,
\intertext{and}
  \mathbb{E}\left[|\calq(t)|^2\right] = |\calq(0)|^2, \qquad \mathbb{E}\left[|\calp(t)|^2\right] = |\calp(0)|^2.
\end{align}
The stochastic process \cref{eq:calq_jump} is a local bounded martingale \footnote{Define the compensated jump processes $\tilde N_k(t) = N_k(t) - \int_0^t \dd{s}\langle L^\dagger_kL_k\rangle$ which is a martingale \cite{Kummerer2004}. The stochastic master equation becomes, $\dd{\rho_J} = \sfl(\rho_J) + \sum_k (L_k\rho_J L^\dagger_k/\langle L^\dagger_kL_k\rangle-\rho_J)\dd{\tilde N_k}$. With this choice, the stochastic differential \cref{eq:calq_jump} results in $\dd{(|\calq(t)|^2)} = \sum_k (\Tr[\rho_J (L^\dagger_kL_k)P_\calq]/\langle L^\dagger_kL_k\rangle-|\calq(t)|^2)\dd{\tilde N_k}$, where $\mathbb{E}[\dd{\tilde N_k}] = 0$, which is a martingale.} and, by virtue of the martingale convergence theorem, converges to one of the two fixed points.
Further fixed points lying in between the boundaries $|\calq|^2 \in (0,1)$ are treated in \cref{sec:incomplete} and need to satisfy
\begin{align}
  |\calq(t)|^2 \langle L_k^\dagger L_k\rangle = \tr[\rho_J (L_k^\dagger L_k) P_\calq], \ \forall t > T, \ \forall k.
  \label{eq:incomplete-jump}
\end{align}
Provided this is not the case, the steady state probability distribution is again given by
\begin{align}
  P^\m{s}(|\calq|^2) 
  = |\calq(0)|^2\delta(|\calq|^2-1) + (1-|\calq(0)|^2)\delta(|\calq|^2).
  \label{eq:steady-state-jump}
\end{align}
Carrying this procedure through iteratively until hitting the lowest level of the decomposition, finally gives the unique solution
\begin{align}
  P^\m{s}(|\calq_1|^2,|\calq_2|^2,\ldots)
  = \sum_j |\calq_j(0)|^2\delta(|\calq_j|^2-1).
  \label{eq:steady-state-full-jump}
\end{align}
Although the steady state probability densities \cref{eq:steady-state-full-diff,eq:steady-state-full-jump} have the same form for both unravelings, this does not mean that their distributions must coincide.
In fact, dramatic differences may occur where localization transitions take place only in the diffusive unraveling while the quantum jump trajectory remains incapable of discerning distinct subspaces (cf. \cref{th:incomplete-diff,th:incomplete-jump} in \cref{sec:incomplete}).

\subsection{Decaying subspace}
\label{sec:decaying-subspace}
The steady state distributions (\cref{eq:steady-state-full-diff,eq:steady-state-full-jump}) of the Fokker-Planck equation have been derived assuming that the decaying subspace is empty.
In the presence of decay, $\cald \neq \emptyset$, any asymptotic quantum trajectory can still reside only in the asymptotic subspace $\calr$ (see \cref{sec:no-trajs-in-D}).
However, probability may flow down from $\cald$ into $\calr$ and the overlap with a minimal subspace, $\mathbb{E}[|\calq(t)|^2]$, is no longer conserved in general.
Conservation of probability in minimal subspaces is restored again once $\cald$ has been drained completely (cf. \cref{eq:infinite-time-projectors1,eq:infinite-time-projectors2}).
We accordingly define the effective asymptotic overlap
\begin{align} 
  |\calq^\infty|^2 \equiv 
  \lim_{t \to \infty} \mathbb{E}[|\calq(t)|^2]
\end{align}
that remains invariant after the transient time of decay has passed.
The previous analysis carries through without change except that the initial overlap $|\calq(0)|^2$ is now replaced by $|\calq^\infty|^2$ which may be considered as an effective initial overlap, since 
$|\calq^\infty|^2 = \tr[\rho(0)P^\infty_\calq]$ (see \cref{eq:effective-weights}).
In particular, the steady state probability distribution \cref{eq:steady-state-full-diff,eq:steady-state-full-jump} for the full decomposition of $\calr$ into minimal subspaces $\calq_j$ becomes (cf. \cref{eq:cala})
\begin{align}
  P^\m{s}\left(|\calq_1|^2,|\calq_2|^2,\ldots\right) 
  = \sum_j |\calq^\infty_j|^2\delta(|\calq_j|^2-1),
\end{align}
where $\sum_j |\calq_j^\infty|^2 = 1$.

\subsection{Decoherence-free subspace}
\label{sec:dfs}
So far we have been concerned with transitions between orthogonal subspaces.
Remarkably, the process of subspace selection can be derived solely based on the structure of the Hilbert space presented in \cref{sec:space-structure} and is, in particular, independent of the size or microscopic details of the individual subspaces.
Once localization has stopped, the system will still undergo stochastic dynamics, strongly dependent on the internal details of the corner of the Hilbert space it is restricted to.
Although the localization process treats all subspaces equally, from a physical standpoint some subspaces may have more desirable properties than others.
In this regard, decoherence-free subspaces turn out to play an exceptional role, which makes it worthwhile to study them separately here.

A decoherence-free subspace is a subspace of the Hilbert space that is effectively decoupled from the influence of the environment, with the defining property that the dissipator in \cref{eq:me} vanishes \cite{Lidar1998}.
Decoherence-free states are thus composed of simultaneous eigenstates of both the Hamiltonian and all the jump operators
\begin{align}
    H \ket{q_n} = \omega_n \ket{q_n}, \quad L_k \ket{q_n} = c_{k,n} \ket{q_n},
  \label{eq:dfs}
\end{align}
where $\omega_n$ are eigenenergies of the Hamiltonian and $c_{k,n}$ are arbitrary complex numbers.
When $c_{k,n} = c_{k,m},\ \forall k$, we say that the states $\ket{q_n}$ and $\ket{q_m}$ belong to the same decoherence-free subspace.
The entirety of these states, the set $\{\ket{q_n}\}$, constitutes the whole collection of decoherence-free subspaces.
First, we can appreciate that a decoherence-free subspace remains decoherence-free even along individual quantum trajectories, \cref{eq:jump-sme,eq:homodyne-sme}, since
\begin{align}
    \dd{\rho^\m{DFS}_\m{c}} = -\I[H,\rho^\m{DFS}_\m{c}],
\end{align}
where we have used the shorthand $c = (J,\xi)$.
This is indeed a very useful feature.
It implies that once a trajectory has decided on the decoherence-free subspace, the entire system will henceforth undergo unitary evolution.
This is to be contrasted with the average dynamics given by the Lindblad equation where, typically, only a fraction of the initial state receives such protection.
This purification of state space properties can be observed in the example in \cref{sec:ring}.
Decoherence-free subspaces belong to the invariant manifold $\calr$ and, as shown below, can always be identified by individual trajectories.

If a decoherence-free subspace exists it is always possible to perform a bipartition of the asymptotic state space according to \cite{Schmolke2024}
\begin{align}
  \calr = \calq \oplus \calp,
  \label{eq:dfs-complement-bipartition}
\end{align}
where all decoherence-free subspaces are collected in $\calq$, and the remainder defines the complement $\calp^\perp=\calq$, which by definition is not decoherence-free.
The temporal change of the probability for the system to be found in $\calq$, reduces to (cf. \cref{eq:calq})
\begin{align}
  \dd{(|\calq(t)|^2)}
  =& \sum_k 
    2|\calq(t)|^2\bigg((1-|\calq(t)|^2)C_k(t) \notag\\
    &-\tr[\rho_\xi(t)(L_k+L_k^\dagger)P_\calp]\bigg)\dd{W_k}.
    \label{eq:dfs-complement}
\end{align}
The above equation follows from the definition of the decoherence-free states (\cref{eq:dfs}) applied to a general mixed state $\rho_\xi(t) = \sum_m u_m \dyad{\Psi_m(t)}$, with $\sum_m u_m = 1$ and defining
\begin{align}
\tr[\rho_\xi(t)(L_k+L_k^\dagger)P_\calq] 
&= 2\sum_{m,n} u_m \m{Re}(c_{k,n})|\braket{q_n}{\Psi_m(t)}|^2\\
&\equiv 2|\calq(t)|^2C_k(t).
\end{align}
A quantum trajectory therefore either goes into $\calq$ with ensuing decoherence-free evolution, or into the complement $\calp$ followed by stochastic behavior.
In the former case, a further bipartition of the remaining space can be made, yielding 
\begin{align}
    \dd{(|\calq_1(t)|^2)} =& 2 |\calq_1(t)|^2 \left(1-|\calq_1(t)|^2\right)\notag\\
    &\times \sum_k [C_{k,1}(t)- C_{k,2}(t)]\dd{W_k}.
    \label{eq:dq_12} 
\end{align}
Here, the probability to be in the decoherence-free subspace $\calq_1$ is denoted by $|\calq_1(t)|^2 = \tr[\rho_\xi(t) P_{\calq_1}]$.
Total probability must be conserved, i.e. $|\calq_1(t)|^2 + |\calq_2(t)|^2 = 1$ with $\tr[\rho_\xi(t)(L_k+L^\dagger_k)P_{\calq_x}] \equiv 2C_{k,x}(t) |\calq_x(t)|^2$, where $x = 1,2$.
The line of arguments following \cref{eq:calq} still applies and ensures that, eventually, one of the two subspaces $\calq_1$ or $\calq_2$ is selected.
This procedure can be repeated until no further partition into distinct decoherence-free subspaces is possible.
Hence, in the presence of multiple distinct decoherence-free subspaces, only one of them is ever selected, in which the system will then remain indefinitely.

Using the shorthand $x \equiv |\calq_1(t)|^2$, the corresponding Fokker--Planck equation can be conveniently expressed as
\begin{align}
  \pdv{t} P(x,t) 
  =& 2\sum_k [C_{k,1}(t)-C_{k,2}(t)]^2 \notag\\
  &\times \pdv[2]{x}\left(x^2(1-x)^2 P(x,t)\right).
\end{align}
In contrast to the general case (\cref{eq:fpe}), the above equation is now closed in the variable $x$.

\subsubsection{Quantum jumps}
The procedure for the quantum jump unraveling is analogous.
From the bipartition \cref{eq:dfs-complement-bipartition}, it follows that the probability for the system to be found in the decoherence-free subspace (\cref{eq:calq_jump}) reduces to
\begin{align}
  \dd{(|\calq(t)|^2)} =& \sum_k 
  |\calq(t)|^2 \bigg[\left(\langle L^\dagger_kL_k\rangle - C^2_k(t)\right)\dd{t}\\
  &+\left(\frac{C^2_k(t)}{\langle L^\dagger_kL_k\rangle-1}\right)\dd{N_k}\bigg],
\end{align}
where we have defined $\tr[\rho_J(t)(L^\dagger_k L_k)P_\calq] \equiv C^2_k(t) |\calq(t)|^2$.
Clearly, when the system goes entirely into the complement with $|\calp(t)|^2=1$, we must have $|\calq(t)|^2 = 0$ and the evolution stops.
When the system evolves into the decoherence-free subspace ($|\calq(t)|^2 = 1$), this implies $\langle L^\dagger_k L_k\rangle = C^2_k(t)$ and thus $\dd{(|\calq(t)|^2)}=0$.
For multiple decoherence-free subspaces we may further decompose $\calq$, yielding the analog of \cref{eq:dq_12}
\begin{align}
  \dd{(|\calq_1(t)|^2)}
  =& |\calq_1(t)|^2\left(1-|\calq_1(t)|^2\right)\notag\\
  & \times \sum_k \left[C^2_{k,1}(t)-C^2_{k,2}(t)\right]\left[\dd{t}-\frac{\dd{N_k}}{\langle L^\dagger_kL_k\rangle}\right].
  \label{eq:dq_12-jump}
\end{align}
Here we can already catch a glimpse of the different symmetry properties of diffusion and jumps.
The diffusive trajectory is able to distinguish between distinct decoherence-free subspaces while the jump trajectory is agnostic to arbitrary phase differences.
We will elaborate on the invariance properties in \cref{sec:incomplete}.

\section{Incomplete localization}
\label{sec:incomplete}
We have observed that under generic conditions (the absence of further nontrivial symmetries) quantum systems subject to continuous monitoring asymptotically converge to one of the minimal invariant subspaces inside the asymptotic subspace $\calr$.
Starting from an arbitrarily delocalized initial state, a quantum trajectory will thus become increasingly confined until it eventually hits upon an atomic component of the Hilbert space structure, a minimal orthogonal subspace, where further confinement is no longer possible and localization is thus considered \emph{complete}.
In this section we discuss the situation where individual trajectories are unable to distinguish between orthogonal subspaces and localization comes to a halt prematurely.
Localization is thus considered \emph{incomplete} in the sense that trajectories are unable to enter entirely an indecomposable substructure but remain delocalized over at least two minimal subspaces instead.
The localization properties of quantum trajectories have profound consequences for their ergodic behavior (cf. \cref{sec:erg-theorem}), they are related to the build-up of quantum correlations between degenerate symmetry sectors (cf. \cref{eq:noiseless-subsystem,sec:scar states}), and influence the purification of individual realizations (see below).
Remarkably, diffusive quantum trajectories are more sensitive to the Hilbert space structure than quantum jumps and are able to discern the most fine-grained nontrivial, nondegenerate structures, with striking physical consequences.

Asymptotic localization terminates prematurely for diffusive trajectories between two orthogonal subspaces $\calq$ and $\calp$, not necessarily minimal, if it holds (cf. \cref{eq:calq,eq:incomplete-diff})
\begin{align} 
  \tr[\tilde\rho_{\xi,\calq}(L_k+L_k^\dagger)_\calq]
 = \tr[\tilde\rho_{\xi,\calp}(L_k+L_k^\dagger)_\calp],
 \label{eq:freeze-diff}
\end{align}
for all times $t > T$ and for all $k$.
Here, we have defined the effective states and operators 
\begin{align}
  \tilde \rho_{\xi,\cala} &\equiv 
  \frac{P_\cala\rho_\xi P_\cala}{\tr[P_\cala \rho_\xi]},\\
  (L_k+L^\dagger_k)_{\cala} &\equiv 
  P_\cala (L_k+L_k^\dagger) P_\cala,
\end{align}
with $\cala = \calq,\calp$.
The following theorem states the necessary and sufficient conditions that prevent further localization of diffusive quantum trajectories.
We consider the situation where the decaying subspace has just been emptied and the trajectory is entirely supported on $\calr$ for the first time.

\begin{theorem}[Incomplete localization, quantum diffusion]
\label{th:incomplete-diff}
Let $\{\calq_i\}_{i\in I}$ be the minimal orthogonal subspaces $\calq_i \in \calr$ that support the diffusive quantum trajectory \cref{eq:homodyne-sme}, where $I$ is the corresponding index set.
Then, there is no subspace selection on the trajectory level between any of the elements of an arbitrary subset $\{\calq_j\}_{j \in J}$, $J \subseteq I$, if and only if at least one of the following holds $\forall k$ and $\forall j,l \in J$.
\begin{enumerate}[(i)]
\item Independent trajectories:
\begin{align} 
    (L_k+L^\dagger_k)_{\calq_j} = z_k\mathds{1}_{d_{\calq_j}}.
    \label{eq:pur-diff}
\end{align}
\item Unitarily equivalent trajectories:
\begin{align}
    u_j(H_{\calq_j})u^\dagger_j &= u_l(H_{\calq_l}) u^\dagger_l,\\
    u_j(L_{k,\calq_j})u^\dagger_j &= u_l(L_{k,\calq_l})u^\dagger_l,
    \label{eq:u-diff}
\end{align}
\end{enumerate}
with $z_k \in \mathbb{R}$ and unitary matrices $u_j,u_l$.
\end{theorem}
Sufficiency can be observed by inserting \cref{eq:pur-diff,eq:u-diff}  into \cref{eq:calq}.
The full proof is given in \cref{sec:proof-no-selection-diff}.

Analogously, asymptotic localization is incomplete for a quantum jump trajectory between two orthogonal subspaces $\calq$ and $\calp$, not necessarily minimal, if (cf. \cref{eq:calq_jump,eq:incomplete-jump})
\begin{align}
    \tr[\tilde\rho_{J,\calq}(L_k^\dagger L_k)_\calq]
    = \tr[\tilde\rho_{J,\calp}(L_k^\dagger L_k)_\calp]
    \label{eq:freeze-jump}
\end{align}
for all times $t > T$ and for all $k$, where
\begin{align}
  \tilde \rho_{J,\cala} &\equiv 
  \frac{P_\cala\rho_J P_\cala}{\tr[P_\cala \rho_J]},\\
  (L^\dagger_k L_k)_{\cala} &\equiv 
  P_\cala (L_k^\dagger L_k) P_\cala,
\end{align}
with $\cala = \calq,\calp$.
To proceed, we need to assume validity of the following conjecture.
\begin{conjecture}\label{pur}
  In a minimal orthogonal subspace $\calq$, absence of purification of a mixed state quantum jump trajectory $\rho_J$ implies
  \begin{align}
    (L^\dagger_kL_k) = z_k \mathds{1}_{d_\calq}, \ \forall k,
  \end{align}
  where $z_k >0$.
\end{conjecture}
Given its validity the following theorem states the necessary and sufficient conditions that prevent further localization of quantum jump trajectories.

\begin{theorem}[Incomplete localization, quantum jumps, conditional on \rconj{pur}]
\label{th:incomplete-jump}
Let $\{\calq_i\}_{i\in I}$ be the minimal orthogonal subspaces $\calq_i \in \calr$ that support the quantum jump trajectory \cref{eq:jump-sme}, where $I$ is the corresponding index set.
Then, provided \rconj{pur} holds, there is no subspace selection on the trajectory level between any of the elements of an arbitrary subset $\{\calq_j\}_{j \in J}$, $J \subseteq I$, if and only if at least one of the following holds $\forall k$ and $\forall j,l \in J$.
\begin{enumerate}[(i)]
    \item Independent trajectories:
    \begin{align} 
        (L^\dagger_kL_k)_{\calq_j} = z_k\mathds{1}_{d_{\calq_j}}.
        \label{eq:pur-jumps}
    \end{align}
    \item[(iia)] Unitarily equivalent trajectories: 
    \begin{align} 
        u_j(H_{\calq_j})u^\dagger_j &= u_l(H_{\calq_l}) u^\dagger_l,\\
        \exp(\I\phi_{k,j})u_j(L_{k,\calq_j})u_j^\dagger &= \exp(\I\phi_{k,l})u_l(L_{k,\calq_l})u^\dagger_l.
        \label{eq:u-jumps}
    \end{align}
    \item[(iib)] Unitarily equivalent trajectories:
    \begin{align}
      &u_j \left(H_{\calq_j}\right)u_j^\dag = \bigoplus_\alpha \left(h_{\calq_{\alpha}} + r_{j,\alpha} \mathds{1}_{d_\alpha}\right),
      \label{eq:block-H}\\
      &\exp(\I\phi_{k,j})u_j(L_{k,\calq_j})u_j^\dagger = \exp(\I\phi_{k,l})u_l(L_{k,\calq_l})u^\dagger_l,
      \label{eq:sim-L}\\
      &P_{j,\beta}\left(\sum_k (L^\dag_kL_k)_{\calq_j}\right)P_{j,\gamma} = 0, \ \forall \beta\neq \gamma
      \label{eq:block-sum}\\
      &P_{j,\beta}\left(L_{k,\calq_j}\right)P_{j,\alpha} = 0 \lor P_{j,\gamma}\left(L_{k,\calq_j}\right)P_{j,\alpha} = 0, \ \forall \alpha, \beta \neq \gamma
      \label{eq:no-coherence-jumps},
    \end{align}
\end{enumerate}
with $z_k \ge 0$ and $\phi_{k,j},\phi_{k,l},r_{j,\alpha} \in \mathbb{R}$, unitary matrices $u_j,u_l$ and orthogonal projectors $P_{j,\alpha}$.
\end{theorem}
The sufficiency follows immediately by insertion of \cref{eq:pur-jumps,eq:u-jumps} into \cref{eq:calq_jump}.
The full proof is given in \cref{sec:proof-no-selection-jumps}.
It is straightforwardly shown that the defining property of getting stuck between subspaces (\cref{eq:freeze-diff,eq:freeze-jump}) directly implies independent, valid trajectories $\tilde \rho_{\m{c},\calq_j}$ in each of the subspaces (see \cref{eq:valid-trajectories-diff,eq:valid-trajectories-jump}) \footnote{The density matrix necessarily must have full support on the composite subspace $\calq \oplus \calp = \bigoplus_j \calq_j$ in order to have a proper notion of decidability between subspaces.
We show in \cref{sec:no-trajs-in-D} that every quantum trajectory must terminate in $\calr$, such that the Hilbert space effectively contracts to $\calh \to \calr$.
Incomplete localization can thus occur only between orthogonal subspaces inside $\calr$ which must moreover belong to the decomposition induced by the Lindblad equation (cf. \cref{sec:incomplete-composite}).
Otherwise, the state might still have partial overlap with $\cald$ and we are not actually comparing evolutions in $\calq$ and $\calp$.
Therefore, even if the operators have the form \cref{eq:u-diff,eq:u-jumps} the evolutions must not be unitarily equivalent when there is still a downflow from the decaying subspace, breaking this symmetry (see \cref{sec:two-qubits} where the trajectory reaches full support on $\calr$ only once it has already localized completely).}.
The incomplete localization \cref{th:incomplete-diff,th:incomplete-jump} are the natural generalization of incomplete localization for two minimal subspaces.
The corresponding \cref{th:mincomplete-diff,th:mincomplete-jump} are presented in \cref{sec:proof-no-selection-diff,sec:proof-no-selection-jumps} and may be help in understanding the more general scenario here.
Note that if \rconj{pur} were not true, then case (i) of \cref{th:incomplete-jump} would have to be replaced by \cref{eq:pur-th-jump} which would generally prevent the derivation of trajectory-independent structural properties on the operators $H$ and $L_k$.
It would moreover, in principle, allow for incomplete localization between subspaces with different purification properties (purification in $\calq_j$ but no purification in $\calq_l$).
We suspect that the invalidity of \rconj{pur} would contradict minimality of the corresponding subspace (cf. \cref{sec:absence-of-purification-in-a-minimal-subspace}).

Incomplete localization can be divided into the same two main cases for both unravelings, (i) and (ii), that generally correspond to very different physical scenarios.
In short, (i) means trivial state-independent measurement backaction which effectively removes the non-linearity in the evolution of the quantum trajectory and thus prevents localization transitions.
Unitary equivalence (ii) on the other hand, basically corresponds to multiple copies of the same evolution inside different subspaces of equal dimension which makes them practically indistinguishable from the point of view of the trajectory.
In general, both unravelings have some common properties but in certain cases may exhibit diverging behavior.
In what follows, we discuss the different cases in more detail.

\subsection{Independent trajectories}
For case (i) of \cref{th:incomplete-diff,th:incomplete-jump} only the structure of the jump operators is relevant.
Both $(L_k+L^\dagger_k)_{\calq_j}$ and $(L^\dagger_kL_k)_{\calq_j}$ determine the measurement backaction of the quantum trajectory which, if trivial, allows for entirely independent evolutions in each minimal subspace with however the same noise.
Independence is meant here in the sense that the dimensions $d_{\calq_j}$ of the subspaces and the Hamiltonians $H_{\calq_j}$ are unconstrained and may thus be arbitrary.

The conditions \eqref{eq:pur-diff} and \eqref{eq:pur-jumps} are intimately related to the unique condition that prevents purification and gives rise to independent, physically valid quantum trajectories in both subspaces driven by classical noise.
The purification theorem for diffusive trajectories was derived by Barchielli and Gregoratti in Ref.~\cite[Theorem 5.12]{Barchielli2009} which we repeat here for convenience.
A diffusive trajectory purifies asymptotically if there does not exist a projector $P_t$ such that 
\begin{align}
P_t(L_k+L_k^\dagger)P_t = z_k(t) P_t,
\label{eq:pur-th-diff}
\end{align}
with $z_k(t) \in \mathbb{R}$ (see also \cref{sec:proof-no-selection-diff}).
The equivalent statement for quantum jumps was derived by Maassen and Kümmerer in Ref.~\cite[Theorem 1]{Maassen2006} and states
\begin{align}
  P_t(L^\dagger_kL_k)P_t = z_k(t) P_t,
  \label{eq:pur-th-jump}
\end{align}
with $z_k(t) \le 0$ (see also \cref{sec:proof-no-selection-jumps}).
Crucially, the projector $P_t$ acts on the support of the state $\rho_\m{c}(t)$.
Conditions \eqref{eq:pur-th-diff} and \eqref{eq:pur-th-jump} imply that the effective measurement action of the operators $L_k$ restricted to the image of $P_t$ are proportional to a shifted anti-Hermitian and unitary operator respectively.
The exceptional role of anti-Hermitian and unitary Lindblad jump operators was already observed in \cref{th:diff-inv,th:jump-inv}, when dealing with the invariant states in \cref{sec:invariant-states}.
If a quantum trajectory hits upon such a subspace the detector effectively decorrelates from the system and the measurement update in \cref{eq:homodyne-sme,eq:jump-sme} becomes state-independent while the noise transitions from quantum to classical (see also \cref{sec:classical-noise}) or vanishes altogether when the system undergoes unitary dynamics in a decoherence-free subspace (cf. \cref{sec:dfs}).
To prevent localization, purification must not take place in both subspaces individually.
Notably, the trivial action of the measurement needs to be proportional to the same number $z_k$ in both subspaces, otherwise the subspaces become distinguishable and localization takes place.

\subsection{Unitarily equivalent trajectories}
Case (ii) forces the operators $H_{\calq_j}$ and $L_{k,\calq_j}$ to be related such that the corresponding subspaces cannot be discerned on the trajectory level.
At most, this entails physically valid, unitarily related, pure state quantum trajectories 
\begin{align}
  u_j(\tilde \rho_{\m{c},\calq_j}) u^\dag_j = u_l\left(\tilde \rho_{\m{c},\calq_l}\right)u^\dagger_l,
\end{align}
in each of the minimal subspaces (see \cref{eq:valid-trajectories-diff,eq:valid-trajectories-jump}).
Here, the measurement update remains however state-dependent and the noise thus quantum, leading to purification locally, inside of each $\calq_j$ but not globally on the combined subspace $\bigoplus_j \calq_j$ (see \cref{sec:unitary-equivalence-diff,sec:unitary-equivalence-jump}).
Purification thus implies localization but the converse need not necessarily be true.

Unitary equivalence of evolutions on the trajectory level implies that all subspaces $\calq_j$ have the same dimension and between any two $\calq_j$ and $\calq_l$ there exists a unitary, $U$, acting on the combined subspace $\calq_j \oplus \calq_l$ with (see \cref{eq:unitary-diff,eq:unitary-jump})
\begin{align}
    U = 
    \begin{pmatrix}
        0 & u_{j,l}\\ u^\dagger_{j,l} & 0
    \end{pmatrix},
    \qquad U^2 = \mathds{1},
\end{align}
where $u_{j,l} \equiv u^\dag_j u_l$ and 
\begin{align}
    U \rho_\m{c} U^\dagger
    = 
    \begin{pmatrix}
        0 & u_{j,l}\\ u^\dagger_{j,l} & 0
    \end{pmatrix}
    \begin{pmatrix}
        \tilde \rho_{\m{c},\calq_j} & 0\\ 0 & \tilde \rho_{\m{c},\calq_l}
    \end{pmatrix}
    \begin{pmatrix}
        0 & u_{j,l}^\dagger \\ u_{j,l} & 0
    \end{pmatrix}
    = \rho_\m{c}.
\end{align}
The unitary $U$ is then a dynamical symmetry of both the quantum trajectory and the Lindblad equation 
\begin{align}
    U\sfl_\m{c}(\rho_\m{c})U^\dagger
    &= \sfl_\m{c}(U\rho_\m{c} U^\dagger),
    \label{eq:traj-symmetry}\\
    U \sfl(\rho)U^\dagger
    &= \sfl(U\rho U^\dagger).
    \label{eq:lind-symmetry}
\end{align}
While the symmetry classifications of Lindbladians have been performed in Refs.~\cite{Baumgartner2008_2,Buca2012,Albert2014,Minganti2018,Lieu2020,Altland2021,Roberts2021,Sa2023,Kawabata2023}, \cref{eq:traj-symmetry} may, in this regard, be seen as a step towards symmetry classifications of quantum trajectories.
$U$ acts as an intertwiner between orthogonal subspaces (see Ref.~\cite[Proposition 16]{Baumgartner2008_2}) that transforms stationary states and their support, into each other
\begin{align}
    \rho^\m{s}_{\calq_j} = U\rho^\m{s}_{\calq_l} U^\dagger, \quad
    P_{\calq_j} = UP_{\calq_l} U^\dagger.
    \label{eq:intertwiner}
\end{align}
Depending on the measurement process, unitary equivalence and dynamical symmetries may, however, arise in fundamentally different ways.

By comparing \cref{th:incomplete-diff,th:incomplete-jump} one can directly observe that the class of systems in which localization transitions are incomplete is larger in the jump case.
This fact is related to the discontinuous nature of the quantum jump unraveling which generally complies with a weaker form of ergodicity than the continuous diffusive trajectories.
It leaves open the possibility of having special residual symmetries in the stochastic evolution of the density matrix, most of which do not survive in the continuous diffusive unraveling.
This difference is particularly evident in (iib) of \cref{th:incomplete-jump}, where each minimal subspace $\calq_j$ further splits into a direct sum of orthogonal subspaces $\calq_{j,\alpha} = P_{j,\alpha} \calq_j$ with corresponding projectors $P_{j,\alpha}$.
Each Hamiltonian $H_{\calq_j}$ is unitarily similar to a block diagonal matrix and, as shown in \cref{sec:proof-iib}, the state must adhere to the same block form.
Transitions between subspaces $\calq_{j,\alpha}$ occur only through discrete jumps that transfer the entire state from one block to another, which is guaranteed by conditions \eqref{eq:block-sum} and \eqref{eq:no-coherence-jumps}.
Note that case (iib) is not included in (iia) since the Hamiltonians in different minimal subspaces are generally not unitarily related. 
Each block can be shifted by a real number $r_{j,\alpha}$ (cf. \cref{eq:block-H}) which leaves the continuous dynamics invariant.
Individual Lindblad jump operators are however still related by a unitary transformation and a phase factor (cf. \cref{eq:sim-L}).
A paradigmatic example for such a situation is a two-level system coupled to a thermal bath of harmonic oscillators with Hamiltonian $H_{\calq_j} = \omega \sigma^z + \bigoplus_{\alpha=1}^2 r_{j,\alpha}$ and Lindblad jump operators $L_{1,\calq_j} = \sqrt{\Gamma n_\m{th}} \sigma^+$ and $L_{2,\calq_j} = \sqrt{\Gamma (1+n_\m{th})} \sigma^-$, where $n_\m{th} = [\exp(\beta \omega)-1]^{-1}$ is the Bose-Einstein distribution at inverse temperature $\beta$.
The Lindblad jump operators act like ladder operators on the eigenstates of the Hamiltonian such that the dynamics is comprised only of discrete jumps between states $\ket{0}$ and $\ket{1}$.
One easily confirms \cref{eq:block-sum,eq:no-coherence-jumps} with $P_{j,1} = \dyad{0}$, $P_{j,2} = \dyad{1}$.

\subsection{Symmetries}
\label{sec:symmetries}
For diffusive trajectories case (ii) is actually equivalent to unitary equivalence on the level of the subspace $\calq_j$ themselves.
The state space structure of the composite subspace is then isomorphic to a noiseless subsystem
\begin{align}
    \bigoplus_j \calq_j
    \simeq \mathds{1} \otimes \calq_j.
    \label{eq:noiseless-subsystem}
\end{align}
Given the full decomposition of the asymptotic subspaces into its irreducible components (cf. \cref{eq:decomposition}) 
\begin{align}
  \calr = \bigoplus^K_{k=1} \calu_k \oplus \bigoplus_{l=1}^M \calx_l,
\end{align}
such degenerate subspaces can reside only inside the second term on the rhs which contains the whole collection of isomorphic subspaces (see \cref{eq:degenerate-subspaces}) implying a dynamical symmetry as in \cref{eq:traj-symmetry,eq:lind-symmetry}, and the intertwining property \cref{eq:intertwiner} \cite{Baumgartner2008_2}.
Since (ii) in \cref{th:incomplete-diff} is a special case of (iia) of \cref{th:incomplete-jump}, we can thus conclude that degenerate subspaces cannot be distinguished by quantum trajectories of both unravelings and hence prevent further localization.
In general, the presence of isomorphic subspaces ($\calx_l \neq 0$) implies that the decomposition of the asymptotic subspace $\calr$ is no longer unique for at least one $l$ and there are projectors (the elements of the commutant, $P_{\calq_j} \in \{H_\calr,L_{k,\calr},L^\dagger_{k,\calr}\}^\prime$) onto minimal subspaces that are not mutually orthogonal nor contained in each other, with
\begin{align}
[P_{\calq_j}, P_{\calq_l}] \neq 0, \ \text{for at least one } j\neq l
\end{align}
forming a non-abelian (sub)algebra \cite{Baumgartner2008_2} that gives rise to noiseless subsystems.
It was shown in Refs.~\cite{Baumgartner2008_2,Baumgartner2012} that a noiseless subsystem represents the unique structure that allows for asymptotic coherences between minimal blocks of the density matrix.
This feature has recently been employed to generate steady state entanglement from dissipation \cite{Li2023,Moharramipour2024,Li2025}.
In fact, we can make the following observation.
\begin{observation}
  Coherences between orthogonal subspaces are guaranteed in case (ii) of \cref{th:incomplete-diff,th:incomplete-jump} for global pure states $\ket{\Psi_\m{c}}$.
\end{observation}
This is a direct consequence of the assumption that the state is distributed over two orthogonal subspaces while the global state $\rho_\m{c} = \dyad{\Psi_\m{c}}$ and also the individual states restricted to the subspaces $\calq_j$ are pure $\tilde \rho_{\m{c},\calq} = \dyad{\tilde \Psi_{\m{c},\calq_j}}$.

In contrast to diffusive trajectories, quantum jumps may exhibit unitarily equivalent evolutions even in the absence of subspace degeneracy ($\calx_l = 0$).
Although condition (ii) of \cref{th:incomplete-jump} implies dynamical symmetries in \cref{eq:lind-symmetry,eq:traj-symmetry} and the intertwining property \cref{eq:intertwiner}, any nonzero phase difference between jump operators ($\exp(\I\phi_{k,j}) \neq \exp(\I\phi_{k,l})$) or any deviation from unitary equivalence between Hamiltonians in (iib) lifts these degeneracies and the corresponding subspaces must instead belong to the unique part of the decomposition.
If the entire decomposition of the asymptotic state space is unique according to
\begin{align}
  \calr = \bigoplus_{k=1}^K \calu_k,
\end{align}
the corresponding (sub)algebra of projectors on minimal subspaces is abelian with \cite{Baumgartner2008_2}
\begin{align}
  [P_{\calq_j},P_{\calq_l}] = 0, \ \forall j,l.
\end{align}
In this case, provided condition (i) of \cref{th:incomplete-diff} does not hold, complete localization is then guaranteed to occur in the diffusive unraveling, while localization may still be incomplete for quantum jumps.

\subsection{Asymptotic behavior of quantum trajectories}
We can now draw some very general conclusions about the asymptotic behavior of quantum trajectories in arbitrary finite dimensions.
Every quantum trajectory will exhibit one of three characteristic behaviors.
\begin{enumerate}[(a)]
    \item Partial localization occurs; trajectories do not purify and evolve independently, there is no measurement backaction just classical noise.
    \item Partial localization occurs; evolutions are unitarily equivalent in minimal orthogonal subspaces and asymptotic purification occurs in each of the subspaces individually but not globally.
    \item Full localization occurs, the trajectory does not purify but is subject to classical noise, or asymptotic purification and full localization occur concurrently; the quantum system enters completely a minimal orthogonal subspace.
\end{enumerate}
In (a), the system may be caught between subspaces that can have arbitrary relative dimensions.
This happens for instance when the system enters a subspace that is not acted upon by the measurement with ensuing decoherence-free evolution (cf. \cref{sec:dfs}).
At most, the detector might effectively decorrelate from the state but still influence the dynamics in a unidirectional manner, inducing classical noise on the system (cf. \cref{sec:classical-noise}).
Subspace degeneracy and noiseless subsystems are always sufficient for unitarily equivalent evolutions in case (b) where the involved subspaces necessarily need to have the same dimension.
In general, the exact conditions on the Hamiltonian and the jump operators are dependent on the particular measurement scheme and therefore on the unraveling.

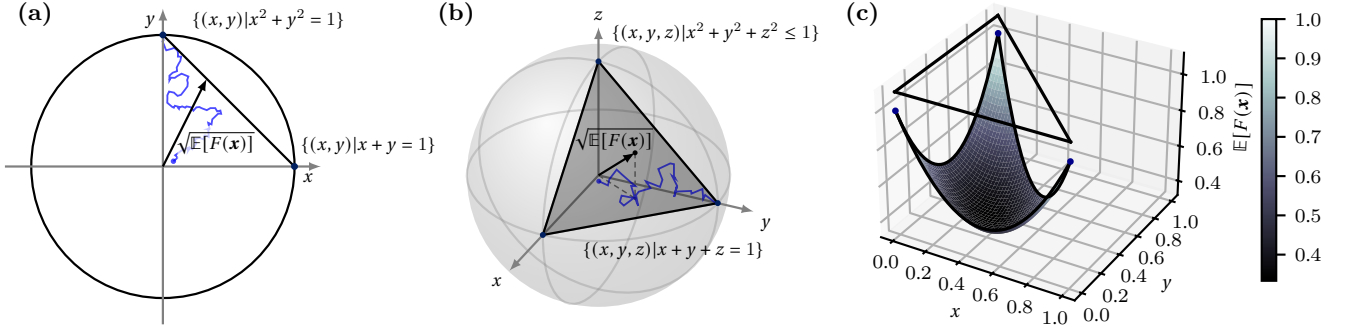
\begin{figure*}[t]
  \centering
  \begin{tikzpicture}
    \node (a) at (0,0) {\includegraphics[scale=0.7]{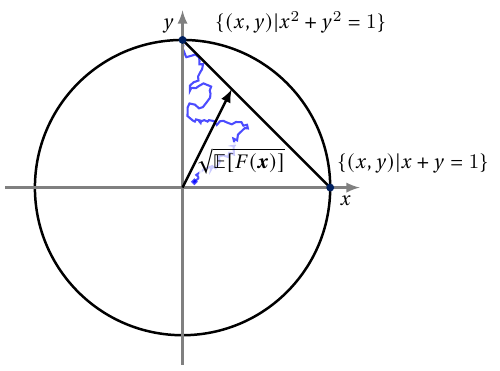}};	
    \node (b) at (5.5,0) {\includegraphics[scale=0.7]{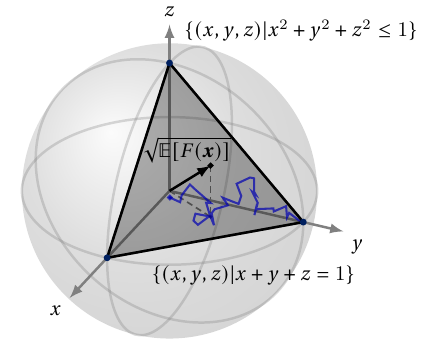}};
    \node (c) at (11.6,0) {\includegraphics{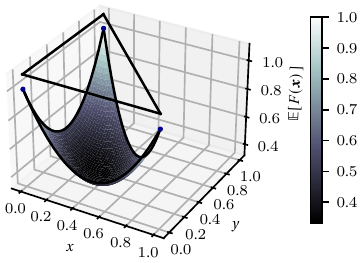}};
    \draw (-2.5,2) node {\textbf{(a)}};
    \draw (3.1,2) node {\textbf{(b)}};
    \draw (8.5,2) node {\textbf{(c)}};
  \end{tikzpicture}
  \caption{Geometric representation of the mean fidelity $\mathbb{E}[F(\overline{\rho}_\m{c},\rho^\m{s})]$ (\cref{eq:fidelity}) in two and three dimensions, quantifying ergodicity between time average, $\overline{\rho}_\m{c}$, and ensemble average, $\rho^\m{s}$, of quantum trajectories.
  Assuming complete localization (cf. \cref{th:incomplete-diff,th:incomplete-jump}), the mean fidelity has an analytical expression $\mathbb{E}[F(\overline{\rho}_\m{c},\rho^\m{s})] = \sum_j w_j^2$, depending only on the effective weights $w_j$, in each minimal subspace.
  Assigning coordinates to the weights, $\vb*{x} = (w_1,w_2,\ldots,w_N)^\m{T}$, ergodicity corresponds to the Euclidian distance from the origin to an $N$-simplex embedded in an $N$-dimensional sphere.
  (a) Bistability. There are two orthogonal subspaces. The $N$-sphere is a circle and the simplex is a line.
  (b) Tristability. There are three orthogonal subspaces. The $N$-sphere is a sphere and the simplex is a triangle.
  On the trajectory level minimal subspaces are selected with probability $\vb*{x}$ (schematically illustrated by the blue line). 
  (c) Mean fidelity, $\mathbb{E}[F(\vb*{x})] = x^2+y^2+(1-x-y)^2$, of a tristable system as a function of the two independent weights $x$ and $y$. 
  Evolution is ergodic at the extremal points ($\mathbb{E}[F(\vb*{x})] = 1$), reflecting the underlying simplex (black triangle), when the system is initialized entirely in one minimal subspace and the long-time average converges to the corresponding extremal stationary state of the Lindblad equation.
  The minimal fidelity is $1/N=1/3$.
  Projecting out one coordinate recovers the fidelity of a bistable system (a), where $\mathbb{E}[F(\vb*{x})] = x^2+(1-x)^2$ (black parabolas).
}
  \label{fig:geometric-fidelity}
\end{figure*}
\section{Ergodicity breaking}
\label{sec:erg-theorem}
Quantum systems subject to continuous monitoring probe statistics of the density operator beyond the mean.
In general, knowledge of the ensemble average is not sufficient to provide information about the behavior of individual realizations, indicating that ergodicity must be violated.
In this section, we investigate in full generality the ergodic properties of continuously monitored quantum systems.
We first present the pathwise ergodic theorem of Kümmerer and Maassen, Ref.~\cite{Kummerer2004}, which states that the long-time average of any quantum trajectory coincides with a randomly chosen stationary state of the Lindblad equation.
In connection with our results on complete and incomplete localization (\cref{sec:transition,sec:incomplete}) we obtain a more refined picture which enables us to quantify how strongly ergodicity is broken on the trajectory level.

Let $\rho(0)$ be a fixed initial state and let the Lindblad equation \cref{eq:me} have a unique stationary state $\rho^\m{s}$ with 
\begin{align}
  \sft^t(\rho^\m{s}) = \rho^\m{s}.
\end{align}
It then holds that the long-time average of any individual trajectory of any unraveling converges exactly to this unique stationary state \cite{Kummerer2004}
\begin{align}
  \lim_{T \to \infty} \frac{1}{T} \int_0^T \dd{t} \rho_\m{c}(t) = \rho^\m{s}.
  \label{eq:unique-erg}
\end{align}
Ergodicity is meant in the precise sense that the infinite-time average of any quantum trajectory will reproduce the unique steady state of the Lindblad equation.
The notion of ergodicity in the classical sense is generally not available in the open quantum system setting.
Due to disconnected symmetry sectors, Lindbladian evolution will generally not explore the whole Hilbert space but will remain itself constrained to a Krylov subspace, the part of the Hilbert space made available by the initial condition \cite{Hochbruck1997,Parker2019,Liu2023,Nandy2024}.

Within this restricted region we compare the time average of individual realizations to the ensemble average for a given initial configuration.
The inherent backaction caused by the quantum measurement usually drives the system into a pure state that will go on to randomly bounce around in Hilbert space.
Naively one might expect that the stochastic nature of quantum trajectories makes the system explore the full available state space as it is typically the case for classical stochastic processes with the paradigmatic example being a Brownian particle immersed in a liquid.
It turns out that this notion is valid only in the absence of steady state degeneracies (cf. \cref{eq:unique-erg}) or when the quantum noise becomes classical (cf. \cref{sec:classical-noise}).

Now, let the Lindblad equation admit multiple steady states. 
Denote by $\rho^\m{s}_j$, the extremal, linearly independent stationary states, each supported on a minimal orthogonal subspace $\calq_j$ inside of $\calr$.
Their convex hull spans the complete set of asymptotic states.
The extremal states are indeed stationary and satisfy
\begin{align}
  \sft^t(\rho^\m{s}_j) = \rho^\m{s}_j.
\end{align}
For a given initial condition, the corresponding asymptotic subspace is then given by the support of $\lim_{t\to\infty}\rho(t)$ (cf. \cref{eq:effective-weights})
\begin{align}
  \calw = \bigoplus_j \calq_j, \text{ s.t. }|\calq_j^\infty|^2 > 0.
\end{align}
The pathwise ergodic theorem states that the long-time average of any trajectory converges to a random choice of the set of stationary states \cite{Kummerer2004}
\begin{align}
  \lim_{T \to \infty} \frac{1}{T} \int_0^T \dd{t} \rho_\m{c}(t)
  = \rho^\m{s}_\m{c},
  \label{eq:erg-theorem}
\end{align}
where the resulting state
\begin{align}
  \rho^\m{s}_\m{c} = \sum_j |\calq_j|^2_\m{c} \rho^\m{s}_j  
\end{align}
is now a random convex combination of extremal states.
The coefficients $|\calq_j|^2_\m{c} \equiv \lim_{t\to \infty}\tr[P_{\calq_j} \rho_\m{c}(t)] \ge 0$ depend on the realization, $\m{c}=(\xi,J)$, of the stochastic process, with $\sum_j |\calq_j|^2_\m{c} = 1$.

We can now establish the connection to the localization theorems (\cref{th:incomplete-diff,th:incomplete-jump}).
Provided complete localization takes place, with probability $|\calq^\infty_j|^2$, the system selects exactly one minimal orthogonal subspace $\calq_j$, where it then remains and undergoes stochastic dynamics indefinitely.
Since the system has support exclusively on $\calq_j$, by virtue of the pathwise ergodic theorem \cref{eq:erg-theorem}, the long-time average can only converge to the corresponding extremal stationary state, $\rho^\m{s}_j$.
More specifically, under complete localization any quantum trajectory converges exactly to one of the extremal stationary states of the Lindblad equation and we thus obtain a refined version of the ergodic theorem \cref{eq:erg-theorem}
\begin{align}
  \lim_{T \to \infty} \frac{1}{T} \int_0^T \dd{t} \rho_\m{c}(t)
  = \rho^\m{s}_j, \quad \text{ w. prob. } |\calq^\infty_j|^2.
\end{align}

When localization is incomplete, we can generally not predict the result of the time average, \cref{eq:erg-theorem}.
There is however one exception which arises when there exists a multidimensional decoherence-free subspace and the decaying subspace is the empty set.
Although trajectories cannot localize between different states of the same decoherence-free subspace (cf. \cref{eq:dq_12,eq:dq_12-jump}), the evolution inside is unitary and the time average can be straightforwardly computed 
\begin{align}
  \lim_{T \to \infty} \frac{1}{T} \int_0^T \dd{t} \rho^\m{DFS}_\m{c}(t)
  = \sum_j \bra{q_j}\rho(0)\ket{q_j} \dyad{q_j},
\end{align}
where $\{\ket{q_j}\}$ is the set states corresponding to the decoherence-free subspace selected by the trajectory.

To quantify general violations of ergodicity we need to assess how much the time average of individual realizations deviates from the ensemble averaged density matrix.
Denote by $\rho^\m{s}$ the asymptotic state of the Lindblad equation that is uniquely determined by the fixed initial condition.
Further denote by
\begin{align}
  \overline{\rho}_\m{c} \equiv 
  \lim_{T \to \infty} \frac{1}{T} \int_0^T \dd{t} \rho_\m{c}(t),
\end{align}
the long-time average of an individual realization.
We employ the mean fidelity between the time and ensemble averaged density matrices \cite{Jozsa1994,Schmolke2024}
\begin{align}
  \mathbb{E}[F(\overline{\rho}_\m{c},\rho^\m{s})]
  = \mathbb{E}\left[\tr(\sqrt{\sqrt{\overline{\rho}_\m{c}}\rho^\m{s}\sqrt{\overline{\rho}_\m{c}}})^2\right],
  \label{eq:fidelity}
\end{align}
as a measure of ergodicity.
The average is necessary because the bare fidelity would generally produce a random variable.
A small fidelity indicates highly non-ergodic dynamics, while unit fidelity corresponds to ergodic evolution where time and ensemble average perfectly coincide.

Evaluating the mean fidelity is now straightforward.
Given an initial condition $\rho(0)$ the decomposition of the asymptotic (time averaged) density matrix is uniquely determined (cf. \cref{eq:asym-states}) and, in full generality, we can define 
\begin{align}
  \rho^\m{s} = \sum_{j=1}^N w_j \rho^\m{s}_j = \bigoplus_{k=1}^K \lambda_k \rho_k \oplus \bigoplus_{l=1}^M \mu_l \sigma_l \otimes \tau_l,
\end{align}
with $N = K+M$, where the weights, $w_j \equiv |\calq^\infty_j|^2$, are equal to the effective support of the initial state in the orthogonal subspaces $\calu_k$ and $\calx_l$ respectively.
The most general expression for the average fidelity is available analytically and takes the form 
\begin{align}
\mathbb{E}[F\left(\overline{\rho}_\m{c},\rho^\m{s}\right)]
= \sum_{j=1}^N w_j^2,
\label{eq:main}
\end{align}
where $N$ is the number of extremal states within the range of the initial condition and $\sum_j w_j = 1$, due to conservation of total probability.
Expression \eqref{eq:main} is a participation ratio that measures how much the initial state is effectively delocalized over the irreducible subspaces of the Hilbert space.
The inverse participation ratio is a prominent measure of localization \cite{Kramer1993,Evers2008}.
The strongest breaking of ergodicity occurs when the weights in each subspace are uniformly distributed with $w_j = 1/N$, and the fidelity assumes its minimum value of $\m{min}_{\rho(0)}\{\mathbb{E}[F\left(\overline{\rho}_\m{c},\rho^\m{s}\right)]\} = 1/N$.
Remarkably, the ergodic behavior depends only on the distribution of the effective support over the orthogonal subspaces and is, in particular, independent of their microscopic details or dimensions.
For a given initial state ergodicity violations can be trivially maximized by choosing a jump operator that commutes with the Hamiltonian, $[H,L] = 0$, and has only simple eigenvalues.
The entire Hilbert space is then decoherence-free and the number of extremal states is equal to the Hilbert space dimension, $N=\dim(\calh)$ and every trajectory results in a common eigenstate of $H$ and $L$.
However, $N$ can still grow exponentially in the number of subsystems in less trivial circumstances, for instance when the Hilbert space is strongly fragmented \cite{Moudgalya2018,Sala2020,Moudgalya2022,Moudgalya2022-review,Li2023,Adler2024,Zhao2025}.

The mean fidelity has a useful geometric interpretation in the space of effective weights $w_j$.
Assigning Cartesian coordinates to the weights, $\vb*{x} = (w_1,w_2,\ldots,w_N)^\m{T}$, it can be represented in $\mathbb{R}^N$.
Its domain, the vector of eligible initial weights $\vb*{x}$, is a $N$-simplex (with $\{\vb*{x}\vert \sum_i x_i = 1\}$). 
It will prove practical to embed the simplex in a $N$-dimensional unit sphere (with $\{\vb*{x}\vert \vb*{x}^\m{T}\vb*{x} \le 1\}$) because ergodicity then corresponds to the squared Euclidian distance from the origin to the simplex (in the convention $F^\prime = \sqrt{F}$, ergodicity coincides with the Euclidian distance).
The geometric representation of the mean fidelity is illustrated in \cref{fig:geometric-fidelity}, in two and three dimensions.
Projecting out coordinates corresponds to a coarse-graining of the Hilbert space structure where several minimal subspaces are collected into a larger one.

In this representation, we can also track the evolution of individual trajectories (blue line in \cref{fig:geometric-fidelity}).
The enclosed convex space between the coordinate axes and the simplex defines the region of transient dynamics that partly takes place inside of $\cald$.
A trajectory in this coordinate space starts somewhere inside of this region.
In the absence of decay, the dynamics takes place entirely on the simplex since, $\sum_jx_j(t) = \tr[\rho_\m{c}(t)P_\calr] = 1, \ \forall t$.
Conversely, a trajectory that starts somewhere off of the simplex undergoes transient dynamics in the decaying subspace as long as $\tr[\rho_\m{c}P_\cald] > 0$ and therefore, $\sum_j x_j(t) < 1$.
In \cref{fig:geometric-fidelity}a,b we schematically illustrate the behavior of a single quantum trajectory.
Initially, the system has partial overlap with both the decaying and asymptotic subspaces.
In the course of time, the trajectory gets projected onto the simplex and moves towards an extremal point, thereby converging to an extremal stationary state, in accordance with complete localization.
Provided localization is incomplete, a trajectory would end up at an arbitrary point on the simplex and remain there.

\section{Emergence of a generalized Born rule}
\label{sec:Born}
Quantum trajectories spontaneously exhibit irreversible transitions where they continuously localize in Hilbert space until either reaching a symmetry or an indecomposable subspace.
On the ensemble level, evolution takes place in all of these subspaces simultaneously.
This fundamentally different behavior gives rise to violations of ergodicity, which have been characterized in the previous section.
What remains is the question of how this general phenomenon can be understood from a physical point of view.
In this section we address the physical interpretation and show that continuous monitoring can be understood in terms of an effective, generalized projective measurement on both the initial state and the time evolution.

With the necessary and sufficient conditions for complete localization (\cref{th:incomplete-diff,th:incomplete-jump}) at hand, we are now in the position to exactly predict when and how often individual trajectories will select minimal invariant subspaces.
In this regard, the steady state probability distributions \cref{eq:steady-state-full-diff,eq:steady-state-full-jump}, may be interpreted as a generalized Born rule (see also \cref{eq:effective-weights,sec:decaying-subspace}).
Before studying the most general scenario, it is instructive to first consider the bare, measurement-only scenario in the limiting case where $H = 0$.
The structure of the state space (\cref{eq:decomposition}), is now determined exclusively by the structure of the operator $L$ \cite{Baumgartner2008_1}.
Standard quantum theory requires Hermitian observables.
Consequently, let $L = \sum_n l_n \Pi_n$, be unitarily diagonalizable with eigenvalues, $l_n$, and projectors, $\Pi_n = P_{\calq_n}$, on the minimal orthogonal eigenspaces $\calq_n$.
Continuous measurement will then gradually map the system onto one of the eigenspaces of $L$ (see \cref{sec:transition,sec:incomplete}) 
\begin{align}
  \lim_{t\to\infty}\sft^t_\m{c}(\rho(0)) = \frac{\Pi_n \rho(0) \Pi_n}{\tr[\rho(0)\Pi_n]}
\end{align}
We may express the effective, infinite-time action of the measurement process in a more suggestive form
\begin{align}
  \rho(0) \to \rho_n = \frac{\Pi_n \rho(0) \Pi_n}{\tr[\rho(0)\Pi_n]}.
  \label{eq:standard}
\end{align}
This process occurs with probability $p_n = \tr[\rho(0) \Pi_n]$, and the standard Born rule is recovered, since $\sum_n \Pi_n = \mathds{1}$.
Continuous monitoring with a single Hermitian Lindblad jump operator hence effectively describes a dragged out projective measurement.
The Hamiltonian can be nonzero yielding the same outcome, as long as the measurement takes place unhindered, viz. $[H,L]=0$ (as observed in Refs.~\cite{Weber2014,Weber2016}).

In general, the continuous measurement competes with the unitary evolution leading to nontrivial stochastic dynamics.
While the measurement---mediated by the set of operators $\{L_k\}$---drives the system into one of the common eigenspaces, the Hamiltonian counteracts this process and rotates the system out again, inducing transitions between eigenspaces if $[H,L_k]\neq 0$ for at least one $k$ and the standard measurement update rule \cref{eq:standard} can no longer be sustained.
The best next thing to commutation is a common symmetry in $\calr$ with $[H_\calr,P_{\calq_j}] = [L_{k,\calr},P_{\calq_j}] = 0$, where $H_\calr = P_\calr H P_\calr$ and $L_{k,\calr} = P_\calr L_k P_\calr$.
The measurement can thus at most succeed in driving the system into an orthogonal subspace, common to both the restricted operators $H_\calr$ and $L_{k,\calr}$.
This selection process is, in a sense, the minimal compromise between the two competing mechanisms and may be interpreted as the result of a generalized measurement update.
We may hence interpret the selected subspace together with the ensuing evolution as the outcome associated with the generalized measurement.

Denote by $\calw = \bigoplus_j \calq_j$, the collection of minimal subspaces in the asymptotic range of the initial state of the Lindblad equation, i.e. all subspaces with $|\calq_j^\infty|^2 > 0$.
Using the shorthand $c = (J,\xi)$ for the conditional evolution, denote further by $\calw_\m{c} \subseteq \calw$ the infinite-time support of an individual quantum trajectory which generally depends on $\m{c}$.
Any initial state is associated with an effective asymptotic mapping that projects the system onto one of various possible quantum trajectory evolutions.
The most general update takes the form
\begin{align}
  \lim_{t \to \infty}\left\|\rho_\m{c}(t) -
  \sft^t_{\m{c},\calw_\m{c}}(\sfp^\infty_{\calw_\m{c}} \rho(0))\right\|
  = 0,
  \label{eq:degenerate-update-rule}
\end{align}
where $\sfp^\infty_{\calw_\m{c}}$ is a realization dependent projector that distributes the initial state over the minimal subspaces $\calq_j$ with potential coherences between them.
On $\calw_\m{c}$ the evolution is then governed by the projected propagator $\sft^t_{\m{c},\calw_\m{c}} = \sfp^\infty_{\calw_\m{c}} \sft^t_\m{c} \sfp^\infty_{\calw_\m{c}}$ with restricted Hamiltonian and jump operators
\begin{align}
H_{\calw_\m{c}} &= \bigoplus_j P_{\calq_j}HP_{\calq_j},\\
L_{k,\calw_\m{c}} &= \bigoplus_j P_{\calq_j}L_kP_{\calq_j}, \quad \forall \calq_j \in \calw_\m{c}.
\end{align}
Although the asymptotic state may be delocalized over several minimal subspaces, the infinite-time support in each minimal subspace is stationary, $|\calq_j|^2_\m{c} = \lim_{t\to\infty}\tr[\rho_\m{c}(t) P_{\calq_j}] = \m{const}.$ \footnote{This follows directly from \cref{sec:transition} and \cref{th:incomplete-diff,th:incomplete-jump}. If localization is complete then $|\calq_j|^2_\m{c} = \delta_{j,l}$ for some $l$. If localization is incomplete, then by the defining property in \cref{eq:incomplete-diff,eq:incomplete-jump} the overlap with the indistinguishable subspaces must become constant.}.
Remarkably, stationarity is always reached asymptotically and, after a transient time, there are hence no more transitions between orthogonal subspaces (cf. \cref{sec:transition,sec:incomplete}).
This transient time thus corresponds to the duration of the generalized measurement process.
The asymptotic weights, $|\calq_j|^2_\m{c}$, are random variables, different for each realization $\m{c}$.
Their distribution is still an open problem and depends on the specific model and the microscopic details of the state space (cf. \cref{sec:two-qubits}).
The above expression is the most general and corresponds to a generalized measurement similar to a standard projective measurement on a degenerate subspace and the superposition \cref{eq:degenerate-update-rule} may lead to the build-up of quantum correlations and coherences.

Provided localization transitions are complete, stronger statements can be made.
In this case, the initial state is mapped onto one and only one minimal orthogonal subspace $\calq_j$ and the emerging update rule accordingly becomes 
\begin{align}
  \rho(0) \rightarrow
  \rho_{\m{c},{\calq_j}}(t) 
  \equiv \frac{\sft^t_{\m{c},{\calq_j}}(\sfp^\infty_{\calq_j}\rho(0))}{\tr[\sfp^\infty_{\calq_j} \rho(0)]}.
  \label{eq:update-rule}
\end{align}
This process is decomposed into two parts,
the asymptotic projection $\sfp^\infty_{\calq_j}$ that carries the initial state to the minimal subspace $\calq_j$ and the projected time evolution $\sft^t_{\m{c},\calq_j}$ that evolves the localized state in the contracted Hilbert space.
The asymptotic projection is defined as 
\begin{align}
  \sfp^\infty_{\calq_j} 
  = \sfp_{\calq_j} \lim_{t\to\infty} e^{\sfl t},
\end{align}
where the quantity $\sfp^\infty \equiv \lim_{t\to\infty}\exp(\sfl t)$ denotes the deterministic asymptotic projection of the Lindbladian (see also Ref.~\cite[Eq.~(2.6)]{Albert2016}).
Note that $\sfp^\infty_{\calq_j}$ reduces to $\sfp_{\calq_j}$, if the decaying subspace is empty.
The generalized measurement update \cref{eq:update-rule} occurs with probability $|{\calq_j}^\infty|^2 = \tr[\sfp^\infty_{\calq_j} \rho(0)]$.
As a main difference to standard projective measurement, here, after the update, there will generally still be ensuing evolution inside the subspace $\calq_j$ with $\sft^t_{\m{c},\calq_j} = \sfp_{\calq_j} \sft^t_\m{c} \sfp_{\calq_j}$, where $\sfp_{\calq_j}(\bullet) = P_{\calq_j} \bullet P_{\calq_j}$.
Indeed, the time evolution itself undergoes a projection $\sft^t_\m{c} \to \sft^t_{\m{c},{\calq_j}}$ which results again in a valid quantum trajectory with a new, restricted Hamiltonian and a new restricted set of Lindblad jump operators, $H_{\calq_j} = \sfp_{\calq_j} H$, and $L_{k,{\calq_j}} = \sfp_{\calq_j} L_k$ respectively, essentially unraveling the extremal stationary state of the Lindblad equation associated with $\calq_j$.
Depending on the properties of the evolution $\sft^t_{c}$, different outcomes $\sft^t_{\m{c},{\calq_j}}$ of the generalized projective measurement may lead to dramatically different behavior.
A particularly prominent case is presented in \cref{sec:ring}.

On the other hand, in \cref{sec:incomplete} we have discussed the conditions under which quantum trajectories are unable to decide between subspaces, localization transitions are incomplete and, as a result, the density matrix converges to a random convex combination of extremal stationary states.
The generalized projection must therefore act on multiple indistinguishable subspaces and we recover \cref{eq:degenerate-update-rule}, a situation similar to Lüders' rule for conventional projective measurements on degenerate eigenspaces.

As before in \cref{eq:erg-theorem}, there is one exception to the stochastic nature of \cref{eq:degenerate-update-rule}.
If the absence of a decaying subspace coincides with the existence of a decoherence-free subspace we get
\begin{align}
  \sft^t_\calq(\sfp_\calq \rho(0)) = 
  \frac{\sfu^t_{\calq}(\sfp_\calq \rho(0))}{\tr[\sfp_\calq \rho(0)]}.
\end{align}
Here, $\sfu^t_\calq(\bullet) = U_\calq \bullet U^\dagger_\calq$ is the residual unitary evolution with $U_\calq = \sfp_\calq U$.
Initial coherences between the $\calq_j$ are preserved and can build up further.

In summary, we thus obtain a measurement update rule similar to standard projective measurement but generalized in several aspects.
Continuous measurement asymptotically acts in one of three possible ways.
\begin{enumerate}[(i)]
  \item Standard measurement postulate: The jump operator is Hermitian, $L=L^\dagger$ and commutes with the Hamiltonian, $[H,L]=0$ , the asymptotic mapping reduces to the standard measurement postulate
  \begin{align}
    \rho(0) \to \frac{\Pi_n \rho(0) \Pi_n}{\tr[\rho(0)\Pi_n]}.
  \end{align}
  $\Pi_n$ projects the state onto one of the eigenspaces that diagonalize $L$. The outcome of the measurement is one of the eigenvalues $l_n$ of $L$. The probability is given by the Born rule, $p_n = \tr[\rho(0)\Pi_n]$.
  \item Generalized update (orthogonal projection): Hamiltonian and jump operators are arbitrary. There is no decaying subspace, $\cald = \emptyset$. The update rule is
  \begin{align}
    \rho(0) \rightarrow
    \frac{\sft^t_{\m{c},\calq}(\sfp_\calq\rho(0))}{\tr[\sfp_\calq \rho(0)]}.
  \end{align}
  $\sfp_\calq$ projects the state onto one of the common invariant subspaces that simultaneously block-diagonalize $H$ and all $L_k$. The outcome of the measurement is an unraveling of a steady state of the Lindblad equation. The probability is given by a generalized Born rule, $p_\calq = \tr[\sfp_\calq\rho(0)]$.
  \item Generalized update (oblique projection): Hamiltonian and jump operators are arbitrary. The decaying subspace exists, $\cald \neq 0$. The update rule is
  \begin{align}
    \rho(0) \rightarrow
    \frac{\sft^t_{\m{c},\calq}(\sfp^\infty_\calq\rho(0))}{\tr[\sfp^\infty_\calq \rho(0)]}.
  \end{align}
  $\sfp^\infty_\calq$ projects the state onto one of the common invariant subspaces that simultaneously block-diagonalize the restricted operators $H_\calr$ and $L_{k,\calr}$. The outcome of the measurement is an unraveling of a steady state of the Lindblad equation. The probability is given by the asymptotic weight of the initial state, $p_\calq = \tr[\sfp^\infty_\calq\rho(0)]$.
  The projection maps the initial state outside of its support, which is always associated to decay.
\end{enumerate}

\section{Examples and special cases}
\label{sec:examples}
In the previous sections, we have completely characterized the asymptotic behavior of quantum trajectories.
Based on this general description, we now investigate the physical implications of the theory in specific examples, highlighting the localization behavior (cf. \cref{sec:transition,sec:incomplete}), the ergodic properties (cf. \cref{sec:erg-theorem}) and the interpretation in terms of a generalized update rule (cf. \cref{sec:Born}).
For every example presented below we also supply the full Hilbert space structure (cf. \cref{sec:sbd}).

\subsection{Strong measurement}
We have so far not made any assumptions about the strength of the measurement.
In fact, while the Hilbert space structure may change in the limit of infinitely strong measurements, our results and derivations remain valid regardless.
It is however instructive to study the implications of the onset of the Zeno regime.
In the conventional quantum Zeno effect an otherwise closed system is repeatedly measured with a certain rate.
When measurements become increasingly frequent, the system gets more and more confined to its initial state until time evolution freezes altogether in the limit of infinitely small time intervals and the system is no longer able to escape \cite{Presilla1996,Misra1977}.
By contrast, in many-body quantum systems where measurement is performed on local observables that act only on a part of the degrees of freedom, freezing does generally not occur \cite{Facchi2002,Facchi2004,Koshino2005,Zanardi2014,Popkov2018,Burgarth2020,Popkov2021}.
Instead, when the measurement is much stronger than the coherent Hamiltonian evolution, the structure of the Hilbert space will be dominated by the orthogonal subspaces of the Lindblad jump operators only.
Denote by 
\begin{align}
  \|H\| \equiv J, \qquad \sum_k\|L^\dagger_kL_k\| \equiv \Gamma,
\end{align}
the characteristic time scales of unitary and dissipative parts and further denote by $\{\Pi_n\}$ the set of projectors onto the common eigenspaces of the collection of jump operators $\{L_k\}$.
Then, the transition probability between eigenspaces of $L_k$
\begin{align}
P^\m{s}_{nm} = \|\Pi_n\rho^\m{s}\Pi_m\| = O(J/\Gamma), \ \forall m\neq n,
\end{align}
becomes small and transitions thus increasingly suppressed for strong measurement, $\Gamma \gg J$ \cite{Zanardi2014}, as the measurement succeeds in trapping the system in one of the eigenspaces at almost all times \cite{Cresser2006,Weber2016,Snizhko2020,Buchhold2021}.
For any finite value of $\Gamma$, subspace selection and localization (\cref{th:incomplete-diff,th:incomplete-jump}) remain unaffected and the steady state distributions \cref{eq:steady-state,eq:steady-state-jump} are still strictly valid but may occur on a different time scale.
Then in the Zeno limit, $\Gamma\to\infty$, a transition takes place in the ergodic properties of quantum trajectories as the Lindblad jump operators take over and impose an orthogonal subspace decomposition upon the Hilbert space, irrespective of the preexisting structure, which results in the emergence of a new, effective evolution with a renormalized, dissipation-projected Hamiltonian and effective dissipator \cite{Zanardi2014,Popkov2018,Burgarth2020,Popkov2021}.
This approach may be used for single-shot autonomous quantum control \cite{Zanardi2014}.
Once a certain state space structure has been established by strong measurement of a second observer \cite{Jacobs2006,Jacobs2014} or strong damping \cite{Popkov2021,Burgarth2020}, the resulting orthogonal subspaces will be asymptotically selected by individual trajectories, driving the quantum system into desired subspaces.

\begin{figure}[t]
  \centering 
  \begin{tikzpicture}
    \node (a) at (0,0.3) {\includegraphics[scale=0.9]{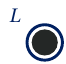}};	
    \node (b) at (4.5,0) {\includegraphics[scale=0.7]{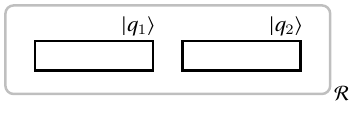}};
    \draw (-0.7,0.6) node {\textbf{(a)}};
    \draw (2,0.6) node {\textbf{(b)}};
  \end{tikzpicture}
  \caption{(a) Single qubit with $H = \omega \sigma^z$ subject to continuous measurement of the population with $L=\sqrt{\Gamma}\sigma^+\sigma^-$. (b) Corresponding Hilbert space structure. The decaying subspace is the empty set, the asymptotic state space occupies the entire Hilbert space. There are two distinct one-dimensional decoherence-free subspaces, $\ket{q_1}$ regular and $\ket{q_2}$ dark, each supplied with a conserved projector $P_{\calq_n}$. The monitoring effectively acts like a projective measurement of the Hamiltonian and individual trajectories will eventually converge to one of its eigenstates $\ket{q_n}$, see \cref{fig:single-qubit}.}
  \label{fig:single-qubit-structure}
\end{figure}

\begin{figure}[t]
  \centering 
  \begin{tikzpicture}
    \node (a) [label={[label distance=-0.1cm]130: \textbf{(a)}}] at (0,0) {\includegraphics{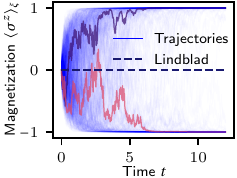}};	
    \node (b) [label={[label distance=-0.1cm]130: \textbf{(b)}}] at (4.2,0) {\includegraphics{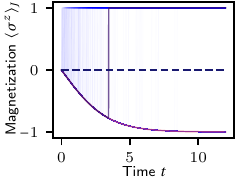}};
    \draw (0.2,1.7) node {\textsf{Diffusion}};
    \draw (4.4,1.7) node {\textsf{Jumps}};
  \end{tikzpicture}
  \caption{Magnetization of a qubit under continuous measurement with $H=\omega \sigma^z$ and $L=\sigma^+\sigma^-$ with initial state $\ket{\Psi(0)} = (\ket{0}+\ket{1})/\sqrt{2}$ for $500$ (a) diffusive and (b) jump trajectories.
  The average evolution given by the Lindblad equation is time-independent.
  Individual trajectories accumulate at the boundaries and select one of the minimal orthogonal decoherence-free subspaces $\ket{q_n}$ with probability given by the initial overlap $w_n = |\braket{\Psi(0)}{q_n}|^2 = 1/2$.
  Localization is thus complete (cf. \cref{sec:incomplete}).
  Two typical realizations are highlighted where the system selects the excited (violet) and ground state respectively (red). The parameters are $\omega=1$ and $\Gamma=0.7$}
  \label{fig:single-qubit}
\end{figure}

\subsection{Continuous measurement of a qubit}
\label{sec:single-qubit}
To illustrate our general results, we first revisit the minimal example of a continuously monitored qubit (cf. \cref{fig:introduction}).
Consider the two-level system with Hamiltonian
\begin{align}
  H = \omega \sigma^z,
\end{align}
where $\omega$ is the level-spacing and $\sigma^z$ denotes the Pauli operator in $z$-direction.
This otherwise closed quantum system is monitored such that it evolves along a quantum trajectory with Lindblad jump operator
\begin{align}
  L = \sqrt{\Gamma}\sigma^+\sigma^-,
\end{align}
where $\Gamma$ denotes the measurement strength.
The slightly contrived form of the measurement operator is necessary to induce subspace selection on both unravelings.
In particular, the choice $L=\sigma^z$ would yield $((\sigma^z)^\dagger \sigma^z) = \mathds{1}$, which is case (i) of \cref{th:incomplete-jump} and prevents localization.

By construction, the Hamiltonian and the measurement operator commute $[H,L] = 0$.
They thus share a common eigenbasis, hence constituting a decoherence-free subspace.
The operator $L$ is unitarily diagonalizable and has distinct eigenvalues $c_1 = 1$ and $c_2 = 0$.
The two eigenstates $\ket{q_1} = \ket{0}$ and $\ket{q_2} = \ket{1}$ are orthogonal and correspond to two distinct one-dimensional decoherence-free subspaces $\calq_1$ and $\calq_2$ respectively (cf. \cref{sec:decaying-subspace}).
We display the structure of the Hilbert space in \cref{fig:single-qubit-structure}.
Consequently, \cref{eq:dq_12,eq:dq_12-jump} apply, guaranteeing that the quantum system asymptotically selects one of the two subspaces,  as demonstrated in \cref{fig:single-qubit}.
The states $\ket{q_n}$ are attractive invariant states of both unravelings (see \cref{sec:invariant-states}).
Since the decaying subspace is empty and localization complete, ergodicity depends only on the distribution of the initial state over the subspaces and is given by (cf. \cref{eq:main})
\begin{align}
  \mathbb{E}[F(\overline{\rho}_\m{c},\rho^\m{s})]
  = w_1^2+(1-w_1)^2,
\end{align}
where $w_1 = \tr[\rho(0)P_{\calq_1}]$, corresponding to the bistable, geometric interpretation in \cref{fig:geometric-fidelity}a.
The generalized update rule, \cref{eq:update-rule}, coincides with the standard projective measurement 
\begin{align}
  \rho(0) \to \rho_{c,n} 
  = \frac{\sfp_n \rho(0)}{\tr[\sfp_n \rho(0)]}
  = \dyad{q_n},
\end{align}
This happens with probability $p_n = \tr[\sfp_n \rho(0)]$, where $\sfp_n(\bullet) = P_{\calq_n} \bullet P_{\calq_n}$ and $P_{\calq_n} = \dyad{q_n}$.
In \cref{fig:single-qubit}, we plot the $z$-polarization of the qubit for a subensemble of diffusive and jump trajectories for the initial state $\ket{\Psi(0)} = (\ket{0}+\ket{1})/\sqrt{2}$, where individual trajectories deviate most strongly from the mean and the largest breaking of ergodicity occurs with $\mathbb{E}[F(\overline{\rho}_\m{c},\rho^\m{s})] = 1/2$.

\begin{figure}[t]
  \centering 
  \begin{tikzpicture}
    \node (a) at (0,0) {\includegraphics[scale=0.9]{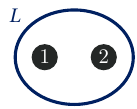}};
    \node (b) at (3.8,0) {\includegraphics[scale=0.7]{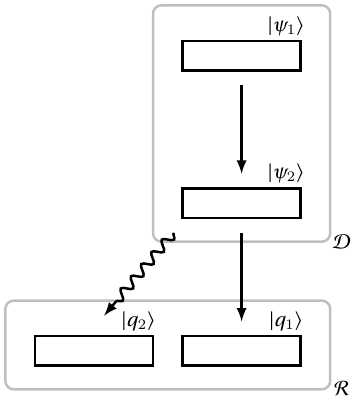}};
    \draw (-1.2,1) node {\textbf{(a)}};
    \draw (3,2.2) node {\textbf{(b)}};
  \end{tikzpicture}
  \caption{(a) Noninteracting qubits subject to common measurement. (b) Hilbert space structure and flow induced by the Lindblad equation of the two qubit system \cref{eq:dark-sync} with $H = \omega_0/2 \sum_k \sigma^z_k$ coupled to a common environment according to $L = \sqrt{\Gamma}(\sigma^-_1+\sigma^-_2)$.
  Each rectangle corresponds to an orthogonal subspace. Arrows indicate unidirectional flow between subspaces. Subspaces without inflow or outflow (no arrows) correspond to conserved projectors (cf. \cref{sec:space-structure}). The decaying subspace is non-empty $\cald = \{\ket{\psi_1},\ket{\psi_2}\}$ (\cref{eq:D}). The asymptotic subspace is spanned by two dark states, $\calr = \calq_1 \oplus \calq_2$ (\cref{eq:R}). $\cald$ gets completely emptied during the evolution and there is a hierarchical flow of probability into the subspace $\calq_1$. On the level of the Lindblad equation (bold dashed lines in \cref{fig:dark-example}a,c), the projector $P_{\calq_2}$ is conserved, $\sfl^\dagger(P_{\calq_2}) = 0$. 
  On the trajectory level (thin lines in \cref{fig:dark-example}a,c), conserved quantities are no longer respected and there is an additional flow from $\ket{\psi_2}$ to $\ket{q_2}$ (wavy arrow).}
  \label{fig:dark-structure}
\end{figure}

\begin{figure*}[t]
  \centering 
  \begin{tikzpicture}
      \node (a) [label={[label distance= -0.2cm]125: \textbf{(a)}}] at (0,0) {\includegraphics{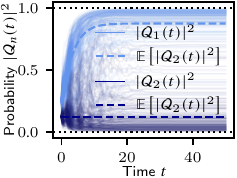}};	
      \node (b) [label={[label distance= -0.2cm]125: \textbf{(b)}}] at (4.2,0) {\includegraphics{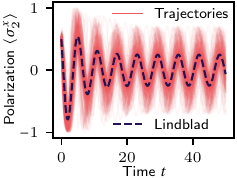}};	
      \node (c) [label={[label distance=-.2 cm]125: \textbf{(c)}}] at (8.4,0) {\includegraphics{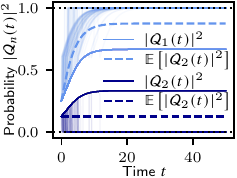}};
      \node (d) [label={[label distance=-.2 cm]125: \textbf{(d)}}] at (12.6,0) {\includegraphics{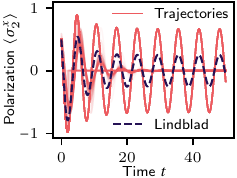}};
      \draw (0.2,1.7) node {\textsf{Diffusion}};
      \draw (4.4,1.7) node {\textsf{Diffusion}};
      \draw (8.6,1.7) node {\textsf{Jumps}};
      \draw (12.8,1.7) node {\textsf{Jumps}};
  \end{tikzpicture}
  \caption{Non-interacting qubits subject to non-local continuous measurement (\cref{eq:dark-sync,fig:dark-structure}).
  The asymptotic subspace $\calr$ is a two-dimensional dark subspace attracting all quantum trajectories.
  (a),(b) Distribution of the overlap with the elements of the dark subspace, $\calq_1$ and $\calq_2$ and evolution of the $x$-polarization of the second qubit for $200$ diffusive trajectories. The two qubits are asymptotically anti-synchronized to each other, for better visibility we only show the second qubit.
  Due to the decaying subspace, probability can be transferred between the dark states, a process that is forbidden on the ensemble level.
  Localization is incomplete and trajectories converge to random convex combinations of the dark states on the trajectory level, resulting in oscillations with random phases and on average increased amplitude.
  The ensemble average is represented by bold dashed lines respectively.
  (c),(d) Distribution of the overlap and evolution of the $x$-polarization for quantum jumps.
  Either at least one jump occurs and the system goes entirely into the collective ground state $\ket{q_1}$ where the evolution stops, or no detection happens and there is a no-click trajectory with probability, $\tr[\rho(0)P_\calr] = 3/8$, resulting in coherent oscillations between the dark states and stable, increased oscillations in the $x$-polarizations.
  The parameters are $\omega_0=1$ and $\Gamma = 1/5$.}
\label{fig:dark-example}
\end{figure*}

\subsection{Coherent oscillations of two qubits}
\label{sec:two-qubits}
With this seemingly simple example we will illustrate the nontrivial influence of the decaying subspace on the behavior of individual trajectories.

Consider two atoms at fixed positions coupled to the quantized electromagnetic field in vacuum (zero temperature environment) \cite{Agarwal1974,Damanet2016}. If the atoms are sufficiently distant from each other, dipolar interactions become negligible and the system Hamiltonian in the interaction picture becomes $H = \omega_0/2\sum_j\sigma^z_j$, where $\omega_0$ denotes the single qubit level spacing. The evolution of the density matrix can then effectively be described by the Lindblad equation \cite{Agarwal1974,Damanet2016}
\begin{align}
  \dot{\rho} = -\I[H,\rho] 
  + \sum_{j,k} \Gamma_{jk} \left(\sigma^-_k \rho \sigma^+_j - \frac{1}{2}\{\sigma^+_j\sigma^-_k,\rho\}\right),
  \label{eq:dark-sync}
\end{align}
where the dissipative term accounts for spontaneous photon emission.
For scalar coupling to the environment $\Gamma \equiv \Gamma_{ij}$, this system has been shown to feature stable oscillations between the local $x$-polarizations in the long-time limit \cite{Karpat2020}.
The dissipator can be written in diagonal form with the single Lindblad jump operator, $L = \sqrt{\Gamma}(\sigma^-_1+\sigma^-_2)$.
The system has two dark states given by 
\begin{align}
  \ket{q_1} = \ket{00}, \quad \ket{q_2} = (\ket{10}-\ket{01})/\sqrt{2},
  \label{eq:R}
\end{align}
where $\ket{q_1} = \calq_1$ is the collective ground state and $\ket{q_2} = \calq_2$ is a Bell state.
Both are eigenstates of the Hamiltonian and get annihilated by the Lindblad operator, $L\ket{q_n} = 0$, $n=1,2$, thus forming a two-dimensional dark subspace.
The two remaining states comprise the complement
\begin{align}
  \ket{\psi_1} = \ket{11}, \quad \ket{\psi_2} = (\ket{10}+\ket{01})/\sqrt{2}
  \label{eq:D}
\end{align}
and belong to the decaying subspace $\cald = \{\ket{\psi_1},\ket{\psi_2}\}$ (cf. \cref{sec:space-structure,sec:decaying-subspace}), which will get emptied out completely in the course of time.
The asymptotic state space and therefore the final destination of any trajectory (cf. \cref{sec:no-trajs-in-D}) is thus entirely dark with $\calr = \calq_1 \oplus \calq_2$ (cf. \cref{eq:asymptotic-subspace}).
The structure of the state space is presented in \cref{fig:dark-structure}.
However, localization is generally incomplete for both unravelings, since trajectories cannot distinguish between orthogonal subspaces belonging to the same decoherence-free subspace (see \cref{eq:dq_12,eq:dq_12-jump}).
As detailed below, this fact together with the presence of a decaying subspace precludes us from predicting the distribution of the asymptotic weights in the minimal subspaces.

To explicitly demonstrate the implications on the behavior of quantum trajectories, we choose as an initial state the coherent product state \cite{Karpat2020}
\begin{align}
  \ket{\Psi(0)} 
  =& (\cos(\theta)\ket{0} + \sin(\theta)\ket{1}) \\
  &\otimes (\cos(\theta)\ket{0} + e^{-i\pi/3}\sin(\theta)\ket{1}),
\end{align}
with $\theta = \pi/4$.
In \cref{fig:dark-example}, we accordingly show the evolution of the overlap $|\calq_n(t)|^2 = \tr[\rho_\m{c}(t)P_{\calq_n}]$ with the minimal subspaces and the $x$-polarizations.
The two unravelings display radically different behavior.
Diffusive trajectories (\cref{fig:dark-example}a) exhibit a continuous distribution in the asymptotic weights that is peaked near the boundaries but with considerable deviations from the mean.
Along the ensemble averaged evolution, the overlap with the Bell state $\ket{q_2}$ is conserved, $\mathbb{E}[|\calq_2(t)|^2]=|\calq_2(0)|^2$, while the subspace $\calq_1$ receives a flow of probability from the decaying subspace (cf. \cref{fig:dark-structure}).
On the trajectory level, the downflow from the decaying subspace effectively enables transitions between the orthogonal subspaces $\calq_1$ and $\calq_2$ (wavy line in \cref{fig:dark-structure}) that are otherwise forbidden.
Convergence to the dark subspace terminates at random times with random weights $|\calq_n|^2_\xi$ in each subspace (cf. \cref{eq:degenerate-update-rule}).
As a result, measurement-induced oscillations occur with an average increased amplitude, at the expense of randomized phases, cf. \cref{fig:dark-example}b.

Jump trajectories, on the other hand, exhibit a discrete, bimodal distribution in the asymptotic weights.
Either, multiple jumps occur and the system is driven into the ground state, where it then remains due to the lack of absorption processes and oscillations are suppressed.
Or, in the complementary event, there is a no-click trajectory (indicated by $\m{nc}$) that occurs with probability $\tr[\rho(0)P_\calr] = 3/8$, builds up coherences between the dark states and induces coherent oscillations in the $x$-polarizations (\cref{fig:dark-example}d).
The asymptotic weight in each of the subspaces becomes $|\calq_1|^2_\m{nc} = 1-|\calq_2|^2_\m{nc} = 2/3$.
Although condition (i) of \cref{th:incomplete-jump} is satisfied in this model with $L_{\calq_1} = L_{\calq_2} = 0$, complete localization can still occasionally take place (here with probability $5/8$).
This is however no contradiction because in these instances the trajectory never actually has full support on $\calq_1 \oplus \calq_2$ due to the unidirectional downflow from $\cald$ into $\calq_1$.
The decaying subspace can never be cut until full localization has already occurred and the subspace $\calq_1$ is thus, until then, never actually orthogonal to $\cald$.

Generally, if localization is incomplete, we cannot predict the distributions in \cref{fig:dark-example}a,c.
However, while the diffusive case remains inaccessible, in this particular example, it is possible to obtain the generalized update rule and to quantify ergodicity breaking in the quantum jump unraveling.
The update rule becomes
\begin{align}
  \rho(0) \to \rho_\m{c} = 
  \begin{dcases}
    \frac{\sfu^t_{\calr}(\sfp_\calr \rho(0))}{\tr[\sfp_\calr \rho(0)]}, &\text{ w. prob. } p_\calr,\\
    \dyad{q_1}, &\text{ w. prob. } 1-p_\calr,
  \end{dcases}
\end{align}
where $p_\calr = \tr[\sfp_\calr \rho(0)]$.
The update rule of the diffusive trajectories is of the general form \cref{eq:degenerate-update-rule}.
The mean fidelity for quantum jumps becomes
\begin{align}
  \mathbb{E}[F(\overline{\rho}_J,\rho^\m{s})]
  =& z\left(\sqrt{xy}+\sqrt{(1-x)(1-y)}\right)^2\\
  &+ (1-z)y,
\end{align}
where $z \equiv \tr[\rho(0)P_\calr]$ is the probability to observe a no-click trajectory, $x = |\calq_1|^2_\m{nc}$ is the asymptotic overlap of the no-click trajectory with the ground state, and, $y = |\calq_1^\infty|^2$ is the effective weight of the Lindblad equation in the subspace $\calq_1$.

\begin{figure}[t]
  \centering
  \begin{tikzpicture}
    \node (a) [label={[label distance= -0.2cm]130: \textbf{(a)}}] at (0,0) {\includegraphics[scale=0.9]{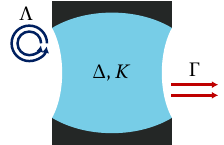}};	
    \node (a) [label={[label distance=.4 cm]150: \textbf{(b)}}] at (4.5,0) {\includegraphics[scale=0.7]{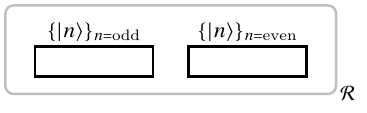}}; 
\end{tikzpicture}
\caption{Detuned driven-dissipative Kerr resonator \cref{eq:Kerr}. (a) The system is subject to coherent two-photon pumping of strength $\Lambda$ and two-photon loss at a rate $\Gamma$. The external driving frequency is detuned from the cavity-mode frequency by $\Delta$. The additional Kerr nonlinearity $K$ effectively introduces photon-photon interactions. (b) For nonzero detuning, the Hilbert space splits into parity sectors of even and odd photon number and the asymptotic state space, $\calh = \calr$, takes up the entire Hilbert space rendering the decaying subspace empty, $\cald = \emptyset$. 
\Cref{th:incomplete-diff,th:incomplete-jump} guarantee complete localization and trajectories select one of the parity sectors with probability given by the initial overlap.}
\label{fig:kerr-oscillator}
\end{figure}

\begin{figure*}[t]
  \centering
  \begin{tikzpicture}
    \node (a) [label={[label distance=-0.2cm]125: \textbf{(a)}}] at (0,0) {\includegraphics{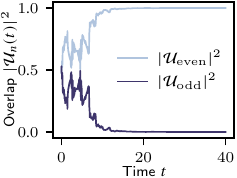}};
    \node (b) [label={[label distance=-0.2cm]125: \textbf{(b)}}] at (4.2,0) {\includegraphics{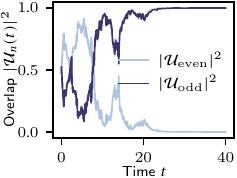}};
    \node (c) [label={[label distance=-0.2cm]125: \textbf{(c)}}] at (8.4,0) {\includegraphics{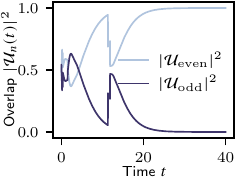}};
    \node (d) [label={[label distance=-0.2cm]125: \textbf{(d)}}] at (12.6,0) {\includegraphics{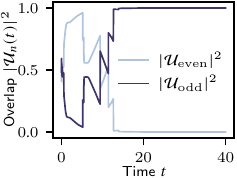}};
    \draw (0.2,1.7) node {\textsf{Diffusion}};
    \draw (4.4,1.7) node {\textsf{Diffusion}};
    \draw (8.6,1.7) node {\textsf{Jumps}};
    \draw (12.8,1.7) node {\textsf{Jumps}};
\end{tikzpicture}
\caption{Localization transitions of the continuously monitored driven-detuned Kerr oscillator (\cref{eq:Kerr,fig:kerr-oscillator}).
The initial state is a superposition of two Fock states $\ket{\Psi(0)} = (\ket{8}+\ket{9})/\sqrt{2}$.
For nonzero detuning, the Hilbert space decomposes into two parity sectors for $n=\m{even}$ and $n=\m{odd}$ photon number and individual quantum trajectories are gradually projected to a definite parity with probability given by the initial overlap $w_n = \tr[\rho(0)P_n] = 1/2$.
(a),(c) Spontaneous localization transition into the even parity subspace.
(b),(d) Spontaneous localization transition into the odd parity subspace.
The parameters are $\Delta=2$, $\Lambda=7/4$, $K=1/3$, $\Gamma=1/2$. The Fock space dimension is $N=20$.}
\label{fig:parity-selection}
\end{figure*}
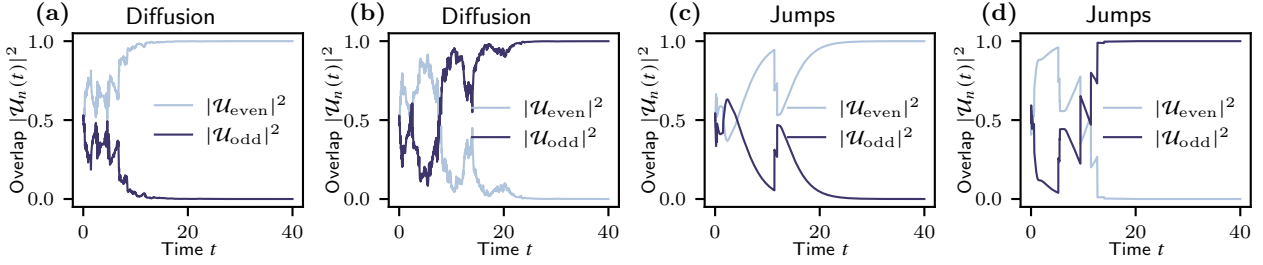

\subsection{Driven dissipative Kerr resonator}
\label{sec:Kerr}
In this example we consider a driven quantum oscillator comprising an, in principle, infinite-dimensional bosonic system and show that our results still apply if the analysis remains restricted to a finite number of Fock states.
Concretely, we study a coherently driven optical cavity with a Kerr nonlinearity subject to two-photon loss.
In the interaction picture, in the frame rotating with two times the driving frequency, the system can be described by the following Hamiltonian \cite{Wolinsky1988,Mirrahimi2014,Bartolo2016,Roberts2020,Landi2024}
\begin{align}
  H = 
  -\Delta a^\dagger a 
  + \frac{1}{2}(\Lambda a^{\dagger 2} + \Lambda^\ast a^2)
  + \frac{K}{2} a^\dagger a^\dagger a a.
\end{align}
Here, $\Delta = \omega_d/2-\omega_\m{c}$ denotes the detuning between the cavity resonance frequency $\omega_\m{c}$ and the frequency of the drive $\omega_d$. The system is coherently driven via a two-photon pumping process with complex amplitude $\Lambda$, where the Kerr nonlinearity, $K$, effectively introduces photon-photon interactions (\cref{fig:kerr-oscillator}a).
If we account for two-photon loss, $L = \sqrt{\Gamma}a^2$, with a rate $\Gamma$, the dissipative evolution of the system can be effectively described by the following Lindblad master equation 
\begin{align}
  \dot{\rho} = 
  -\I[H,\rho] + \Gamma \left(a^2\rho (a^\dagger)^2-\frac{1}{2}\{(a^\dagger)^2a^2,\rho\}\right).
  \label{eq:Kerr}
\end{align}
For zero detuning, $\Delta = 0$, this model exhibits bistability with two extremal stationary states spanning the asymptotic subspace, while all other states experience decay.
The steady states are given by the Cat-states $\ket{c^\pm} = \mathcal{N}(\ket{\alpha} \pm \ket{-\alpha})$, forming a two-dimensional decoherence-free subspace, where $\alpha = \sqrt{\Lambda/\Gamma}$ ($\ket{\pm \alpha}$ are the usual coherent states and $\mathcal{N}$ is the normalization) \cite{Wolinsky1988,Mirrahimi2014,Bartolo2016,Roberts2020}.
Since the restriction of $L$ to both Cat-state subspaces is the same, $\bra{c^+}L\ket{c^+} = \bra{c^-}L\ket{c^-} = \alpha^2$, we have $L_\calr \propto \mathds{1}$ which is case (i) of \cref{th:incomplete-diff,th:incomplete-jump}, and trajectories are unable to discern between the subspaces.
The Cat-states are invariant states of the quantum trajectories (cf. \cref{sec:invariant-states}) but are attractive only when the system is initialized with definite parity \cite{Mirrahimi2014}.
Therefore, we here consider finite detuning, $\Delta \neq 0$, which eliminates the decaying subspace.
Bistability still persists but the steady state manifold transitions from a decoherence-free subspace to a unique stationary state supported in the even and odd photon number parity sector respectively \cite{Roberts2020}.
The Hilbert space thus decomposes into a direct sum of these two subspaces and the asymptotic state space is given by $\calr = \calu_\m{odd} \oplus \calu_\m{even}$, where $\calu_\m{odd/even} = \{\ket{n}\}_{n=\m{odd}/\m{even}}$ are the subspaces of Fock states with even and odd photon number respectively.

\cref{th:incomplete-jump,th:incomplete-diff} now guarantee full localization and, as shown in \cref{fig:parity-selection}, both the diffusive and the quantum jump unraveling will select one of the two parity sectors where the time average will then converge to one of the two steady states of the Lindblad equation.
In \cref{fig:kerr-oscillator}b we show the Hilbert space structure of the detuned Kerr oscillator model under two-photon loss.
Denote by $P_\m{odd}$ and $P_\m{even}$ the projectors onto the subspaces $\calu_\m{odd}$ and $\calu_\m{even}$ respectively.
Then, the effective asymptotic projection of the generalized update rule \cref{eq:update-rule} projects the system onto a quantum trajectory with a definite, conserved parity according to 
\begin{align}
    \rho(0) \rightarrow
    \rho_\m{c} = 
    \begin{dcases}
        \frac{\sft^t_{\m{c},\m{even}}(\sfp_{\m{even}} \rho(0))}{\tr[(\sfp_{\m{even}} \rho(0))]}, \text{ w. prob. } p_\m{even} \\
        \frac{\sft^t_{\m{c},\m{odd}}(\sfp_{\m{odd}} \rho(0))}{\tr[(\sfp_{\m{odd}} \rho(0))]}, \text{ w. prob. } p_\m{odd}
    \end{dcases},
\end{align}
with probabilities $p_\m{even} = \tr[\sfp_\m{even}\rho(0)]$ and $p_\m{odd} = \tr[\sfp_\m{odd}\rho(0)]$.
For any finite number of Fock states, $N$, in the effective finite-dimensional Hilbert space, the mean fidelity between time and ensemble averaged density matrices is thus given by 
\begin{align}
\mathbb[F(\rho^\mathrm{s},\overline{\rho}_\m{c})] 
= w^2_\m{even}(1-w_\m{even})^2,
\end{align}
with $w_\m{even} = |\calu_\m{even}(0)|^2$,
and corresponds to the geometrical representation in \cref{fig:geometric-fidelity}a,c.
This shows that violations of ergodicity may remain unaffected by the Hilbert space dimension.

\begin{figure}[t]
  \centering
  \includegraphics[scale=0.7]{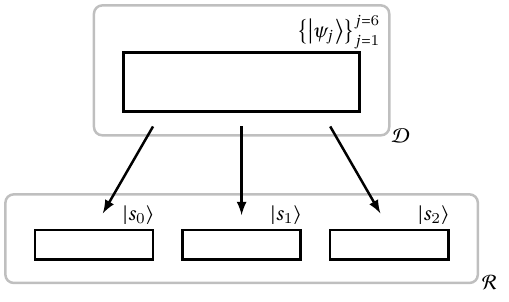}
  \caption{State space structure of the $N=2$ spin-1 system (\cref{eq:scar-H,eq:scar-L}).
  All the chaotic states of the non-integrable Hamiltonian lie in the decaying subspace.
  There is no substructure to $\cald$.
  Probability flows down along three decay channels, one for each scar state.
  The scar states comprise a $N+1$-dimensional decoherence-free subspace, which prevents localization between the scars (cf. \cref{th:incomplete-diff,th:incomplete-jump}), but leads to the build-up of quantum coherences (see \cref{fig:scars}) due to the existence of the decaying subspace.
  Adding global dephasing (\cref{eq:Lz}) partly lifts the degeneracy for quantum jumps and entirely for quantum diffusion, leading to partial and complete localization respectively.
  The commutant is trivial, $\{H,L_k,L^\dagger_k\}^\prime = z\mathds{1}$, nevertheless the system features $N+1$ extremal stationary states.}
  \label{fig:scar-structure}
\end{figure}
\subsection{Stabilizing many-body scar states}
\label{sec:scar states}
In this example we illustrate that continuous measurement can induce regular dynamics in a chaotic system by singling out special integrable parts of the Hilbert space.

Consider a closed system with a non-integrable Hamiltonian $H$.
Although the level spacing might follow a Wigner-Dyson distribution, there can be exceptional rare scar states embedded inside the chaotic subspaces that do not adhere to the overall statistics \cite{Bernien2017,Moudgalya2018,Turner2018,Serbyn2021}.
These regular states clearly do not conform with the eigenstate thermalization hypothesis that would predict thermalization of local observables in generic non-integrable models.
If only a tiny fraction of the total Hilbert space is occupied by scars, their presence leaves no noticeable imprint on the level spacing distribution but, on the other hand, allows at most for a weak form of the eigenstate thermalization hypothesis to hold \cite{Biroli2010,Kim2014,Naoto2017,Serbyn2021}.
\begin{figure*}[t]
  \centering
  \begin{tikzpicture}
    \node (a) [label={[label distance= -0.2cm]125: \textbf{(a)}}] at (0,0) {\includegraphics{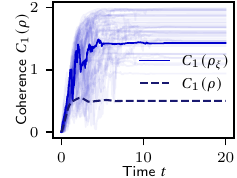}};	
    \node (b) [label={[label distance=-.2 cm]125: \textbf{(b)}}] at (4.2,0) {\includegraphics{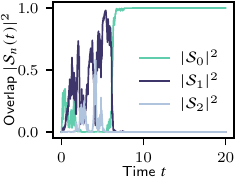}}; 
    \node (c) [label={[label distance=-.2 cm]125: \textbf{(c)}}] at (8.4,0) {\includegraphics{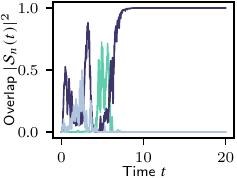}}; 
    \node (d) [label={[label distance=-.2 cm]125: \textbf{(d)}}] at (12.6,0) {\includegraphics{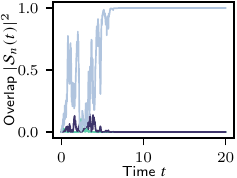}};
    \node (e) [label={[label distance= -0.2cm]125: \textbf{(e)}}] at (0,-3.6) {\includegraphics{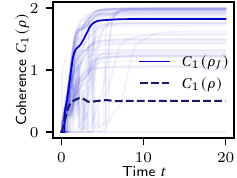}};	
    \node (f) [label={[label distance= -0.2cm]125: \textbf{(f)}}] at (4.2,-3.6) {\includegraphics{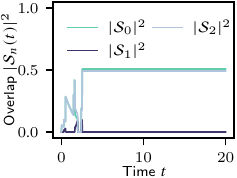}};	
    \node (g) [label={[label distance=-.2 cm]125: \textbf{(g)}}] at (8.4,-3.6) {\includegraphics{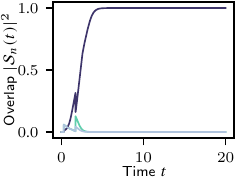}};
    \draw (0.3,1.7) node {\textsf{Diffusion}; $L_z = 0$};
    \draw (4.5,1.7) node {\textsf{Diffusion}; $L_z \neq 0$};
    \draw (8.7,1.7) node {\textsf{Diffusion}; $L_z \neq 0$};
    \draw (13,1.7) node {\textsf{Diffusion}; $L_z \neq 0$};
    \draw (0.3,-1.9) node {\textsf{Jumps}; $L_z = 0$};
    \draw (4.5,-1.9) node {\textsf{Jumps}; $L_z \neq 0$};
    \draw (8.7,-1.9) node {\textsf{Jumps}; $L_z \neq 0$};
\end{tikzpicture}
  \caption{Two spin-1 particles subject to continuous measurement (\cref{eq:scar-H,eq:scar-L}).
  The measurement creates a stable invariant manifold exclusively spanned by the $N+1$ many-body scar states in the spectrum of the otherwise chaotic Hamiltonian.
  The scar states, $\ket{s_n}$, span a $N+1$-dimensional decoherence-free subspace whose substructures is not discernable on the trajectory level (\cref{th:incomplete-diff,th:incomplete-jump}) but allows for asymptotic coherences (see the structure in \cref{fig:scar-structure}).
  (a),(e) Evolution of the $l_1$-norm of coherence in the scar state basis (cf. \cref{eq:l1-norm}) for $50$ realizations (thin lines) and for the Lindbladian dynamics (bold dashed line).
  (b),(c),(d) Overlap with the individual scar states for diffusive trajectories (\cref{eq:scar-tower}). 
  Adding global dephasing (\cref{eq:Lz}) lifts the degeneracy, trajectories asymptotically purify and get mapped to exactly one of the scar states. Localization is complete.
  (f),(g) Overlap with the individual scar state subspaces $\cals_n$ for individual jump trajectories.
  The global dephasing makes only half (more precisely $\lfloor (N+1/2) \rfloor$) of the scar states distinguishable. Trajectories cannot decide between $\ket{s_1}$ and $\ket{s_2}$. Localization is partially incomplete.
  The parameters are $h=1$, $D=1.3$ and $\Gamma=1$.
  }
  \label{fig:scars}
\end{figure*}
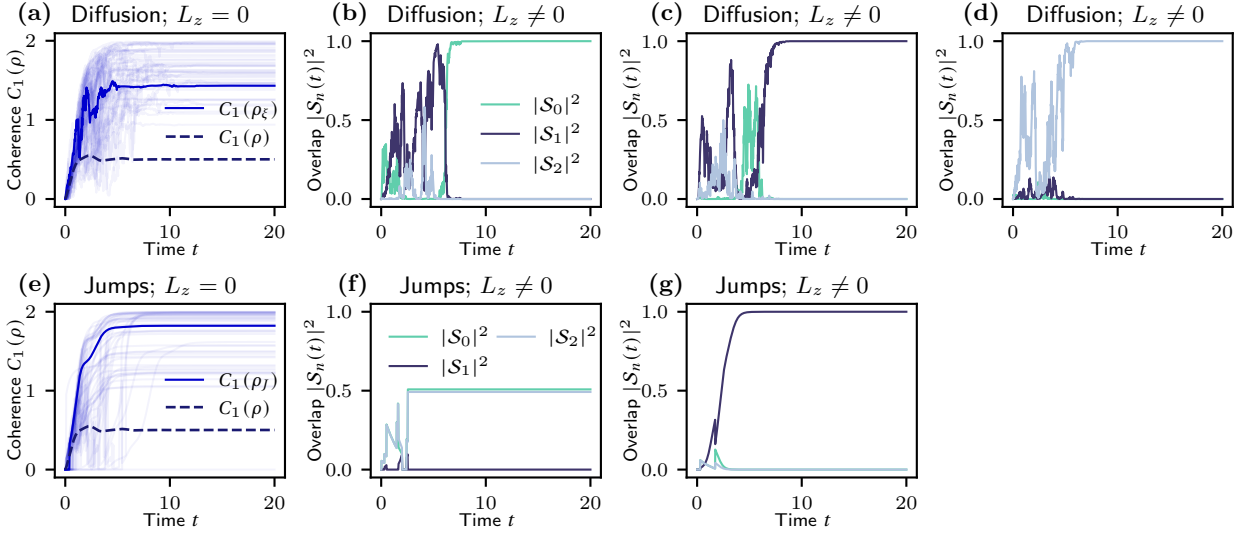
Many-body quantum scar states can be embedded into the decoherence-free subspaces of a Lindbladian by identifying a suitable dissipative process in such a way that all chaotic states get annihilated \cite{Wang2024}.
In terms of the state space structure, this amounts to engineering the decaying subspace and creating a unidirectional flow into designated basins containing the targeted scar states.

We consider the general spin-1 model put forth in Refs.~\cite{Schecter2019,Wang2024}, determined by the Hamiltonian and Lindblad jump operators
\begin{align}
  H &= 
  \sum_j (S_j^xS_{j+1}^x+S_j^yS_{j+1}^y)
  + h \sum_j S^z_j 
  + D \sum_j (S^z_j)^2,
  \label{eq:scar-H}\\
  L_k &= S_k^x(S^x_kS^x_{k+1}+S^y_kS^y_{k+1})
  \label{eq:scar-L}
\end{align}
where $S_k^a$, $a = (x,y,z)$ are the spin operators, $h$ is the strength of an external magnetic field, and $D$ is the magnitude of a possible self interaction.
The bare closed system has been shown to be non-integrable for any choice of parameters in arbitrary spatial dimensions \cite{Schecter2019}.
The Lindblad jump operators $L_k$ create a unique invariant manifold, that consists only of the ladder of evenly spaced scar states $\ket{s_n}$.
This tower of scar states can be obtained by repeated action of the ladder operator, $Q^\dagger = \sum_j (-1)^j(S^+_j)^2$, onto the collective ground state, $\ket{s_0} = \ket{-1,\ldots,-1}$, according to
\begin{align}
  \ket{s_n} = (Q^\dagger)^n \ket{s_0},
  \label{eq:scar-tower}
\end{align}
where $n = 0,1,\ldots,N$.
Every scar state $\ket{s_n}$ belongs to a one-dimensional orthogonal subspace $\cals_n$.
The scar states effectively form a decoherence-free subspace, that prevents full localization of trajectories and generally leads to coherent oscillations between them (cf. \cref{sec:space-structure,sec:incomplete}).

We illustrate these general results explicitly for $N=2$ coupled spin-1 particles.
In \cref{fig:scar-structure} we show the corresponding Hilbert space structure induced by the Lindblad equation.
The $N+1$-dimensional decoherence-free subspace allows for steady state coherences of the Lindblad equation.
To quantify the quantum coherence between the scar states in the asymptotic regime, we employ the $l_1$-norm of coherence \cite{Baumgratz2014}
\begin{align}
  C_1(\rho) = \sum_{n \neq m} \bra{s_n} \rho \ket{s_m}.
  \label{eq:l1-norm}
\end{align}
We choose an initial pure state that is delocalized over the decaying subspace and has zero overlap in $\calr$.
In particular, we project the uniform superposition in the $z$-basis onto the decaying subspace
\begin{align}
  \ket{\Psi(0)} = \mathcal{N}\bigg(P_\cald \quad \sum_{\mathclap{s_1,s_2=\{-1,0,1\}}} \ket{s_1,s_2}\bigg),
\end{align}
where $\mathcal{N}$ accounts for the normalization.
In \cref{fig:scars}a,e we plot $C_1(\rho)$ as a function of time for individual trajectories and the Lindblad equation.
On average, individual trajectories build up more coherence than the Lindbladian evolution.
Although this clearly indicates non-ergodic behavior, incomplete localization prevents prediction of the mean fidelity between time and ensemble averaged dynamics and the generalized update rule remains stochastic (cf. \cref{eq:degenerate-update-rule}).

To lift the degeneracy, we now additionally introduce global dephasing with
\begin{align}
  L_z &= \sqrt{\Gamma} \sum_j S^z_j,
  \label{eq:Lz}
\end{align}
where $\Gamma$ is a real parameter controlling the strength.
The jump operator $L_z$ (not featured in the original model of Ref.~\cite{Wang2024}) does not deform the asymptotic state space but changes its internal symmetry.
Trajectories will enter the whole invariant manifold and exhibit unitary evolution in the decoherence-free subspace.
However, as shown in \cref{fig:scars}b,c,d, nonzero dephasing, $\Gamma\neq 0$, breaks the subspace equivalence and induces complete localization for all diffusive trajectories.
On the other hand, the jump operator $L_z$ is not sufficient to lift entirely the equivalence of quantum jump trajectories because jump operators restricted to minimal orthogonal subspaces, $L_{z,n} = \bra{s_n}L_z\ket{s_n}$, are still related by a phase, $L_{z,n} = -L_{z,N+2-n}$, $n \in [1,\lfloor(N+1)/2\rfloor]$ (cf. case(i) of \cref{th:incomplete-jump}).
In the particular case of $N=2$ spins, quantum jump trajectories are not able to decide between the subspaces $\cals_0$ and $\cals_2$, see \cref{fig:scars}f,g.

Whereas on the level of the Lindblad equation the steady state will always be a classical mixture of the scar states (excluding the case where $\rho(0) = \dyad{s_n}$), the diffusive quantum trajectories will single out exactly one of the scar states for any arbitrary initial state.
All scar states thus become attractive fixed points of the diffusive quantum trajectory individually and each scar state constitutes an invariant state (cf. \cref{th:diff-inv}).
The generalized update rule (\cref{eq:update-rule}) then takes a form similar to conventional projective measurement where the initial state is mapped to a scar state according to
\begin{align}
    \rho(0) \rightarrow
    \rho_{\xi,n} = 
    \frac{\sfp^\infty_{s_n}\rho(0)}{\tr[\sfp^\infty_{s_n}\rho(0)]} = 
    \dyad{s_n},
\end{align}
with probability $p_n = |\cals^\infty_n|^2 = \tr[\sfp^\infty_{s_n}\rho(0)]$.
The projections $\sfp^\infty_n$ are however oblique projectors that will generally map the state outside of its support.
\Cref{eq:main} thus strictly applies only to the diffusive unraveling and violations of ergodicity are captured by the mean fidelity
\begin{align}
  \mathbb{E}[F(\rho^\m{s},\overline{\rho}_\xi)] 
  = \sum_{n=0}^{N} w_n^2,
\end{align}
where, $w_n = |\cals_n^\infty|^2$, are the asymptotic conserved weights in each scar state subspace.
In contrast to the Kerr oscillator model (\cref{sec:Kerr}), violations of ergodicity grow with the system size.

\begin{figure}[t]
  \centering
  \begin{tikzpicture}
    \node (a) at (0,0) {\includegraphics[scale=0.9]{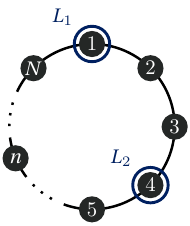}};	
    \node (a) at (4.15,0) {\includegraphics[scale=0.7]{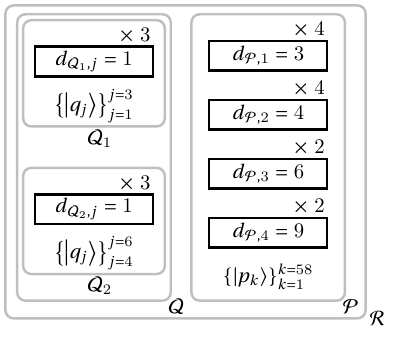}}; 
    \draw (-1.5,1.75) node {\textbf{(a)}};
    \draw (1.5,1.75) node {\textbf{(b)}};
\end{tikzpicture}
\caption{(a) Heisenberg $XX$-model on a ring (cf. \cref{eq:XX-Hamiltonian}) with two-site dephasing, $L_1=\sqrt{\Gamma}\sigma^z_u$, $L_2=\sqrt{\Gamma}\sigma^z_v$. Provided conditions \eqref{eq:sync5} and \eqref{eq:syncN} are satisfied, there is a unique decoherence-free mode.
(b) Hilbert space structure of the $N=6$ ring with $L_1=\sqrt{\Gamma}\sigma^z_1$, $L_2= \sqrt{\Gamma}\sigma^z_4$ (since $\cald = \emptyset$, \cref{alg:sbd} outputs the full decomposition).
The total Hilbert space can be decomposed into the collection of decoherence-free subspaces, $\calq = \calq_1 \oplus \calq_2$, and its complement, $\calp = \bigoplus_{j=1}^{12} \calp_j$. There is a total of $18$ minimal orthogonal subspaces: $3$ one-dimensional subspaces in $\calq_1$ and $\calq_2$ respectively and $12$ minimal subspaces in $\calp$ with dimensions $d_{\calp,k} = 3,4,6,9$. The initial state $\ket{\Psi(0)} = \ket{010\ldots 0}$ has overlap with two states in $\calq_1$ and a four-dimensional minimal subspace in $\calp$, with $|\calq_1(0)|^2 = |\calp(0)|^2 = 1/2$. Individual diffusive trajectories either select $\calq$ and evolve unitarily with maximally entangled oscillations or $\calp$ with irregular, stochastic dynamics, cf. \cref{fig:ring-evolution}.}
\label{fig:ring}
\end{figure}

\subsection{Heisenberg $XX$-ring}
\label{sec:ring}
In this section we emphasize the role of the microscopic details of the Hilbert space structure on the asymptotic evolution of continuously monitored quantum systems.
As we will see, the coexistence of different microscopic structures can have a dramatic effect on the evolution of individual realizations.

\begin{figure*}[t] 
    \centering
    \begin{tikzpicture}
    \node (a) [label={[label distance= -0.2cm]132: \textbf{(a)}}] at (0,0) {\includegraphics{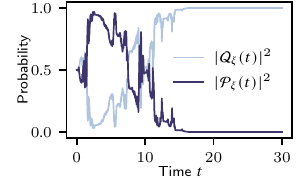}};
    \node (a) [label={[label distance= -0.2cm]132: \textbf{(b)}}] at (5.5,0) {\includegraphics{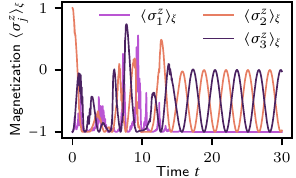}};
    \node (a) [label={[label distance= -0.2cm]132: \textbf{(c)}}] at (11,0) {\includegraphics{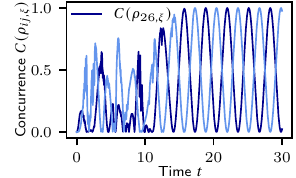}};
    \node (a) [label={[label distance= -0.2cm]132: \textbf{(d)}}] at (0,-3.1) {\includegraphics{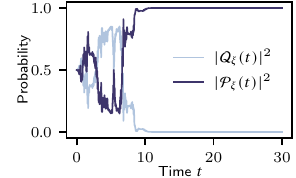}};
    \node (a) [label={[label distance= -0.2cm]132: \textbf{(e)}}] at (5.5,-3.1) {\includegraphics{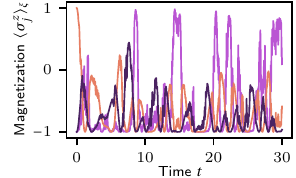}};
    \node (a) [label={[label distance= -0.2cm]132: \textbf{(f)}}] at (11,-3.1) {\includegraphics{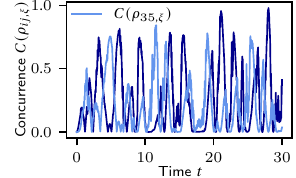}};
    \end{tikzpicture}
    \caption{Heisenberg $XX$-ring (\cref{fig:ring,eq:XX-Hamiltonian}), with $N=6$ sites subject to homodyne measurement on two sites $L_1 = \sqrt{\Gamma}\sigma^z_1$, $L_4 = \sqrt{\Gamma}\sigma^z_4$, guaranteeing a unique decoherence-free mode, \cref{eq:syncN}. The system is initially prepared in the state $\ket{\Psi(0)}=\ket{010\ldots 0}$ with equal weights in the complement and decoherence-free subspace $|\calp(0)|^2=|\calq(0)|^2=1/2$. 
    (a) The continuous monitoring projects the system either onto the decoherence-free subspace or (d) the complement with probability $1/2$ respectively, cf. \cref{eq:ring-update}.
    Accordingly, a spontaneous irreversible localization transition occurs where the individual $z$-magnetizations (only the first three are shown due to symmetry) (b) exhibit noise-free unitary evolution in the decoherence-free subspace, or (e) stochastic, irregular dynamics in the complement.
    (c) Formation of Bell states in the decoherence-free subspace between sites $(2,6)$ and $(3,5)$ as measured by the concurrence.
    (d) Inside the complement, entanglement remains noisy and unpredictable, rendering it effectively inaccessible. The parameters are $\omega = J = 1$ and $\Gamma = 0.4$.}
    \label{fig:ring-evolution}
\end{figure*}

Consider the $N$-site Heisenberg $XX$-model on a ring
\begin{align}
  H =
  \omega\sum_{k=1}^N  \sigma^z_k 
  +J \sum_{j=1}^N (\sigma^+_j\sigma^-_{j+1} + \sigma^-_j\sigma^+_{j+1}),
  \label{eq:XX-Hamiltonian}
\end{align}
with homogeneous qubit level spacing $\omega$, nearest neighbor interaction strength $J$, and periodic boundary conditions $(\sigma^{\pm}_N = \sigma^{\pm}_1)$ (cf. \cref{fig:ring}).
Let the $z$-polarization of either one or two qubits at arbitrary sites $u$ and $v$ be monitored continuously.
On the level of the Lindblad equation, this induces dephasing with $L_1 = \sqrt{\Gamma}\sigma^z_u$ and $L_2 = \sqrt{\Gamma}\sigma^z_v$.
Since the jump operators are unitary, localization will be completely absent for jump trajectories (cf. case (i) of \cref{th:incomplete-jump}) and we therefore study only the diffusive case here.
For single-site measurement we can, without loss of generality, choose to monitor the first site with $L_1 = \sqrt{\Gamma}\sigma^z_1, \ L_2 = 0$, because the system is invariant under rotation.
As detailed in \cref{sec:dfs}, if a decoherence-free subspace exists it belongs to the asymptotic state space and will thus always be found by individual trajectories.
We are interested in particular in decoherence-free subspaces that protect coherent oscillations between energy eigenstates.
The unperturbed system \cref{eq:XX-Hamiltonian} can be mapped to free fermions via the Jordan-Wigner transformation \cite{Takahashi1999}, and the protected decoherence-free eigenmodes of the local $z$-polarizations can then be found analytically by means of perturbation theory in Liouville space \cite{Schmolke2022,Benatti2011,Li2014}.
Following Ref.~\cite{Schmolke2022}, we obtain the first order corrections (the decay rates) to the $z$-polarization eigenfrequencies, $\Lambda_{\alpha\beta} = \lambda_\alpha - \lambda_\beta$, where $\lambda_\alpha = 2\cos(2\pi \alpha/N)$ are the eigenvalues of the Jordan-Wigner transformed Hamiltonian.
In particular, for single site dephasing there are 
\begin{align}
\#\Lambda 
= \frac{m(m-1)}{2}, \
\text{with} \ 
\begin{dcases}
  m = \lfloor N/2 \rfloor,  \ N = \m{odd},\\
  m = N/2 -1, \ N = \m{even}
\end{dcases} .
\label{eq:ring-degeneracy}
\end{align}
protected eigenmodes that are immune to dissipation.
The number of decoherence-free modes can be reduced to a unique one if additional dephasing is placed on a second qubit.
In general, the configurations that guarantee survival of a single eigenmode in the entire chain are 
\begin{align}
  N = 5, \ u = v = n, \ \alpha = 1, \ \beta = 2,
  \label{eq:sync5}
\end{align}
and
\begin{align}
  \frac{N}{6} \in \mathbb{N}, \ u = n, \ v = n+3, \ \alpha = \frac{N}{6}, \ \beta = 2\alpha,
  \label{eq:syncN}
\end{align}
where $n \in [1,N]$ is an arbitrary site.
The corresponding unique eigenfrequency is
\begin{align}
  \Lambda_{\alpha \beta} = 
  \begin{dcases}
    \sqrt{5}J, &N = 5,\\
    2J, &N > 5.
  \end{dcases}
\end{align}
In the five-qubit ring, single site dephasing is sufficient, which is in accordance with \cref{eq:ring-degeneracy}.
In the many-body case, $N>5$, the system size has to be divisible by $6$ and the two noise sources have to be placed three sites apart from each other.
In \cref{fig:ring-evolution} we demonstrate the time evolution for the $N=6$ qubit ring.
Using \cref{alg:sbd} in \cref{sec:sbd}, we can find the full structure of the Hilbert space for the exact open system (without Jordan-Wigner transformation) which is depicted in \cref{fig:ring}.
There are two distinct decoherence-free subspaces $\calq_1$ and $\calq_2$ each containing three states.
The collection $\calq = \calq_1 \oplus \calq_2$ defines the whole decoherence-free subspace.
The complement $\calp$ contains $12$ minimal subspaces of various dimensions.
We are interested in localization transitions into the decoherence-free subspace that contains the unique protected eigenmode.
We therefore choose the initial product state, $\ket{\Psi(0)} = \ket{010\ldots 0}$, which has overlap with the protected eigenmode inside of the decoherence-free subspace $\calq_1$ and with a four-dimensional subspace inside the complement, $\calp$, with $|\calq(0)|^2 = |\calp(0)|^2 = 1/2$ (see the structure in \cref{fig:ring}).
On the level of the Lindblad equation the initial overlap in each of these subspaces is conserved and the system evolves in both subspaces simultaneously.
On the trajectory level however, the system always completely enters only one subspace (cf. \cref{fig:ring-evolution}a,d).
The coexistence of substructures with different microscopic details has drastic implications for the fate of individual trajectories.
The trajectories reveal the structures of the Hilbert space and purify their characteristics.
The continuous measurement effectively projects the system onto the decoherence-free subspace with probability $\tr[\rho(0)P_\calq] = 1/2$, leading to noise-free unitary evolution of the local $z$-polarizations (cf. \cref{fig:ring-evolution}b) and entangled oscillations between two pairs of qubits as indicated by the concurrence $C(\rho_{ij})$, which we plot in \cref{fig:ring-evolution}c.
Performing local continuous measurements thus produces maximally entangled Bell states from a product of local states; a feat that is impossible for either Hamiltonian or measurement alone.
With probability $\tr[\rho(0)P_\calp] = 1/2$, the system is projected onto the complement where the evolution remains stochastic at all times (cf. \cref{fig:ring-evolution}e), rendering the entanglement virtually inaccessible (cf. \cref{fig:ring-evolution}f).

Restricted to the coarse-grained level of $\calq$ and $\calp$, the generalized update rule, \cref{eq:update-rule}, becomes
\begin{align}
  \rho(0) \rightarrow \rho_{\xi}(t) =
  \begin{dcases}
    \frac{\sfu^t(\sfp_\calq\rho(0))}{\tr[\sfp_\calq \rho(0)]}, \text{ w. prob } p_\calq,\\
    \frac{\sft^t_{\xi,\calp}(\sfp_\calp \rho(0))}{\tr[\sfp_\calp \rho(0)]}, \text{ w. prob } p_\calp
  \end{dcases}.
  \label{eq:ring-update}
\end{align}
With probability $p_\calq = \tr[\sfp_\calq\rho(0)]$ the system is projected onto the decoherence-free subspace undergoing unitary evolution with $\sft^t_{\xi,\calq}(\bullet) = \sfu^t(\bullet) = U \bullet U^\dagger$, where $U = \exp(-iH_\calq t)$ and $H_\calq = \sfp_\calq H$.
With probability $p_\calp = \tr[\sfp_\calp \rho(0)]$ the system is projected onto the complement which is, by construction, not decoherence-free and evolves along the quantum trajectory $\sft^t_{\xi,\calp} = \sfp_\calp \sft^t_\xi \sfp_\calp$ with Hamiltonian $H_\calp = \sfp_\calp H$ and Lindblad jump operators $L_{k,\calp} = \sfp_\calp L_k$.

The average ergodic behavior is agnostic to the microscopic details of the subspaces but cares only for the general structure.
Accordingly, within the coarse-grained resolution of the Hilbert space, the mean fidelity becomes 
\begin{align}
  \mathbb{E}[F(\overline{\rho}_\xi,\rho^\m{s})]
  = w_{\calq}^2 + (1-w_{\calq}^2),
\end{align}
where $w_\calq = \tr[\rho(0)P_\calq]$ is the initial overlap in the decoherence-free subspace.

\subsection{Classical noise}
\label{sec:classical-noise}
The nontrivial behavior of quantum trajectories and the probing of higher-order statistic of the density operator is inherently related to the non-linear measurement-backaction, constituting the essential driving force behind localization transitions and purification.
The only possible way for the backaction to vanish is case (i) of \cref{th:incomplete-diff,th:incomplete-jump}, which gives rise to independent trajectories in each subspace.
Here, we shed more light on the physical meaning of this particular case.

A special case arises in the diffusive unraveling of the Lindblad equation when the jump operators are such that $(L_k+L_k^\dagger) = z_k\mathds{1}$ \cite{Barchielli2009}.
Denote by $L_k = -iV_k+\mathds{1}a_k$, the jump operator with $a_k + a_k^\ast \equiv z_k$ and suitable Hermitian operators $V_k$.
Plugging this in into \cref{eq:homodyne-sme}, the stochastic master equation in \Ito form becomes
\begin{align}
  \dd{\rho_\xi} 
  =& -\I[H,\rho_\xi] \dd{t} + \sum_k (-\I)[V_k,\rho_\xi] \dd{W_k} \notag \\ 
  &+ \left(V_k\rho_\xi V^\dagger_k - \frac{1}{2}\left\{V_k^\dagger V_k,\rho_\xi\right\}\right) \dd{t}.
\end{align}
If $V_k = 0 \ \forall k$, then $L_k \propto \mathds{1}$, and there is no measurement anymore.
This corresponds to a decoherence-free subspace with noise-free evolution (cf. \cref{sec:dfs}).
To provide a physical interpretation to the case where $V_k \neq 0$ for at least one $k$, we first need to convert the above equation to its equivalent Stratonovich form \cite{VanKampen1981}.
Using the standard conversion rules \cite{VanKampen1992,Gardiner2010}, we obtain \cite{Schmolke2024}
\begin{align}
  \dot{\rho}_\xi = -\I\big[H+\sum_kV_k\xi_k(t),\rho\big],
  \label{eq:noise}
\end{align}
where $\xi_k(t)$ are independent white noise processes with zero mean $\mathbb{E}[\xi_k(t)] = 0$ and delta auto-correlation function $\mathbb{E}[\xi_j(t)\xi_j(t^\prime)] = \delta_{j,k} \delta(t-t^\prime)$.
The continuous monitoring thus becomes equivalent to the evolution of a closed quantum system with Hamiltonian $H$, subject to classical white noise acting on the operators $V_k$.
Evidently, there is no measurement backaction anymore.
This may be regarded as a key difference between quantum noise and classical noise with profound implications on the asymptotic behavior of the density matrix in Hilbert space.
For the quantum jump unraveling a similar situation arises when $(L^\dagger_kL_k) = z_k \mathds{1}$. 
The jump operator can then be written as $L_k = c_k U_k$, with $|c_k|^2 \equiv z_k$ and a suitable unitary $U_k$. 
This choice renders the stochastic master equation \cref{eq:jump-sme}
\begin{align}
  \dd{\rho_J} = -\I[H,\rho_J] \dd{t}
  + \sum_k\left[U_k\rho_J U^\dagger_k - \rho_J\right] \dd{N_k}.
  \label{eq:noise-jump}
\end{align}
Continuous measurement in the quantum jump unraveling with jump operators proportional to a unitary is thus equivalent to coherent evolution interrupted by unitary operations applied at random times.
Here too, the noise becomes classical and independent of the state (the probability to observe a jump, $\langle L^\dagger_k L_k\rangle = z_k$, remains constant)

In contrast to state-dependent quantum noise, classical noise is agnostic to the state; there is no backaction and, as we will now show, the evolution is always ergodic.
To see this it is sufficient to again consider the evolution equation for the overlap with an orthogonal subspace (cf. \cref{eq:calq,eq:calq_jump}) which for the classical noise in \cref{eq:noise,eq:noise-jump} is stationary at all times, $\dd{\left(|\calq(t)|^2\right)} = 0$.
There is hence no localization transition in Hilbert space and symmetries and conserved quantities are respected by the evolution.
Still, the pathwise ergodic theorem (cf. \cref{eq:erg-theorem}) applies to the classical noise trajectories \cref{eq:noise,eq:noise-jump}, implying that the long-time average of individual realizations converges to one of the asymptotic states of the Lindblad equation.
Since the weight in each minimal orthogonal subspace remains constant, the only such state accessible is the asymptotic state of the Lindblad equation which is uniquely determined because the initial condition is kept fixed.
Classical noise is therefore always ergodic in the sense that \cite{Schmolke2024}
\begin{align}
  \lim_{T \to \infty} \frac{1}{T} \int_0^T \dd{t} \rho_c(t)
  = \rho^\m{s},
\end{align} 
where $\rho^\m{s}$ is the asymptotic state of the Lindblad equation. It can thus be employed to unravel a Lindblad equation into classical noise trajectories and thereby generate individual ergodic realizations irrespective of the Hilbert space structure.

We furthermore suspect that ergodic dynamics is a generic feature of general classical noise, Markovian or non-Markovian.
If the stochastic differential equation corresponds to an unraveling of the Lindblad equation, ergodicity follows by the same reasoning as above.
In general, the average over the ensemble of realizations will however result in a non-Markovian master equation for the density matrix \cite{Chenu2017,Gu2019,Kiely2021}.
Since the structure theorems of \cref{sec:space-structure} remain essentially the same in a non-Markovian setting \cite{Zhang2016}, conserved quantities remain conserved and we expect a modified version of the pathwise ergodic theorem to apply, where individual realizations will exhibit ergodicity with respect to the corresponding non-Markovian master equation.
Unravelings of non-Markovian master equations have been discussed in Refs.~\cite{Breuer2004,Breuer2007,Barchielli2010,Luoma2020}

\section{Finding all minimal orthogonal subspaces}
\label{sec:sbd}
Many effects presented here are based on the existence of an orthogonal decomposition of the Hilbert space according to \cref{eq:decomposition}.
Although the strength of our results lies in the fact that prior knowledge of the state space structure is not required, for practical applications and engineering purposes it can be very useful to have a deterministic procedure that outputs the full decomposition into minimal orthogonal subspaces.
This task is of independent interest for a variety of research areas.
For instance, knowledge of the symmetry sectors is crucial to analyze the level spacing statistics and chaotic behavior of open and closed quantum systems \cite{Santos2004,Santos2020,Sa2020} because the spectra corresponding to different orthogonal subspaces are independent and need to be considered separately \cite{Guhr1998,Santos2004}.
The task of finding the finest possible decomposition and therefore the full structure of the Hilbert space \cref{eq:decomposition} is generally not straightforward, because symmetries might be hidden, not immediately admitting an obvious physical interpretation.
Symmetries might moreover be entirely absent although the steady state is degenerate.
Our approach relies on systematic numerical procedures to identify the finest simultaneous block decomposition of a set of complex matrices that can be harnessed for this problem \cite{Murota2010,Maehara2011,deKlerk2011}.

Here, we implement an algorithm to find the Hilbert space structure for finite-dimensional Lindblad equations.
The algorithm consists of two basic parts.
First, in \cref{alg:sbd}, one finds the finest simultaneous block decomposition of the $(2n+1)$-tuple of complex matrices $(H,L_1,L^\dagger_1,L_2,L^\dagger_2,\ldots,L_n,L^\dagger_n)$.
From an algebraic perspective, \cref{alg:sbd} identifies all Evans irreducible invariant subspaces \cite{Evans1977,Zhang2024}.
Each of the resulting blocks will evolve independently along its own effective Lindblad equation and is guaranteed to have at least one stationary state.
Every individual block corresponds to an orthogonal subspace that might still have a decaying subspace attached to it.
Second, in \cref{alg:steady-states}, the decaying subspace (if it exists) is separated from the asymptotic subspace in each of the orthogonal subspaces.
This way, all the extremal steady states of the Lindblad equation are identified.
Inspired by complete localization of trajectories we propose an additional algorithm (\cref{alg:trajs}) that consistently evolves pure state quantum trajectories exploiting the fact that they naturally converge to the extremal steady states as established in \cref{sec:transition,sec:incomplete}.

\begin{figure}[t]
  \centering 
  \includegraphics[scale=0.48]{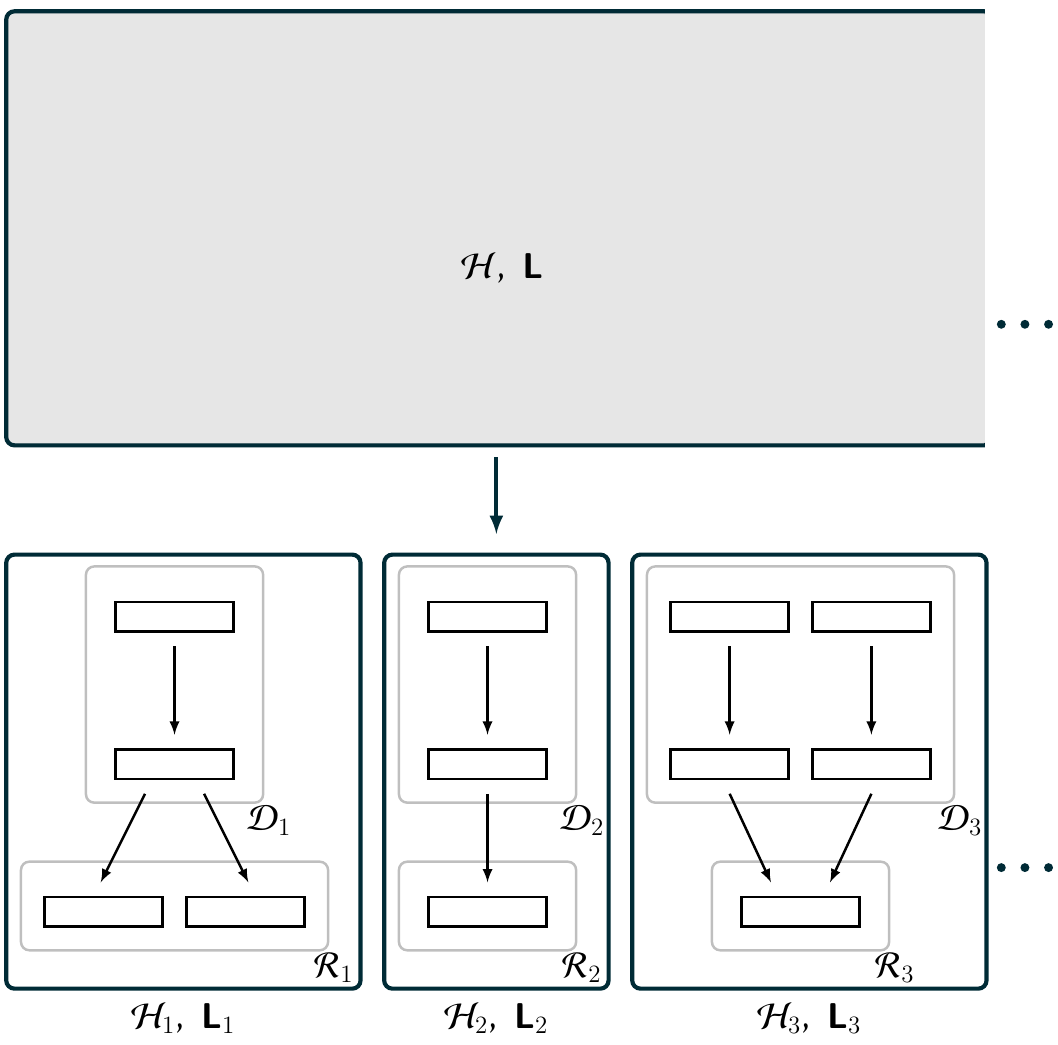}
  \caption{Schematic representation of the Hilbert space structure obtained from \cref{alg:sbd}. 
  Before applying the algorithm, the total Hilbert space $\calh$ and the total Liouvillian $\sfl$ generally appear to be unstructured (upper panel).
  After applying \cref{alg:sbd} (lower panel), the total Hilbert space decomposes into a direct sum, $\calh = \bigoplus_\alpha \calh_\alpha$ of orthogonal subspaces $\calh_\alpha$, each with an asymptotic subspace $\calr_\alpha$ that may still have a decaying subspace $\cald_\alpha$ attached to it. 
  Subspaces are indicated by rectangles and arrows indicate flow between them.
  Total decaying and asymptotic subspaces decompose into $\cald = \bigoplus_\alpha \cald_\alpha$ and $\calr = \bigoplus_\alpha R_\alpha$. This decomposition is minimal iff the decaying subspace is empty, $\cald_\alpha = \emptyset,\ \forall \alpha$. On the level of operators, this induces a simultaneous block decomposition of Hamiltonian $H = \bigoplus_\alpha H_\alpha$ and Lindblad operators $L_k = \bigoplus_\alpha L_{k,\alpha}$ and thus the Liouvillian $\sfl = \bigoplus_\alpha \sfl_\alpha$ giving rise to independent Lindbladian evolution in each block. To separate the decaying from the asymptotic subspace inside every $\calh_\alpha$ and to identify all the extremal stationary states, \cref{alg:steady-states} should be applied subsequently.}
  \label{fig:sbd-strucutre}
\end{figure}

The problem of finding all minimal orthogonal subspaces of $\calh = \cald \oplus \calr$, can be transferred to finding the minimal simultaneous block diagonal form on the space of bounded linear operators $\mathfrak{B}(\calh)$.
A set of operators $\{A_k\}$ admits a common block diagonal form if there exists a nontrivial Hermitian operator $X$ such that
\begin{align}
  [A_k,X] = 0, \ \forall k.
  \label{eq:linear-system}
\end{align}
The unitary that diagonalizes $X$ is the transformation that puts the $A_k$ into common block form.
To find $X$, we need to solve the linear system \cref{eq:linear-system}.
For practical purposes it is useful to allow for some numerical tolerance $\epsilon$.
Suppose we can find a Hermitian operator that almost commutes with the operators $A_k$, such that
\begin{align}
  \|[A_k,X]\| \le \epsilon, \ \forall k.
  \label{eq:error-controlled}
\end{align}
Then, after the unitary transformation, the off-diagonal elements between blocks will be of order $\epsilon$ \cite{Maehara2011}.
The remaining challenge is to find a solution to \cref{eq:error-controlled} that produces the finest block decomposition possible.
Remarkably, Maehara and Murota have proven that sampling a random solution $X$ is sufficient with probability one \cite{Murota2010,Maehara2011}.
We repeat the algorithm of Ref.~\cite{Maehara2011} tailored to our problem here for completeness.
\begin{algorithm}[H]
  \caption{All orthogonal subspaces}
  \label{alg:sbd}
  \begin{algorithmic}[1]
  \Require $A = (H,L_1,L_2,\ldots,L_n)$
  \Ensure  $T$ simultaneously block diagonalizing $A$ and $A^\dagger$.
  \State Construct $\textsf{\textbf{A}}_k = \lcom{A_k,\mathds{1}_d}$
  and $\textsf{\textbf{B}}_k = \lcom{A^\dagger_k,\mathds{1}_d}$
  \State Construct $\textsf{\textbf{M}} = \sum_k \textsf{\textbf{A}}^\dagger_k\textsf{\textbf{A}}_k+\textsf{\textbf{B}}^\dagger_k\textsf{\textbf{B}}_k$
  \State Find normalized eigenvectors $\textsf{\textbf{M}} v_j = \Lambda_j v_j$, s.t. $\Lambda_j < \epsilon^2$
  \State Draw randomly $c_j\in\mathbb{C}$ with $\sum_j |c_j|^2 = 1$ and construct $W = \sum_j c_j v_j$
  \State Reshape $Y = \m{vec}^{-1}(W)$, proceed with $X = \frac{1}{2}(Y+Y^\dagger)$
  \State Perform the diagonalization $X = T^{-1}D_XT$
  \State \textbf{return} $T$
  \end{algorithmic}
\end{algorithm}

Here, $\lcom{A,B} = A \otimes B^\m{T} - B \otimes A^\m{T}$, denotes the vectorized commutator and $\m{vec}(\bullet)$ is the vectorization operation \cite{Gyamfi2020}.
\Cref{alg:sbd} outputs a unitary transformation $T$, that simultaneously block-diagonalizes the Hamiltonian and the Lindblad jump operators
\begin{align}
  T^{-1}H T = \bigoplus_{\alpha=1}^n H_\alpha, \quad
  T^{-1}L_k T = \bigoplus_{\alpha=1}^n L_{k,\alpha},
  \label{eq:algorithm-blocks}
\end{align}
where the number of orthogonal blocks $n$ in the decomposition corresponds to the number of distinct eigenvalues of $X = \sum_\alpha x_\alpha \Pi_\alpha$.
The degeneracy $g_\alpha(X)$ of each eigenvalue $x_\alpha$ determines the dimension of each block $d_\alpha = g_\alpha(X)$, with $\sum_\alpha d_\alpha = \m{dim}(\calh)$.
The block dimension is bounded by $1 \le d_\alpha \le d$, where the lower bound corresponds to a one-dimensional decoherence-free subspace and the upper bound is saturated if for instance there is only one block that spans the whole Hilbert space supporting a unique, full-rank stationary state.
For every $d_\alpha < d$, steady state degeneracy of the Lindblad equation is guaranteed.
After application of \cref{alg:sbd}, the total Hilbert space decomposes into a direct sum, $\calh = \bigoplus_\alpha \calh_\alpha$, of orthogonal subspaces $\calh_\alpha$, each with an asymptotic subspace $\calr_\alpha$ that may still have a decaying subspace $\cald_\alpha$ attached to it (see \cref{fig:sbd-strucutre}). 
By construction, the projectors $\Pi_\alpha$ onto the eigenspaces of $X$ are elements of the commutant, $\Pi_\alpha \in \{H,L_k,L^\dagger_k\}^\prime$.
The common block decomposition, \cref{eq:algorithm-blocks}, thus immediately implies a decomposition of the Liouvillian into orthogonal blocks
\begin{align}
  \sfl =& \bigoplus_{\alpha=1}^n \sfl_\alpha,\\ 
  \sfl_\alpha 
  =& -\I\lcom{H_\alpha,\mathds{1}_{d_\alpha}}\\
  &+ \sum_k \left(L_{k,\alpha} \otimes L^\ast_{k,\alpha}
  -\frac{1}{2}\lcom{L^\dagger_{k,\alpha}L_{k,\alpha},\mathds{1}_{d_\alpha}}_+\right),
\end{align}
where $\lcom{A,\mathds{1}}_+ = A\otimes \mathds{1} + \mathds{1} \otimes A^\ast$ is the vectorized anti-commutator.
\Cref{alg:sbd} is guaranteed to find the complete decomposition into all existing orthogonal subspaces.
Every block $\alpha$ will undergo a valid Lindbladian evolution of its own, each admitting at least one steady state \cite{Evans1977,Baumgartner2008_2}.
\Cref{alg:sbd} thus finds the finest decomposition into independent Lindbladian evolutions which is valid at all times.
If the decaying subspace is the empty set, $\cald_\alpha = \emptyset$, then each $\calh_\alpha = \calr_\alpha$ is a minimal orthogonal subspace supporting a unique steady state.
If $\cald_\alpha \neq \emptyset$, there may still be multiple steady states (for instance \cref{fig:sbd-strucutre} block $(\calh_1,\sfl_1)$) and the decomposition $\calr_\alpha = \bigoplus_j \calr_{\alpha,j}$ is not yet minimal.

The decaying subspace impacts the transient dynamics but knowledge of its structure is not immediately necessary to identify the asymptotic states of the Lindblad equation.
However, if the decaying subspace is empty, \cref{alg:sbd} already outputs the minimal orthogonal block decomposition and therefore the full structure of the asymptotic state space.
On the other hand, even if it outputs the identity on the full Hilbert space, $T = \mathds{1}_d$, there can still be multiple steady states \footnote{This can be observed in \cref{fig:sbd-strucutre}. Consider the special case where, $\calh = \calh_1$, is the full Hilbert space. There is no trivial conserved projector and the commutant is trivial, $\{H,L,L^\dagger\}^\prime = z\mathds{1}$, but there are two steady states supported on the orthogonal subspaces $R_{1,1}$ and $R_{1,2}$. This happens because they are both attached to the same decaying subspace $\cald_1$ and cannot be separated into orthogonal blocks without first removing $\cald_1$. See also examples in \cref{sec:scar states,sec:two-qubits}.}.
The presence or absence of $\cald$ can also have a significant impact on the asymptotics of quantum trajectories, as discussed in \cref{sec:decaying-subspace}.
We can make some basic observations. 

\begin{observation}[Absence of a decaying subspace]
\label{obs:decay}
  In the unique splitting of the Hilbert space \cref{eq:decomposition} into decaying and asymptotic subspaces, $\calh = \cald \oplus \calr$, the following are equivalent:
  \begin{itemize}
    \item[1.] The decaying subspace is the empty set, $\cald=\emptyset$.
    \item[2.] The dynamics induced by $\sfl$ is rank non-decreasing.
    \item[3.] There exists at least one faithful stationary state.
    \item[4.] There exists at least one full rank initial state whose rank remains preserved throughout the evolution.
  \end{itemize}
\end{observation}
The structure of the decaying subspace generally controls the unidirectional flow into the minimal orthogonal subspaces and is responsible for the distribution of the probability over them (see \cref{fig:sbd-strucutre}).
Therefore, the task of stabilizing desired invariant states of quantum trajectories (cf. \cref{th:diff-inv,th:jump-inv}) is intimately related to engineering of the decaying subspace by finding suitable choices of the Hamiltonian and Lindblad jump operators.
In fact, slightly adapting the proofs in \cref{sec:invariant-jumps,sec:invariant-diff}, yields a complementary statement to 
\robs{obs:decay}.
\begin{theorem}[Existence of a decaying subspace {\cite{Baumgartner2008_2,Ticozzi2012,Albert2016}}]
  \label{th:decaying-subspace}
  In the unique decomposition of the Hilbert space, $\calh = \cald \oplus \calr$, \cref{eq:decomposition}, into a decaying subspace, $\cald$, and an asymptotic subspace, $\calr$, the decaying subspace is not the empty set if and only if the Hamiltonian and the Lindblad jump operators satisfy 
  \begin{align}
    L_k &= 
    \begin{bmatrix}
      L_{k,11} & L_{k,12} \\ 0 & L_{k,22}
    \end{bmatrix}, \ \forall k,\\
    H_{12} &= -i \frac{1}{2} \sum_k L^\dagger_{k,11}L_{k,12},
  \end{align}
  where the upper left block corresponds to $\calr$ and the lower right block to $\cald$.
\end{theorem}
A sufficient condition for $\cald = \emptyset$ is that the Lindblad jump operators are normal $[L_k,L^\dagger_k] = 0,$ $\forall k$.

Methodically and efficiently obtaining the extremal set of stationary states presents a formidable challenge \cite{Thinga2021}.
Various approaches exist to compute the unique steady state of a finite dimensional Lindblad equation \cite{Nation2015,Nation2015_2,Weimer2021,Campaioli2024,Melo2025}.
However, a fundamental challenge in direct methods for multiple steady states is restricting the solution space of the Liouvillian $\sfl \m{vec}(\rho) = 0$ to only the physical subset $\{\sfl\m{vec}(\rho) = 0 \vert \tr[\rho] = 1, \rho^\dag = \rho, \rho \ge 0\}$.
An additional complication arises because as soon as there is more than one stationary state, there are infinitely many convex combinations that are also a stationary solution.
The only sensible way to count them is thus to consider the convex hull of the restricted solution space.
Based on \robs{obs:decay}, we present an elementary algorithm that exploits the accessible knowledge about the Hilbert space structure to systematically identify all the minimal subspaces.
Once these are found, standard methods can be applied to obtain the unique stationary state supported on each of them.

\begin{algorithm}[H]
  \caption{All stationary states}
  \label{alg:steady-states}
  \begin{algorithmic}[1]
    \Require $\sfl_\alpha$ in block form
    \Ensure Asymptotic state spaces $\calr_{\alpha,j}$, steady states $\rho^\m{s}_{\alpha,j}$
    \State In each block evolve $A^\infty_\alpha = \lim_{t \to \infty} \sft^t_\alpha(\mathds{1}_{d_\alpha})$
    \State Get the asymptotic support $\calr_\alpha = \m{supp}(A^\infty_\alpha)$
    \State Project out the decaying subspace $\sfl_{\calr_\alpha} = \sfp_{\calr_\alpha}\sfl_\alpha \sfp_{\calr_\alpha}$
    \State Apply Alg.~\ref{alg:sbd} to each $\sfl_{\calr_\alpha}$ and find the minimal blocks $\sfl_{\calr_{\alpha_j}}$
    \State Find the unique steady state $\rho^\m{s}_{\alpha_j}$ in each $\sfl_{\calr_{\alpha_j}}$
    \State \textbf{return} $\calr_{\alpha,j}$, $\rho^\m{s}_{\alpha,j}$
  \end{algorithmic}
\end{algorithm}

The key to \cref{alg:steady-states} is to remove the decaying subspace in each orthogonal subspace $\calq_\alpha$ and the perform \cref{alg:sbd} again in the respective asymptotic state spaces $\calr_\alpha$ to finally resolve all minimal subspaces $\calq_{\alpha_j}$.
This is done by simply evolving the identity matrix using the Lindblad equation \footnote{We use the identity instead of the maximally mixed state to avoid small overlaps for large systems.}.
The identity has support on the entire Hilbert space and will thus flow down into $\calr_\alpha$ without missing any subspace.
Its asymptotic support thus corresponds to the asymptotic subspace, i.e. $\calr_\alpha = \m{supp}(A^\infty_\alpha)$.
Projecting onto $P_{\calr_\alpha}$ thus removes the decaying subspace $\cald_\alpha$ and \cref{alg:sbd} yields the minimal decomposition.

Based on the considerations of \cref{sec:transition}, an alternative method to finding the extremal stationary states of the Lindblad equation and their corresponding subspaces can be employed.
One may numerically simulate multiple quantum trajectories and compute their time averaged states which will, generically, correspond to an extremal stationary state if the subspaces are not degenerate in the sense of \cref{th:incomplete-diff,th:incomplete-jump}.
In particular we have proven that the time average of two trajectories of the same unraveling is either identical or orthogonal $\rho^\m{s}$ in the absence of such symmetries.
We propose the following algorithm based on the systematic evolution of quantum trajectories.

\begin{algorithm}[H]
  \caption{Trajectory-based steady state finder}
  \label{alg:trajs}
  \begin{algorithmic}[1]
    \Require $\sfl$
    \Ensure Asymptotic state spaces $\calr_j$, steady states $\rho^\m{s}_j$
    \State Choose a complete basis $\{\ket{n}\}$ of $\calr$
    \State Evolve each $\ket{n}$ and get $\overline{\dyad{n}} = \rho^\m{s}_j$
    \State Construct the projector on the joint support of steady states $P_1 = \m{Proj}(\bigcup_j \m{supp}(\rho^\m{s}_j))$
    \State Remove the joint support $\ket*{n^{(1)}} = P^\perp_1 \ket*{n}/\|P^\perp_1\ket{n}\|$
    \State Perform steps 2 to 4 with initial states $\ket*{n^{(1)}}$
    \State Proceed iteratively until $P^\perp_m \ket*{n^{(m)}} = 0$
    \State Reduce the set $\{\rho^\m{s}_j\}$ and keep only $\tr[\rho^\m{s}_i\rho^\m{s}_j] = 0, \ \forall i \neq j$
    \State If $\tr[\rho^\m{s}_i\rho^\m{s}_j] \neq 0$, use Alg.~\ref{alg:sbd} to get an orthogonal pair
    \State \textbf{return} $\calr_j$, $\rho^\m{s}_j$
  \end{algorithmic}
\end{algorithm}

This procedure is not guaranteed to resolve all extremal stationary states if localization transitions are generally incomplete (cf. \cref{sec:incomplete}) because it is not clear which of the states are actually of minimal dimension.
To remedy this one would have to check minimality of every subspace $\calr_j = \m{supp}(\rho^\m{s}_j)$ with \cref{alg:sbd}.
\Cref{alg:trajs} may nevertheless provide an efficient alternative to existing methods in many practical applications.

The algorithm starts with an arbitrary orthonormal basis $\{\ket{n}\}$ of the Hilbert space.
Each basis state is evolved once, will localize and converge to a stationary state in the time mean, i.e. (cf. \cref{sec:erg-theorem})
\begin{align}
  \overline{\dyad{n}} 
  = \lim_{t\to \infty} \frac{1}{T}\int_0^T \dd{s} \dyad{n_\m{c}(s)}
  = \rho^\m{s}_j.
\end{align}
Once the trajectory has found a steady state, we remove the corresponding support on all states $\ket{n}$ and renormalize.
The new initial states $\{\ket{n^{(1)}}\}$ then have overlap only with minimal subspaces that have not yet been identified.
This removes redundancies, preventing trajectories from repeatedly localizing to the same subspace and guarantees that only a finite number of trajectories (the amount of iterations is at most the number of steady states) is required to chart the entire Hilbert space.
After performing this procedure we obtain a set of mutually identical or orthogonal stationary states $\{\rho^\m{s}_j\}$ where it only remains to remove duplicates.
If there are multiple steady states with non-zero overlap that are not identical they must correspond to a symmetry of the unraveling (cf. \cref{th:incomplete-diff,th:incomplete-jump}) and \cref{alg:sbd} may be applied to obtain all the corresponding minimal subspaces and then find the unique stationary states in each of them via established methods \cite{Nation2015,Nation2015_2,Weimer2021,Campaioli2024,Melo2025}.

For a quantum system of Hilbert space dimension $\m{dim}(\calh) = d$, this requires storage of $d$ entries in the pure state time evolution of a quantum trajectory, $\ket{\Psi_\m{c}(t)}$, while both, simultaneous block diagonalization and diagonalization of the generator $\sfl$, are typically performed in Liouville space which involves dealing with matrices of dimension $d^2\times d^2$ \cite{Gyamfi2020,Nation2015,Nation2015_2,Weimer2021,Campaioli2024,Melo2025}.

\section*{Conclusion}
Modern trajectory theory describes the physical evolution of quantum systems subject to continuous monitoring and provides fundamental insights into the nature of quantum measurement.
At the same time, quantum trajectories represent a computationally efficient tool to simulate many-body open quantum dynamics.
In light of their fundamental and practical importance, it is therefore crucial to gain a general understanding of their long-time behavior.
We have completely characterized the asymptotics of quantum trajectories in the quantum jump and diffusive (homodyne) unravelings for arbitrary finite-dimensional quantum systems.
As a first step, we have generalized the concept of stationary states from Lindbladian dynamics to individual quantum trajectories and shown that, by engineering the measurement operators, any pure state can be stabilized.
More generally, we have shown that quantum trajectories are susceptible to the Hilbert space structure and, in the presence of steady state degeneracy of the Lindblad equation, undergo irreversible localization transitions in Hilbert space, where the evolution will become effectively constrained to a subspace of the total available state space.
We have identified the necessary and sufficient conditions for complete and partial localization to occur, which can be entirely related to properties of the Hamiltonian and the Lindblad jump operators.

In general, any quantum trajectory can exhibit only one of three characteristic behaviors with varying degree of confinement and purification.
First, purification may stop prematurely and the detector effectively decorrelates from the system resulting in unitary evolution in a decoherence-free subspace, or in classical noise while localization may be either complete or incomplete.
Second, evolutions become unitarily equivalent in different subspaces (e.g. due to the presence of abelian symmetries) and typically build up coherences or correlations but prevent further confinement, while purification occurs in each of the subspaces individually but not globally.
Third, if quantum backaction persists, the state purifies and full localization occurs where the system will end up confined to one of the irreducible components of the Hilbert space.
This classification leads to the natural emergence of a generalized Born rule, where the continuous observation effectively acts like a prolonged projection onto one of several possible time evolutions.
As a result, quantum trajectories may deviate significantly from their average evolution and violate ergodicity.
The degree of ergodicity breaking can be computed exactly for arbitrary finite-dimensional quantum systems, given full localization takes place and depends only on the effective distribution of the initial state over the Hilbert space.
This knowledge can be used to control the transition probabilities into subspaces with desired properties.
One useful application is decoherence-free subspaces, which are always identified by individual trajectories, leading to single-shot decoherence-free evolution.
We have investigated the general behavior in a range of different examples and explicitly demonstrated the occurrence of irreversible localization transitions in a variety of contexts such as parity selection in a Kerr-oscillator, coherences between many-body scar states, and Bell state generation from local measurement.
Finally, we have introduced a structural and a trajectory-based algorithm to find the finest decomposition of the Hilbert space, granting access to symmetries, conserved quantities and all the steady states of the Lindblad equation, thus providing information about the asymptotic behavior of its unraveled quantum trajectories. 

\acknowledgements
I am greatly indebted to Milton Aguilar for invaluable advice, discussions and support.
Numerical simulations were performed with QuTip \cite{Johansson2012,Johansson2013}.
I acknowledge support from the Vector foundation.

\appendix 

\section{Asymptotic states of diffusive trajectories---Proof of \cref{th:diff-inv}}
\label{sec:invariant-diff}
\begin{thm-hand}[1]
  A state $\rho$ is invariant under the diffusive unraveling, \cref{eq:homodyne-sme}, if and only if the Hamiltonian commutes with the state, $[H,\rho] = 0$, and it holds that
  \begin{align}
    L_k &=
    \begin{bmatrix}
      A_k& L_{k,12} \\
      0 & L_{k,22}
    \end{bmatrix}, \ \forall k,\\
    H_{12} &= -\I\frac{1}{2} \sum_k A_k^\dagger L_{k,12},
  \end{align} 
  with $A_k = S_k + \mathds{1}a_k$, where $S_k$ is a anti-Hermitian operator that commutes with the state, $[S_k,\rho] = 0, \forall k$
  and $a_k \in \mathbb{C}$, is an arbitrary complex number.
  The block matrices $H_{22}$ and $L_{k,22}$ are arbitrary.
\end{thm-hand}
Let $\rho$ be a density operator such that $\forall k$:
\begin{align}
	L_k \rho + \rho L^\dagger_k - \tr[\rho(L_k+L^\dagger_k)] \rho
	& = 0
	\label{eq:meas-diff}\\
	- i [H , \rho] 
	+ \sum_k\left(L_k \rho L^\dagger_k - \frac{1}{2} \left(L_k^\dagger L_k \rho + \rho L_k^\dagger L_k\right)\right) 
	& = 0
	\label{eq:lindblad}
\end{align}
We define $r_k \equiv \tr[\rho(L_k+L_k^\dagger)]$, with $r_k \in \mathbb{R}$.
The Lindblad equation is invariant under the transformation
\begin{align}
	\tilde L_k &\equiv L_k-a_k\mathds{1},
	\label{eq:shift}\\
	\tilde H &\equiv H - \frac{1}{2\I} \sum_k (a^\ast_k L_k - a_kL^\dagger_k) + b \mathds{1},
\end{align}
with $a_k \in \mathbb{C}$ and $r_k = a_k + a_k^\ast$, where $b \in \mathbb{R}$ is arbitrary.
\Cref{eq:meas-diff} becomes
\begin{align}
	\tilde L_k\rho + \rho \tilde L^\dagger_k = 0,
\end{align}
which implies
\begin{align} 
	\tilde L_k \rho \tilde L_k^\dagger 
	= - \tilde L_k \tilde L_k \rho 
	= - \rho \tilde L^\dagger_k \tilde L^\dagger_k.
\end{align}
\Cref{eq:lindblad} can thus be written as
\begin{align}
	\sum_k (\tilde L^\dagger_k \tilde L_k + \tilde L_k\tilde L_k + 2 i \tilde H) \rho + \rho (\tilde L^\dagger_k \tilde L_k + \tilde L^\dagger_k\tilde L^\dagger_k - 2 i \tilde H) = 0.
\end{align}
Define the operator $K = \sum_k (\tilde L^\dagger_k  \tilde L_k + \tilde L_k \tilde L_k) + 2 i \tilde H$. Then, $\rho$ is the simultaneous solution of the algebraic Lyapunov system and $\rho$ needs to lie in the intersection of the solution spaces of
\begin{align}
		\tilde L_k \rho + \rho \tilde L^\dagger_k = 0 \quad \land \quad
		K \rho + \rho K^\dagger  = 0.
    \label{eq:Lyapunov}
\end{align}
Without loss of generality, we can consider the problem in the eigenbasis of $\rho$ and write
\begin{align}
	\rho = 
	\begin{bmatrix}
		D & 0 \\
		0 & 0
	\end{bmatrix},
\end{align}
where $D$ is a diagonal, real, and invertible. 
In this basis let
\begin{align}
	B = 
	\begin{bmatrix}
		B_{11} & B_{12} \\
		B_{21} & B_{22}
	\end{bmatrix}.
\end{align}
Then the condition $B \rho + \rho B^\dagger = 0$ implies
\begin{align}
	B_{11} D + D B_{11}^\dagger & = 0, \qquad
	B_{21} D = 0.
\end{align}
Because $D$ is invertible, it follows that $B_{21} = 0$. 
We can write $\tilde{B}_{11} \equiv B_{11} D$ such that $\tilde{B}_{11} + \tilde{B}_{11}^\dagger = 0$. 
Thus, $\tilde{B}_{11}$ is anti-Hermitian. 
Therefore, the operators $\tilde L_k$ and $K$ have the form
\begin{align}
	\tilde L_k =
	\begin{bmatrix}
		\tilde S_k  D^{-1} & \tilde L_{k,12} \\
		0 & \tilde L_{k,22}
	\end{bmatrix}
	\qquad
	K =
	\begin{bmatrix}
		Z  D^{-1} & K_{12} \\
		0 & K_{22}
	\end{bmatrix}
	\label{eq:LandK}
\end{align}
with $\tilde S_k=-\tilde S_k^\dagger$ and $Z=-Z^\dagger$, anti-Hermitian. 
In the eigenbasis of $\rho$ the Hamiltonian assumes the form
\begin{align}
	\tilde H =
	\begin{bmatrix}
		\tilde H_{11} & \tilde H_{12} \\
		\tilde H_{12}^\dagger & \tilde H_{22}
	\end{bmatrix},
\end{align}
the explicit form of $K$ is
\begin{widetext}
\begin{align}
	K =
	\begin{bmatrix}
		\sum_k \left((\tilde S_k  D^{-1})^{2} + (\tilde S_k  D^{-1})^\dagger (\tilde S_k  D^{-1})\right) + 2 i \tilde H_{11} 
		& \sum_k \left((\tilde S_k  D^{-1}) \tilde L_{k,12} + \tilde L_{k,12} \tilde L_{k,22}\right) + 2 i \tilde H_{12} 
		\label{eq:skew}\\
		\sum_k \left(\tilde L_{k,12}^\dagger (\tilde S_k  D^{-1})\right) + 2 i \tilde H_{12}^\dagger 
		& \sum_k \left((\tilde L_{k,22})^{2} + \tilde L_{k,12}^\dagger \tilde L_{k,12} + \tilde L_{k,22}^\dagger \tilde L_{k,22}\right) + 2 i \tilde H_{22}
 	\end{bmatrix}.
\end{align}
\end{widetext}
Comparison with \cref{eq:LandK} yields
\begin{align}
	\sum_k \left((\tilde S_k  D^{-1})^{2} D + (\tilde S_k  D^{-1})^\dagger (\tilde S_k  D^{-1}) D\right) + 2 i \tilde H_{11} D & = Z \\
	\sum_k \left(\tilde L_{k,12}^\dagger (\tilde S_k  D^{-1})\right) + 2 i \tilde H_{12}^\dagger & = 0
\end{align}
Since $Z$ is anti-Hermitian, \cref{eq:Lyapunov} implies
\begin{align}
	- i [\tilde H_{11},D] =
    \frac{1}{2}\sum_k\left(\tilde S_k[D^{-1},\tilde S_k] - [D^{-1},\tilde S_k]\tilde S_k\right),
\end{align}
where we used the fact that $\tilde S_k$ is anti-Hermitian. 
This yields $\diag(\tilde S_k[D^{-1},\tilde S_k] - [D^{-1},\tilde S_k]\tilde S_k) = 0, \forall k$.
But we also have $\diag([D^{-1},\tilde S_k]\tilde S_k) = -\diag(\tilde S_k,[D^{-1},\tilde S_k])$, so $\diag(\tilde S_k[D^{-1},\tilde S_k]) = 0$ and $\tilde S_k[D^{-1},\tilde S_k]$ is a hollow matrix.
Consider the positive-semidefinite operator $P = \sum_k [D^{-1},\tilde S_k][D^{-1},\tilde S_k]^\dagger$
\begin{align}
    \tr[P] = 2\sum_k \tr(D^{-1}\tilde S_k[D^{-1},\tilde S_k]) = 0.
\end{align}
Therefore, $P=0$ and $[D^{-1},\tilde S_k] = 0, \forall k$, because the individual elements, $[D^{-1},\tilde S_k][D^{-1},\tilde S_k]^\dagger \ge 0, \forall k$, are also positive and it follows that $[D,\tilde S_k] = 0, \forall k$. 
Consequently, the state commutes with the Hamiltonian, $[\tilde H_{11},D] = [H_{11},D] = 0$, and thus $[\tilde H,\rho] = [H,\rho] = 0$.

Since the operator $\tilde L_{k,11} = \tilde S_kD^{-1}$ is itself anti-Hermitian and commutes with the state, $[\tilde L_{k,11},\rho] = 0$, we can define a new anti-Hermitian $\tilde L_{k,11} = S_k$, with $S_k^\dagger=-S_k$ and $[S_k,\rho]=0$.
$S_k$ can otherwise be arbitrary because $\tilde S_k$ was arbitrary.
Transforming back then gives
\begin{align}
  L_{k,11} = S_k + a_k \mathds{1},
\end{align}
with, $a_k \in \mathbb{C}$, an arbitrary complex number that determines, $r_k = a_k+a_k^\ast$.

In general, the form of the original Hamiltonian and the Lindblad jump operators is
\begin{align}
  L_k 
	= \begin{bmatrix}
        A_k & L_{k,12} \\
        0 & L_{k,22}
    \end{bmatrix}, \qquad
	H_{12} 
	= -\I\frac{1}{2}\sum_k A_k^\dagger L_{k,21},
\end{align}
where $L_{k,11} = A_k = S_k + \mathds{1}a_k$ where, $a_k \in \mathbb{C}$, is a free parameter and $S_k$ is a anti-Hermitian commuting with the state, $[S_k,\rho]=0$.

\section{Asymptotic states of jump trajectories---Proof of \cref{th:jump-inv}}
\label{sec:invariant-jumps}
\begin{thm-hand}[2]
  A state $\rho$ is invariant under the quantum jump unraveling, \cref{eq:jump-sme}, if and only if the Hamiltonian commutes with the state, $[H,\rho] = 0$, and it holds that
  \begin{align}
    L_k &= 
    \begin{bmatrix}
      C_k & L_{k,12} \\
      0 & L_{k,22}
    \end{bmatrix}, \ \forall k,\\
    H_{12} &= -\I\frac{1}{2}\sum_k C_k^\dagger L_{k,12},
  \end{align}
  with $C_k = c_k U_k$, where $U_k$ is a unitary operator that commutes with the state, $[U_k,\rho] = 0, \forall k$
  and $c_k \in \mathbb{C}$, is an arbitrary complex number.
  The block matrices $H_{22}$ and $L_{k,22}$ are arbitrary.
\end{thm-hand}

Let $\rho$ be a density operator such that $\forall k$:
\begin{align}
	&L_k \rho L_k^\dagger = \tr[L_k\rho L_k^\dagger]\rho, 
	\label{eq:measurement}\\
	&- i [H , \rho] +
	\sum_k\left(\tr[L_k\rho L^\dagger_k]\rho - \frac{1}{2} (L_k^\dagger L \rho + \rho L_k^\dagger L_k)\right) = 0.
	\label{eq:liouvillian}
\end{align}
We define 
\begin{align}
\kappa_k \equiv \tr[L_k \rho L_k^\dagger], 
\end{align}
with $\kappa_k \in \mathbb{R}_+$. 
Without loss of generality, we can write 
\begin{align}
    \rho = 
    \begin{bmatrix}
        D & 0 \\ 0 & 0
    \end{bmatrix}
    \quad 
    L_k = 
    \begin{bmatrix}
        L_{k,11} & L_{k,12} \\ L_{k,21} & L_{k,22}
    \end{bmatrix},
\end{align}
where $D$ is a diagonal, real, positive-definite matrix and $L_k$ is arbitrary.
\Cref{eq:measurement} implies 
\begin{align}
    L_{k,11}DL^\dagger_{k,11} &= \kappa_k D, 
    \label{eq:first}\\
    L_{k,11} D L^\dagger_{k,21} &= 0, 
    \label{eq:second}\\
    L_{k,21} D L^\dagger_{k,21} &= 0.
\end{align}
Let $\tilde L_{k,11} \equiv L_{k,11}/c_k$, where, $c_k \in \mathbb{C}$ is an arbitrary complex number that determines the jump probability, $\kappa_k = |c_k|^2$.
It then follows from \cref{eq:first}
\begin{align}
    \tilde L_{k,11} D \tilde L_{k,11}^\dagger = D.
\end{align}
Since $D$ is invertible and positive-definite by construction, we can equivalently write
\begin{align}
    \left[D^{-1/2}\tilde L_{k,11} D^{1/2}\right] \left[D^{-1/2}\tilde L_{k,11}D^{1/2}\right]^\dagger = \mathds{1}.
\end{align}
We can compute 
\begin{align}
    \left|\det[D^{-1/2} \tilde L_{k,11} D^{1/2}]\right|^2 = 1.
\end{align}
The matrix $D^{-1/2}\tilde L_{k,11} D^{1/2}$ is thus invertible and we obtain
\begin{align}
        [D^{-1/2} \tilde L_{k,11} D^{1/2}]^{-1} = [D^{-1/2} \tilde L_{k,11} D^{1/2}]^\dagger.
\end{align}
The operator
\begin{align}
    U_k = D^{-1/2} \tilde L_{k,11} D^{1/2}
\end{align}
is hence a unitary.
This results in the general form
\begin{align}
    L_{k,11} = c_k D^{1/2}U_kD^{-1/2},
\end{align}
where $U_k$ is an arbitrary unitary operator.
In particular, $L_{k,11}$ is invertible: $L^{-1}_{k,11} = c_kD^{1/2}U^\dagger_k D^{-1/2}$. 
Together with \cref{eq:second}, this implies $L_{k,21} = 0, \forall k$.
Up to this point, the most general form of $L_k$ that satisfies \cref{eq:measurement} reads 
\begin{align}
    L_k = 
    \begin{bmatrix}
        c_k D^{1/2}U_kD^{-1/2} & L_{k,12} \\
        0 & L_{k,22}
    \end{bmatrix},
    \label{eq:Lprelim}
\end{align}
with $L_{k,12}$ and $L_{k,22}$ arbitrary.

\Cref{eq:liouvillian} may be rewritten as a homogeneous Sylvester equation
\begin{align}
    K\rho + \rho K^\dagger = 0,
\end{align}
with $K = \kappa_k \mathds{1} - \sum_k L^\dagger_k L_k - 2\I H$.
Inserting $L_k$ and $H$ in block form yields
\begin{align}
    &-2\I[H_{11},D] \notag\\
    &+ \sum_k \left(2\kappa_k D - L^\dagger_{k,11}L_{k,11}D + DL_{k,11}L^\dagger_{k,11}\right) = 0,\\
    &\sum_k L^\dagger_{k,21}L_{k,11} = 2\I H^\dagger_{12}. 
\end{align}
Now using \cref{eq:Lprelim}, we obtain
\begin{align}
    &-2\I[H_{11},D] 
    + \sum_k \bigg(2\kappa_k D - D^{-1/2}U_k^\dagger DU_kD^{1/2} \notag\\
    &- D^{1/2}U_k^\dagger DU_kD^{-1/2}\bigg) = 0,\\
    &H_{12} = -\I\frac{1}{2}\sum_k c_kL^\dagger_{k,11}L_{k,21}.
\end{align}
Using $D = D^{-1/2}DD^{1/2} = D^{1/2}DD^{-1/2}$, the first equation can be written as 
\begin{align}
    2\I[H_{11},D] 
    =& \sum_k \kappa_k \bigg[D^{-1/2}(D-U_k^\dagger DU_k)D^{1/2} \notag \\
    &+ D^{1/2}(D-U_k^\dagger DU_k)D^{-1/2}\bigg].
\end{align}
In the eigenbasis of $\rho$, we have $\diag([H_{11},D]) = 0$, implying that $\diag(D-U_k^\dagger DU_k) = 0, \forall k$.
Consider the positive-semidefinite matrix $P = \sum_k [D,U_k][D,U_k]^\dagger$
\begin{align}
    \tr[P] = 2 \sum_k\tr[D(D- u^\dagger_k DU_k)].
\end{align} 
Since $D$ is diagonal and $\sum_k(D- U^\dagger_k DU_k)$ is hollow, the product is hollow and thus traceless, $\tr[P] = 0$.
The only traceless positive-semidefinite matrix is the zero matrix.
It thus follows, $P=0$, yielding $[D,U_k] = 0, \forall k$, because the individual elements, $[D,U_k][D,U_k]^\dagger \ge 0, \forall k$, are also positive-semidefinite.
As a consequence, the density matrix commutes with the Hamiltonian, $[H_{11},D] = 0$ and thus $[H,\rho] = 0$.

In general, the form of the Hamiltonian and the Lindblad jump operators is therefore given by
\begin{align}
  L_k 
= \begin{bmatrix}
      C_k & L_{k,12} \\
      0 & L_{k,22}
  \end{bmatrix},\qquad
	H_{12} 
	= -\I\frac{1}{2}\sum_k C_k^\dagger L_{k,21},
\end{align}
where $L_{11} = C_k = c_kU_k$ where, $c_k \in \mathbb{C}$, is a free parameter and $U_k$ is a unitary commuting with the state, $[U_k,\rho]=0$.

\section{Invariant states---special cases}
\label{sec:invariant-appendix}
Creating invariant states according to \cref{th:diff-inv,th:jump-inv} is basically equivalent to engineering the decaying subspace $\cald$ (see \cref{sec:space-structure}).
An invariant state can live only inside the asymptotic state space $\calr$, which by construction corresponds to the upper left block of the block decomposition in \cref{eq:block form} and \cref{th:diff-inv,th:jump-inv}.
The conditions \cref{eq:lindblad,eq:liouvillian} impose a structure on $H$ and the $L_k$ (see also \cref{th:decaying-subspace}) such that probability cannot not leak out of $\calr$.
The steady state manifold hosting the invariant states is therefore an attractive subspace.
However, invariant states of rank more than two are generally not attractive fixed points of the evolution.

\subsection{Full rank invariant state}
\label{sec:full-rank}
By virtue of \robs{obs:decay}, if a full rank invariant state $\rho$ exists, the decaying subspace is the empty set and, from \cref{th:diff-inv}, we obtain in particular for the diffusive unraveling
\begin{align}
  \label{eq:diff-full-rank}
    [H,\rho] = 0, \quad L_k = S_k+\mathds{1}a_k, \quad [S_k,\rho] = 0, \ \forall k
\end{align}
where $S_k$ is anti-Hermitian and $a_k \in \mathbb{C}$ is arbitrary.
A full rank invariant state of the diffusive unraveling thus forces the Lindblad jump operators to be normal and the corresponding Lindblad equation becomes a unital map.
For the jump unraveling, we analogously obtain from \cref{th:jump-inv}
\begin{align}
  \label{eq:jump-full-rank}
    [H,\rho] = 0, \quad L_k = c_k U_k, \quad [U_k,\rho] = 0, \ \forall k,
\end{align}
where $U_k$ is unitary and $c_k \in \mathbb{C}$ is arbitrary.
The form of the Lindblad jump operators in \cref{eq:jump-full-rank,eq:diff-full-rank} corresponds to classical noise trajectories (\cref{sec:classical-noise}), where the measurement update vanishes and purification and localization cannot take place anymore (\cref{th:incomplete-diff,th:incomplete-jump}).
The commutation relations imply a common block decomposition of $\rho$ and $A \in \{H,S_k,U_k\}$
\begin{align}
  D = \bigoplus_j p_j \mathds{1}_{m_j}, \quad
  A = \bigoplus_j a_{jj}
\end{align}
where the multiplicity, $m_j$, of each eigenvalue, $p_j$, determines the dimension of each block.
The block matrices $a_{jj} \in \mathbb{C}^{m_j\times m_j}$ are arbitrary square Hermitian ($H$), anti-Hermitian ($S_k$) and unitary ($U_k$) matrices respectively.
The more degeneracy in $\rho$, the more freedom there is in $A$.
In the limiting case where, $\rho = \mathds{1}_d/d$, is the maximally mixed state, $A$ is completely arbitrary and there may not be any further invariant states.
On the other hand, if $\rho$ has only simple eigenvalues, the commutation \cref{eq:jump-full-rank,eq:diff-full-rank} becomes transitive and the Hamiltonian and Lindblad jump operators are simultaneously diagonalizable, $[H,L_k] = 0, \forall k$, and any state $\rho$ diagonal in the same basis is also invariant.

\subsection{Block diagonal jump operators}
Now assume that $\rho$ is arbitrary and the Lindblad jump operators $L_k$ are block diagonal in the basis of $\rho$, i.e. $L_{k,12} = 0$.
This implies $H_{12} = 0$.
Thus, both the Hamiltonian and the Lindblad jump operators are in simultaneous block diagonal form.
The full rank case of \cref{sec:full-rank} is then recovered in the upper left block corresponding to $H_{11}$ und $L_{k,11}=c_k U_k$.
In the lower right block the presence or absence of invariant states depends on the microscopic details of $H_{22}$ and $L_{k,22}$.
The Hilbert space decomposes into two orthogonal subspaces $\calh = \calq\oplus\calp$, where $\calq$ contains full rank invariant states and $\calp$ is of arbitrary structure. There is no decaying subspace.
If the $L_k$ are normal operators, the Lindbladian is also unital.

\section{No asymptotic trajectories in the decaying subspace}
\label{sec:no-trajs-in-D}
We here show that in the infinite-time limit, there is no trajectory with support on the decaying subspace, i.e.
\begin{align}
  \lim_{t \to \infty} |\cald(t)|^2 
  = \lim_{t \to \infty} \tr[\rho_\m{c}(t)P_\cald] = 0.
\end{align}
Existence of a decaying subspace implies the specific structure of the Hamiltonian and the Lindblad jump operators in \cref{th:decaying-subspace}.

Given a finite-dimensional master equation of Lindblad form, there is a unique decomposition of the Hilbert space
\begin{align}
  \calh = \cald \oplus \calr.
\end{align}
The decaying subspace is defined as (cf. \cref{eq:decaying-subspace})
\begin{align}
  \cald = \left\{\ket{\psi} \in \calh \big\vert \lim_{t \to \infty}\bra{\psi}\sft^t(\rho) \ket{\psi} = 0\right\}.
  \label{eq:defining-property}
\end{align}
Absence of asymptotic trajectories in the decaying subspace is already implied by the pathwise ergodic theorem, \cref{sec:erg-theorem}.
In order for the long-time average to reproduce a stationary state of the Lindblad equation, the asymptotic trajectory cannot spend an extensive amount of time outside of $\calr$, the domain of the stationary state.

We want to show this more concretely here.
We first establish that, for Lindbladian evolution, the existence theorem, \cref{th:decaying-subspace}, implies the defining property \cref{eq:defining-property}.
By assumption, Hamiltonian and jump operators thus have the form 
\begin{align}
  L_k &= 
  \begin{bmatrix}
    L_{k,11} & L_{k,12} \\ 0 & L_{k,22}
  \end{bmatrix}, \ \forall k,\\
  H_{12} &= -i \frac{1}{2} \sum_k L^\dagger_{k,11}L_{k,12}.
\end{align}
By denoting $A \in \{H,L_k\}$, the upper left block, $A_{11} = P_\calr A P_\calr$, and lower right block, $A_{22} = P_\cald A P_\cald$, hosts operators restricted to $\calr$ and $\cald$ respectively.
The temporal evolution of the overlap in $\cald$ for Lindbladian dynamics readily follows
\begin{align}
  \dv{t}\left(|\cald(t)|^2\right) = 
  - 2\sum_k \tr[\rho_{22}(t) \left(L^\dagger_{k,12}L_{k,12}\right)] \le 0,
\end{align}
which is non-increasing.
In contrast to previous analysis with orthogonal subspaces, care must be taken since, $[P_\cald,L_k]\neq0$ and $[P_\cald,H]\neq0$.
For diffusive quantum trajectories (\cref{eq:homodyne-sme}), this immediately yields 
\begin{align}
  \dd{\left(|\cald(t)|^2\right)} =& 
  - 2\sum_k \tr[\rho_{\xi,22}(t) \left(L^\dagger_{k,12}L_{k,12}\right)] \dd{t} \notag\\
  &+ \bigg(|\calr(t)|^2\tr[\rho_\xi(t)\left(P_\cald L_k+L^\dagger_k P_\cald\right)] \notag\\
  &-|\cald(t)|^2\tr[\rho_\xi(t)\left(P_\calr L_k+L^\dagger_k P_\calr\right) ]\bigg) \dd{W_k}.
\end{align}
Similar to \cref{eq:calq} for orthogonal subspaces, we obtain Brownian motion with state-dependent diffusion, except that now there is also a drift term that induces a deterministic flow away from $\cald$, eventually filling $\calr$ entirely.
There is only one fixed point, it is attractive and stable and given by $|\calr(t)|^2=1$.

Proceeding analogously, for quantum jumps (\cref{eq:jump-sme}) we obtain
\begin{align}
  \dd{\left(|\cald(t)|^2\right)} =&
  \sum_k|\cald(t)|^2 \langle L^\dagger_k L_k\rangle \dd{t} - \tr[\rho_J(t)(L^\dagger_kL_k)P_\cald] \dd{t} \notag\\
  &-i\tr[\rho_J(t)(P_\cald H-HP_\cald)] \dd{t}\notag\\
  &+\sum_k \left[\frac{\tr[\rho_J(t)(L^\dagger_kP_\cald L_k)]}{\langle L^\dagger_kL_k\rangle}-|\cald(t)|^2\right] \dd{N_k},
\end{align}
with $\tr[\rho_J(t)(L^\dagger_kP_\cald L_k)] = \tr[\rho_{J,22}(t)(L^\dagger_{k,22}L_{k,22})]$.
This form bears less similarities with the corresponding expression for orthogonal subspaces \cref{eq:calq_jump} and the interpretation is more involved.
Clearly, $|\calr(t)|^2=1$, is a fixed point still.
On the other hand $|\cald(t)|^2=1$ implies $\tr[\rho_J(t)(P_\cald H-HP_\cald)]=0$ and also $\langle L^\dagger_kL_k\rangle = \tr[\rho_{J,22}(t)(L^\dagger_{12}L_{12}+L^\dagger_{22}L_{22})] \ge \tr[\rho_J(t)(L^\dagger_kP_\cald L_k)]$, such that the measurement update becomes
\begin{align}
  \dd{\left(|\cald(t)|^2\right)} = 
  \sum_k \left[\frac{\tr[\rho_J(t)(L^\dagger_kP_\cald L_k)]}{\langle L^\dagger_kL_k\rangle}-1\right] \dd{N_k} \le 0.
\end{align}
The system cannot stay in $\cald$ indefinitely and it is only a matter of time until it jumps out, decreasing the overlap with the decaying subspace.
$|\calr(t)|^2=1$ remains the unique fixed point, eventually attracting all quantum trajectories.

\section{Incomplete localization---outline}
In full generality, the decomposition of the Hilbert space into minimal orthogonal blocks 
\begin{align}
    \calh &= \cald \oplus \calr,\\
    \mathcal{R} &= \bigoplus_{k=1}^K \calu_k \oplus \bigoplus_{l=1}^L \calx_l, \qquad 
    \calx_{l=1}^L \simeq \mathbb{C}^{m(l)} \otimes \calv_l,
\end{align}
holds (see \cref{sec:space-structure}), where $\cald$ denotes the decaying subspace 
\begin{align}
    \cald = \left\{\ket{\psi} \in \calh \big\vert \lim_{t \to \infty}\bra{\psi}\sft^t(\rho) \ket{\psi} = 0\right\},
\end{align}
which gets completely emptied asymptotically and $\calr$ is the asymptotic subspace 
\begin{align}
  \calr = \left\{\ket{\psi} \in \calh, \ \exists \rho \big\vert \lim_{t \to \infty}\bra{\psi}\sft^t(\rho) \ket{\psi} > 0\right\}.
\end{align}
We are interested in the asymptotic behavior of quantum trajectories where the decaying subspace is already completely vacated.
It thus suffices to consider only the orthogonal subspaces inside of $\calr$.
We have shown that diffusive and quantum jump trajectories eventually get trapped completely in one of the minimal orthogonal subspaces except if the cases (i) and (ii) of \cref{th:incomplete-diff,th:incomplete-jump} hold.
In the following we provide the corresponding proofs.
We consider the situation where the quantum trajectory gets stuck between arbitrary orthogonal subspaces $\calq$ and $\calp$ and cannot further localize in Hilbert space.
$\calq$ and $\calp$ may be minimal themselves or further comprised of minimal orthogonal subspaces.
The combined subspace $\calq\oplus\calp$ is hence the whole support of the trajectory.
The Hamiltonian and Lindblad jump operators are thus in block diagonal form with
\begin{align}
  H_\calq &= P_\calq H P_\calq, &H_\calp &= P_\calp H P_\calp,\\
  L_{k,\calq} &= P_\calq L_k P_\calq, &L_{k,\calp} &= P_\calp L_k P_\calp,
\end{align}
their restrictions to $\calq$ and $\calp$ respectively.

\subsection{Outline of the proof}
The proof of \cref{th:incomplete-diff,th:incomplete-jump} proceeds in three main steps that are detailed below.
\begin{enumerate}
    \item We first show (diffusion) and argue (jumps) respectively that the absence of purification inside a minimal subspace $\calq$ implies that the measurement backaction is trivial (cf. \cref{sec:absence-of-purification-in-a-minimal-subspace}), viz.
    \begin{align}
      \begin{drcases}
        \text{diffusion:} \ (L_k+L_k^\dag)_\calq \\
        \text{Jumps:} \ (L^\dag_kL_k)_\calq
      \end{drcases}
      = z_k \mathds{1}_{d_\calq}, \ \forall k.
    \end{align}
    In the case of quantum jumps this statement is a conjecture (cf. \rconj{pur} and also \rconj{appendix-pur}).
    \item Next, we proof the incomplete localization theorems, \cref{th:mincomplete-diff,th:mincomplete-jump}, for two minimal subspaces $\calq$ and $\calp$, where three cases need to be distinguished
    \begin{itemize}
      \item Purification in both subspaces, leading to unitary equivalence, case (ii) of \cref{th:mincomplete-diff,th:mincomplete-jump} (cf. \cref{sec:unitary-equivalence-diff,sec:unitary-equivalence-jump}).
      \item No purification in either subspace leading, to independent classical noise trajectories, case (i) of \cref{th:mincomplete-diff,th:mincomplete-jump} (cf. \cref{sec:absence-of-purification-in-both-subspaces-diff,sec:absence-of-purification-in-both-subspaces-jumps}).
      \item Purification in one subspace but not the other, which also leads to case (i) (cf. \cref{sec:purification-in-one-subspace-diff,sec:purification-in-one-subspace-jumps}).
    \end{itemize}
    \item Once this established, we use this result to prove incomplete localization between two arbitrary orthogonal subspaces $\calq$ and $\calp$ (cf. \cref{sec:incomplete-composite}).
\end{enumerate}
In general, we prove that the elements of the generator, the set $\{H,L_k\}$, must obey certain structural properties in order to stop localization for any arbitrary trajectory.
Whenever we can, we prove these properties relying only on an individual realization $\rho_\m{c}$ of the stochastic master equations \eqref{eq:homodyne-sme} and \eqref{eq:jump-sme}.
For this purpose, for pure state trajectories, the pathwise ergodic theorem grants us at least a complete but not necessarily orthogonal basis.
On several occasions we need to invoke every possible state in order to make general, structural statements independent of any particular realization.
This happens for both unravelings in \cref{sec:incomplete-composite}, as well as for quantum jumps in \cref{sec:absence-of-purification-in-a-minimal-subspace} and \cref{sec:proof-iib,sec:purification-in-one-subspace-jumps}.
Everything else, \cref{sec:absence-of-purification-in-a-minimal-subspace} and the entire \cref{sec:proof-no-selection-diff} for diffusion and \cref{sec:unitary-equivalence-jump} for jumps, are proven requiring only a single trajectory.

\section{Incomplete localization between two minimal subspaces $\calq$ and $\calp$}
\label{sec:proof-incomplete-minimal}
\subsection{Absence of purification in a minimal subspace}
\label{sec:absence-of-purification-in-a-minimal-subspace}
Before we can study the localization properties of quantum trajectories supported on two minimal subspaces, we first need to investigate the purification behavior on a single minimal subspace.
For this purpose, consider therefore a quantum trajectory $\rho_\m{c}$ that has full support on the minimal orthogonal subspace $\calq$ where asymptotic purification does not take place.
For the diffusive unraveling, we employ the asymptotic purification theorem by Barchielli and Gregoratti, Ref.~\cite{Barchielli2009}.
\begin{theorem}[{\cite[Theorem 5.12]{Barchielli2009}}]
  \label{th:pur-diff}
  The diffusive quantum trajectory $\rho_\xi$ (\cref{eq:homodyne-sme}) asymptotically purifies if there does not exist a projector $P_t$ of dimension at least two such that $\forall k$
  \begin{align}
      P_t(L_k+L^\dagger_k)P_t = z_k(t) P_t
      \label{eq:Barchielli}
  \end{align}
  holds, where $z_k \in \mathbb{R}$ are real numbers.
\end{theorem}
For the quantum jump unraveling we employ the analogous theorem by Maassen and Kümmerer, Ref.~\cite{Maassen2006}.
\begin{theorem}[{\cite[Theorem 1]{Maassen2006}}]
    \label{th:pur-jumps}
    The quantum jump trajectory asymptotically purifies if there does not exist a projector $P_t$ of dimension at least two such that $\forall k$
    \begin{align}
        P_t(L^\dagger_k L_k)P_t = z_k(t)P_t
        \label{eq:Maassen}
    \end{align}
    holds, where $z_k \ge 0$ are positive real numbers. 
\end{theorem}
Denote by
\begin{align}
    M_{k,\calq} = 
    \begin{dcases}
        (L_k+L^\dagger_k)_\calq, &\text{ diffusion}\\
        (L^\dagger_kL_k)_\calq, &\text{ jumps}
    \end{dcases}
\end{align}
the effective measurement operator.
Crucially, the time-dependent projector $P_t$ is such that it projects on the support of the state $\rho_\m{c}(t)$.
\Cref{eq:Barchielli,eq:Maassen} imply that the action of $M_{k,\calq}$ restricted to the subspace supporting the state, $P_t\calh$, is effectively proportional to the identity.
The measurement backaction is thus trivial and thereby prevents any further purification.
According to \cref{th:pur-diff,th:pur-jumps} we can thus find a projector $P_t$ of rank $d_P = \rank(\rho_\m{c})$ that satisfies the following properties
\begin{align}
    P_t(M_{k,\calq})P_t &= z_k(t) P_t,
    \label{eq:P} \\
    \label{eq:P-rho} P_t\rho_\m{c}(t)P_t &= \rho_\m{c}(t).
\end{align}
We may express the state in its eigenbasis according to
\begin{align}
    \rho_\m{c}(t) = \sum_m p_m(t) \dyad{\psi_m(t)}.
\end{align}
Since $P_t$ projects at least on the support of $\rho_\m{c}(t)$, it must hence be of the form
\begin{align}
    P_t = \sum_m \dyad{\psi_m(t)}.
\end{align}
From \cref{eq:P} it follows that $M_{k,\calq}$ is diagonal in the instantaneous eigenbasis of $\rho_\m{c}(t)$
\begin{align}
    \bra{\psi_m(t)} M_{k,\calq} \ket{\psi_m(t)} &= z_k(t),\\
    \bra{\psi_m(t)} M_{k,\calq} \ket{\psi_n(t)} &= 0, \ \forall n\neq m.
\end{align}
Due to the pathwise ergodic theorem (\cref{sec:erg-theorem}) the set of states $\{\ket{\psi_m(t)}\}_{m=1}^{m=d_P}$ will cover the whole subspace $\calq$.
Two cases need to be distinguished.
Either the trajectory gaplessly sweeps the whole subspace or some subspaces remain that are entirely disconnected from each other.
In the former case we have $\bigcap_t \ker(P_t) = \emptyset$ and
\begin{align}
  \nexists \calq = \calq_1 \oplus \calq_2, \text{ s.t. } P_t = P^{(1)}_t + P^{(2)}_t, \ \forall t.
  \label{eq:irreducible-assumption}
\end{align}
There does not exist a nontrivial orthogonal decomposition of the subspace $\calq$ such that the projector $P_t$ takes a corresponding block form at all times with $P_{\calq_\alpha} P_t^{(\beta)} = \delta_{\alpha,\beta}P_t^{(\beta)}$.
Since $P_t$ is of rank at least two, any state $\ket{\psi_m(t)}$ must moreover be an eigenstate of the Hermitian operator $M_{k,\calq}$.
Now for every projector $P_{t_1}$ at time $t_1$ we can find an overlapping projector $P_{t_2}$ at time $t_2$ and a state in their intersection $\ket{\psi} \in \Im(P_{t_1}) \cap \Im(P_{t_2})$ and with \eqref{eq:P} it follows that
\begin{align}
    \bra{\psi}P_{t_1} M_{k,\calq}P_{t_1} \ket{\psi}
    &= z_k(t_1)\\
    &= z_k(t_2)\\
    &= \bra{\psi}P_{t_2} M_{k,\calq}P_{t_2} \ket{\psi}.
\end{align}
Therefore, for all times $t_1$ and $t_2$, it must hold that $z_k(t_1) = z_k(t_2) \equiv z_k$ is actually time-independent.
Since there are no disconnected subspaces it follows that the effective measurement operator is proportional to the identity
\begin{align}
    M_{k,\calq} = z_k \mathds{1}_{d_\calq}.
\end{align}
We now proceed by treating the two unravelings sepearately.
We first discuss the diffusive case, where it remains sufficient to keep considering only a single trajectory.

\emph{Diffusion.---}In the complementary case to \cref{eq:irreducible-assumption}, there exists a nontrivial orthogonal decomposition of the subspace such that the trajectory assumes a block form at all times and we have
\begin{align}
    \calq = \bigoplus_\alpha \calq_\alpha, \quad P_t = \sum_\alpha P_t^{(\alpha)}, \quad \rho_\xi(t) = \bigoplus_\alpha \rho_{\xi,\alpha}(t).
    \label{eq:reducible-assumption}
\end{align}
For the sake of clarity, we focus on the case where $\alpha=1,2$, the general case then follows straightforwardly.
Denote by $P$ the projector on the subspace $\calq_1 = P\calq$ and denote by $\calq_2 = P^\perp\calq$ its complement.
Sustaining the block form at all times means
\begin{align}
  P \rho_\xi(t) P^\perp &= 0, \text{ and} &P \dot{\rho}_\xi(t) P^\perp &= 0.
  \label{eq:diagonal-evolution}
\intertext{Furthermore, we find}
    P^\perp \dot{\rho}_{\xi,1}(t) P^\perp &= 0, \text{ and} &P \dot{\rho}_{\xi,2}(t) P &= 0,
    \label{eq:no-state-transfer}
\end{align}
which means that there is no state transfer between the subspaces and the evolutions in $\calq_1$ and $\calq_2$ are actually independent.
Since quantum trajectories are rank non-increasing and the total rank of the state $\rho_\xi$ has to be preserved by assumption, the rank is preserved in $\calq_1$ and $\calq_2$ separately.
If $\m{rank}(P_t^{(i)}) > 1$, $\alpha=1,2$, we can apply the same reasoning as before in each subspace individually to obtain
\begin{align}
    (L_k+L_k^\dag) = z_{k,1} \mathds{1}_{d_{\calq_1}} \oplus z_{k,2} \mathds{1}_{d_{\calq_2}}.
\end{align}
Using the property \eqref{eq:P}, then determines that
\begin{align}
    (L_k+L_k^\dag) = z_{k} \mathds{1}_{d_{\calq}}.
\end{align}
If $\m{rank}(P_t^{(\alpha)}) = 1$ in both subspaces, the states $\rho_{\xi,\alpha}$ are both pure and from condition \eqref{eq:diagonal-evolution} we obtain
\begin{align}
  P \sfl_\xi(\rho_{\xi,1}) P^\perp
  =& \bigg(\I P\rho_{\xi,1} P H P^\perp + \sum_k PL_k P \rho_{\xi,1} P L^\dag_kP^\perp \notag\\
  & - \frac{1}{2} P \rho_{\xi,1} P L^\dag_k L_k P^\perp\bigg)\dd{t} \notag\\
  & + P\rho_{\xi,1} P L^\dag_k P^\perp \dd{W_k}.
  \label{eq:no-leak-diff}
\end{align}
where we have dropped the explicit time dependence for the sake of clarity.
Stochastic and deterministic parts must vanish independently.
The pure state $\rho_{\xi,1}$ goes through a complete basis in $\calq_1$ and it hence follows for all $k$ that $P^\perp L_k P = 0$, and checking the complementary condition ($P^\perp \sfl_\xi(\rho_{\xi,2}) P = 0$) that $P L_k P^\perp = 0$, which makes also the second and the third term vanish and thus leaves only $P H P^\perp = 0$.
This implies a common block decomposition of the Hamiltonian and the jump operators with $H = H_1 \oplus H_2$ and $L_k = L_{k,1} \oplus L_{k,2}$ which is incompatible with the assumption of $\calq$ being minimal.
Due to the continuous nature of the diffusive trajectory this follows without explicitly enforcing the absence of purification condition \cref{eq:Barchielli}.
The only case remaining is where one of the states is a rank one projector and the other has rank greater than one.
Assume without loss of generality that $\rho_{\xi,1}$ is pure. 
In the other subspace, $\calq_2$, we thus obtain $(L_{k,2}+L_{k,2}) = z_{k,2} \mathds{1}_{d_{\calq_2}}$, by the usual arguments.
Together with \eqref{eq:P}, this removes the time dependence of the scalar $z_{k,1}(t)$ in $\calq_1$ and amounts to
\begin{align}
    P_t (L_k+L_k^\dag) P_t = 
    \begin{pmatrix}
        z_k P_t^{(1)} & 0 \\ 0 & z_k P_t^{(2)}
    \end{pmatrix},
\end{align}
where we have defined $z_k \equiv z_{k,2}$.
The one-dimensional projector $P_t^{(1)} = \rho_{\xi,1}$ continuously goes through a complete basis in $\calq_1$ and thus eventually establishes $(L_{k,1}+L_{k,1}) = z_k \mathds{1}_{d_{\calq_1}}$, which finally results in
\begin{align}
    (L_k+L_k^\dag) = z_{k} \mathds{1}_{d_{\calq}}.
    \label{eq:result}
\end{align}
For more than two orthogonal subspaces in the decomposition \cref{eq:reducible-assumption}, the same reasoning can be applied to all of them and the same result, \cref{eq:result}, directly follows.

\emph{Jumps.---}The jump case is however more involved because probability can in principle be transferred discontinuously between subspaces.
At this point, to obtain results independent of the individual realizations, we have to supply that the decomposition \eqref{eq:reducible-assumption} must be a structural property and thus holds for all states $\rho_J$.
Therefore, we require the conditions in \cref{eq:diagonal-evolution} to apply for all states, i.e.
\begin{align}
  \forall \rho = P \rho P: \notag\\
  P \sfl_J(\rho) P^\perp 
  =& \bigg(\I P\rho P H P^\perp - \frac{1}{2} \sum_k P \rho P L^\dag_k L_k P^\perp\bigg)\dd{t}\notag\\
  &  + \frac{PL_k P \rho P L^\dag_kP^\perp}{\langle L^\dag_k L_k\rangle} \dd{N_k}, 
  \label{eq:no-leak-jump}
\end{align}
From the stochastic contribution in \cref{eq:no-leak-jump} we obtain
\begin{align}
  \forall \rho = P \rho P: PL_kP = 0 \ \lor \ P^\perp L_k P = 0,
  \label{eq:PLP}\\
  \forall \rho = P^\perp \rho P^\perp: P^\perp L_k P^\perp = 0 \ \lor \ P L_k P^\perp = 0,
  \label{eq:PpLPp}
\end{align}
at least one condition in each line must be fulfilled at the same time.
Moreover, invoking the absence of purification \cref{eq:Maassen}, it holds
\begin{align}
  P(L^\dagger_kL_k)P^\perp = P^\perp (L^\dagger_k L_k)P = 0,
  \label{eq:no-pur}
\end{align}
which in turn implies $PHP^\perp = 0$.
Writing the effective measurement operator as a block matrix
\begin{align}
  (P+P^\perp)(L^\dagger_kL_k)(P+P^\perp) = 
  \begin{pmatrix}
    L_{k,11} & L_{k,12}\\ L_{k,21} & L_{k,22}
  \end{pmatrix}
\end{align}
and going through all possible combinations in \cref{eq:PLP,eq:PpLPp} while enforcing \cref{eq:no-pur} leaves as only possibility compatible with minimality of the subspace, jump operators of block form
\begin{align}
  L_k =
\begin{pmatrix}
  0 & L_{k,12}\\ L_{k,21} & 0
\end{pmatrix}.
\end{align}
If the rank of at least one of the projectors is greater than one, $\m{rank}(P_t^{(1)}) > 1 \lor \m{rank}(P_t^{(2)}) > 1$, we obtain
\begin{align}
  (L^\dagger_kL_k) = z_k\mathds{1}_{d_\calq}.
\end{align}
However, if both states $\rho_{J,\alpha}$ are pure and $\m{rank}(P_t^{(\alpha)}) = 1$, we obtain instead
\begin{align}
  P_t(L^\dagger_kL_k)P_t = 
  \begin{pmatrix}
    \bra{\psi_1} A_1 \ket{\psi_1} & 0\\ 0 & \bra{\psi_2} A_2 \ket{\psi_2}
  \end{pmatrix}
  = z_k(t)P_t,
  \label{eq:A-psi}
\end{align}
where we have defined $A_1 \equiv (L_k^\dagger L_k)_{11}$, $A_2 \equiv (L_k^\dagger L_k)_{22}$.
Besides the trivial choice $A_1 = \mathds{1}_{d_{\calq_1}}$ and $A_2 = \mathds{1}_{d_{\calq_2}}$, a natural solution is unitary equivalence between the operators and states, i.e. $A_1 = uA_2 u^\dagger$, $\ket{\psi_1} = u\ket{\psi_2}$.
However, the structure of $(L^\dag_kL_k)$ is not uniquely determined by \cref{eq:A-psi} and there still remains some freedom.
Typically, unitarily related states can however not be sustained by the trajectory without having independent evolutions inside orthogonal subspaces (cf. \cref{sec:proof-no-selection-jumps}).
On the other hand every choice where
\begin{align}
  H = 
  \begin{pmatrix}
    h & 0\\ 0 & h
  \end{pmatrix}, \text{ and }
  L_k = 
  \begin{pmatrix}
    0 & l_k\\ l_k & 0
  \end{pmatrix},
\end{align}
is bound to fail since there always exists a Hermitian matrix
\begin{align}
  X = 
  \begin{pmatrix}
    A & B\\ B & A
  \end{pmatrix},
\end{align}
with diagonal matrices $A = a \mathds{1}$ and $B = b \mathds{1}$ that satisfies $[X,H] = [X,L] = 0$ and $H$ and $L$ thus have a common block diagonal form.
We therefore conjecture that to prevent purification in a minimal subspace $\calq$, the only compatible solution is $(L_k^\dagger L_k) = z_k \mathds{1}_{d_\calq}$ (cf. \rconj{pur}).
\begin{conjecture}\label{appendix-pur}
  In a minimal orthogonal subspace $\calq$, absence of purification of a mixed state quantum jump trajectory $\rho_J$ implies
  \begin{align}
    (L^\dagger_kL_k) = z_k \mathds{1}_{d_\calq}, \ \forall k,
  \end{align}
  where $z_k >0$.
\end{conjecture}

For completeness, a general decomposition into multiple subspaces \cref{eq:reducible-assumption}, yields by assumption
\begin{align}
  \calq = \bigoplus_\alpha \calq_\alpha, \quad \sum_\alpha P^{(\alpha)}_t, \quad \rho_J = \sum_\alpha p_\alpha \dyad{\psi_\alpha},
\end{align}
with $\calq_\alpha = P_\alpha \calq$ and $P_\alpha P^{(\beta)}_t = \delta_{\alpha,\beta} P^{(\beta)}$.
Sustaining the block form of the state translates to the conditions
\begin{align} 
  \forall \rho = P_\alpha\rho P_\alpha: \ P_\beta\rho P_\gamma = 0, \ P_\beta\dot{\rho} P_\gamma = 0, \ \forall \beta \neq \gamma,
\end{align}
from which it follows that the Hamiltonian has a block diagonal decomposition
\begin{align}
  \quad H = \bigoplus_\alpha h_\alpha,
\end{align}
and for the jump operators we obtain the conditions
\begin{align}
  P_\beta L_k P_\gamma = 0 \lor P_\gamma L_k P_\alpha &= 0, \ \forall \gamma,\alpha \neq \beta,\\
  P_\beta (L^\dagger_kL_k)P_\gamma &= 0, \ \forall \beta \neq \gamma,\\
  P_t (L^\dagger_k L_k)P_t &= z_k(t)P_t,
\end{align}
with $P_t = \sum_\alpha \dyad{\psi_\alpha}$.

\subsection{Incomplete localization of diffusive trajectories between two minimal subspaces}
\label{sec:proof-no-selection-diff}
We proof the following theorem.
\begin{theorem}
    \label{th:mincomplete-diff}
  Let the diffusive quantum trajectory \cref{eq:homodyne-sme} have support exclusively on two minimal subspaces $\calq \oplus \calp$.
  Then, there is no subspace selection on the trajectory level between $\calq$ and $\calp$ if and only if at least one of the following holds $\forall k$.
  \begin{enumerate}[(i)]
  \item Independent trajectories:
  \begin{align} 
      (L_k+L^\dagger_k)_{\calq} = z_k\mathds{1}_{d_{\calq}},\quad (L_k+L^\dagger_k)_{{\calp}} = z_k\mathds{1}_{d_{\calp}}.
      \label{appendieq:pur-diff}
  \end{align}
  \item Unitarily equivalent trajectories:
  \begin{align}
      H_{\calq} = uH_{\calp} u^\dagger, \quad L_{k,{\calq}} = uL_{k,{\calp}}u^\dagger,
      \label{appendixeq:u-diff}
  \end{align}
  \end{enumerate}
  with $z_k \in \mathbb{R}$ and unitary operator $u$.
\end{theorem}
Sufficiency follows immediately by insertion.
Consider an arbitrary diffusive quantum trajectory $\rho_\xi(t)$ in finite dimensions.
Denote by 
\begin{align}
  \overline{\rho}_\xi = \lim_{t \to \infty} \frac{1}{t} \int_0^t \dd{s} \rho_\xi(s),
\end{align}
the infinite time averaged state.
From the pathwise ergodic theorem (\cref{sec:erg-theorem}) it follows that this state is a stationary state of the Lindblad equation, i.e. $\sfl(\overline{\rho}_\xi) = 0$.
Decompose the support of $\overline{\rho}_\xi$ into two minimal orthogonal subspaces $\calq \oplus \calp$, then $\overline{\rho}_\xi$ has full rank on $\calq \oplus \calp$.
Asymptotic dynamics can thus take place only on the combined subspace $\calq \oplus \calp$.
Now, localization does not occur between $\calq$ and $\calp$ if and only if the following holds
\begin{align}
  \label{eq:diff-freeze}
    |\calp(t)|^2&\tr[\rho_\xi(t) (L_k+L_k^\dagger)P_\calq] = \notag \\
    & |\calq(t)|^2 \tr[\rho_\xi(t)(L_k+L_k^\dagger)P_\calp],
\end{align}
for all times $t > T$, where $T$ is random variable that is generally different for every realization.
This relation follows directly from the evolution equation of the overlap \cref{eq:calq}.
The trajectory is thus unable to decide between the two subspaces $\calq$ and $\calp$ and gets stuck in between.
If this happens then $\forall t > T$ the probabilities for the trajectory to be found in the respective subspace become stationary, $0< |\calq(t)|^2 = \m{const.} <1$ and $0<|\calp(t)|^2 = \m{const.}<1$, and it still holds that $|\calq(t)|^2+|\calp(t)|^2=1$ (the subspaces span the whole support of $\rho_\xi$).
In this stationary regime, we define the effective trajectories inside the subspaces $\calq$ and $\calp$
\begin{align}
    \rho_{\xi,\cala} &\equiv P_\cala \rho_\xi(t) P_\cala, \\
    \tilde \rho_{\xi,\cala} &\equiv \frac{\rho_{\xi,\cala}}{\tr[\rho_{\xi,\cala}]} = \frac{\rho_{\xi,\cala}}{|\cala|^2_\xi},
\end{align}
where $\cala = \calq,\calp$ and, in accordance with the definitions in \cref{sec:erg-theorem,sec:Born}, we denote by $|\cala|^2_\xi \equiv |\cala(t)|^2, \ t>T$, the random variable, constant for every realization.
\Cref{eq:diff-freeze} can thus be rewritten as
\begin{align} 
    \tr[\tilde\rho_{\xi,\calq}(L_k+L_k^\dagger)_\calq]
   = \tr[\tilde\rho_{\xi,\calp}(L_k+L_k^\dagger)_\calp].
   \label{eq:first-order-diff}
\end{align}   
and the evolution of $\tilde \rho_{\xi,\calq}$ and $\tilde \rho_{\xi,\calp}$ in each subspace may thus be considered an independent trajectory
\begin{align}
  \dd{\tilde \rho}_{\xi,\calq} 
  =& -\I[H_\calq,\tilde \rho_{\xi,\calq}] \dd{t} \notag \\
  &+ \sum_k \left(L_{k,\calq}\tilde \rho_{\xi,\calq} L_{k,\calq}^\dagger - \frac{1}{2}\left\{L_{k,\calq}^\dagger L_{k,\calq},\tilde \rho_{\xi,\calq}\right\}\right) \dd{t} \notag \\
    &+ \Big[L_{k,\calq}\tilde \rho_{\xi,\calq}+\tilde \rho_{\xi,\calq} L_{k,\calq}^\dagger \notag \\
    & \quad \ - \langle L_{k,\calq}+L_{k,\calq}^\dagger\rangle_\calq \tilde \rho_{\xi,\calq}\Big]\dd{W_k},
    \label{eq:valid-trajectories-diff}
\end{align}
since \cref{eq:diff-freeze} is equivalent to 
\begin{align}
  \tr[\rho_\xi (L_k+L^\dagger_k)P_\calq]
  =\langle L_k+L_k^\dagger \rangle|\calq(t)|^2,
\end{align}
and therefore
\begin{align}
  \langle L_k + L_k^\dagger \rangle
  = \tr[\tilde \rho_\xi (L_k+L^\dagger_k)P_\calq]
  = \langle L_{k,\calq}+L_{k,\calq}^\dagger\rangle_\calq,
  \label{eq:meas-record-diff}
\end{align}
and analogously for $\dd{\tilde \rho}_{\xi,\calp}$.

\subsubsection{Purification in both subspaces}
\label{sec:unitary-equivalence-diff}
Before proceeding, we need to distinguish two cases.
Either, as we will now assume, the trajectories $\tilde \rho_{\xi,\calq}$ and $\tilde \rho_{\xi,\calp}$ asymptotically purify individually or at least one of them does not.
The latter case is deferred to \cref{sec:absence-of-purification-in-both-subspaces-diff,sec:purification-in-one-subspace-diff}.

We employ the asymptotic purification theorem, \cref{th:pur-diff}.
If the asymptotic purification theorem holds in both subspaces we can assume that 
\begin{align}
    \tilde \rho_{\xi,\calq} = \dyad{\psi_t}, \qquad 
    \tilde \rho_{\xi,\calp} = \dyad{\varphi_t}
\end{align}
are pure states for all times $t > T$.
In general, $\calq$ and $\calp$ might have different dimensions.
Assume without loss of generality that $d_\calq \ge d_\calp$.
Then, to compare evolutions we elevate the state $\ket{\psi_t}$ to
\begin{align}
  \dyad{\tilde \varphi_t}
  = \begin{pmatrix}
    \dyad{\varphi_t} & 0\\ 0 & 0
  \end{pmatrix}
  \in \mathbb{C}^{d_\calq \times d_\calq},
\end{align} 
such that the states $\dyad{\tilde \varphi_t}$ and $\dyad{\psi_t}$ live on a $d_{\calq}$-dimensional Hilbert space.
At each point in time there thus exists a unitary transformation $u_t$ that maps the states into each other and we can write
\begin{align}
  \dyad{\psi_t} = u_t \dyad{\tilde \varphi_t}u^\dag_t.
  \label{eq:unitary-relation}
\end{align}
Taking the stochastic differential on both sides then yields
\begin{align}
  \dd{(\dyad{\psi_t})} 
  =& (\dd{u_t})\dyad{\tilde \varphi_t}u^\dag_t + u_t \dd{(\dyad{\tilde \varphi_t})}u^\dag_t \notag\\
  &+ u_t \dyad{\tilde \varphi_t}(\dd{u^\dag_t})
  +(\dd{u_t})\dd{(\dyad{\tilde \varphi_t})}u^\dag_t \notag\\
  &+ u_t \dd{(\dyad{\tilde \varphi_t})}(\dd{u^\dag_t})
  +(\dd{u_t}){\dyad{\tilde \varphi_t}}(\dd{u^\dag_t}),
\label{eq:ito-unitary}
\end{align}
where we have accounted for the possibility that the unitary may have a stochastic component according to
\begin{align}
  \dd{u_t} = (A_t\dd{t} + \sum_k B_{t,k}\dd{W_k}) u_t,
\end{align}
with $B_t = -B^\dag_{t,k}$ and $A_t+A^\dag_t = B^2_{t,k}$.
\Cref{eq:ito-unitary} is the evolution equation of a valid quantum trajectory. 
In order for the right hand side to be consistent, it turns out that the last three terms $(\dd{u_t})\dd{(\dyad{\tilde \varphi_t})}u^\dag_t + u_t \dd{(\dyad{\tilde \varphi_t})}(\dd{u^\dag_t}) +(\dd{u_t}){\dyad{\tilde \varphi_t}}(\dd{u^\dag_t})$ need to vanish identically or equivalently $B_{t,k} = 0$.
Using \cref{eq:unitary-relation} to replace $\dyad{\tilde \varphi_t}$ and the fact that both trajectories have identical noise histories and measurement records (cf. \cref{eq:first-order-diff,eq:meas-record-diff}), we can identify an effective Hamiltonian $\tilde H_\calp$ and effective jump operators $\tilde L_{k,\calp}$ such that relation \eqref{eq:ito-unitary} can be cast in the form
\begin{align}
  \dd{\rho_\calq}
  = \sfl_{\xi,\calq}(\rho_\calq) 
  = \tilde \sfl_{\xi,\calp}(\rho_\calq),
  \label{eq:cast-diff}
\end{align}
where
\begin{align}
  \tilde \sfl_{\xi,\calp}(\rho_\calq) 
  =& -\I [\tilde H_\calp,\rho_\calq] \dd{t} \notag \\
  &+ \sum_k \left(\tilde L_{k,\calp} \rho_\calq \tilde L^\dag_{k,\calp}
  - \frac{1}{2} \{(\tilde L^\dag_k\tilde L_k)_\calp,\rho_\calq\}\right)\dd{t} \notag \\
  &+\Big[\tilde L_{k,\calp} \rho_\calq + \rho_\calq \tilde L^\dag_{k,\calp} \notag \\
  &\quad - \langle \tilde L_{k,\calp}+\tilde L^\dag_{k,\calp} \rangle_\calq \rho_\calq\Big]\dd{W_k}.
\end{align}
For the sake of brevity we have used the shorthand $\rho_\calq = \dyad{\psi_t}$.
The effective operators read
\begin{align}
  \tilde H_\calp = H_\calp +\I (\dd{u_t})u^\dag_t \oplus \vb{0}, \quad \tilde L_{k,\calp} = u_tL_{k,\calp}u^\dag_t \oplus \vb{0}.
\end{align}
$\dyad{\psi_t}$ is the uniqe solution of the quantum stochastic differential equations $\sfl_{\xi,\calq}$ and $\tilde \sfl_{\xi,\calp}$.
Since they produce the same unique solution, the tuples $(\tilde H_\calp,\tilde L_{k,\calp})$ and $(H_\calq, L_{k,\calq})$ must be related by a gauge transformation that transforms the operators but leaves the equation invariant.
The invariance of diffusive trajectories has been considered by Wiseman and Di\'osi in Ref.~\cite{Wiseman2001} (see also Ref.~\cite{Chia2011}).
For the diffusive, homodyne unraveling with real-valued mutually independent Wiener processes $\dd{W_k}$ we consider here, they are the same as for the Lindblad equation.
Crucially, these transformations take time independent operators to time independent operators and therefore $u_t$ must not be time dependent.
Because the effective operators $(\tilde H_\calp,\tilde L_{k,\calp})$ are not uniquely determined by \cref{eq:cast-diff}, all that can be said at this point is that the trajectory in $\calq$ and the inflated state in $\calp$ must be related by a time-independent unitary according to 
\begin{align}
  \dyad{\psi_t} = u\dyad{\tilde \varphi_t}u^\dag.
\end{align}
But since 
\begin{align}
    \dyad{\psi_t} = u\dyad{\tilde \varphi_t}u^\dag = 
    \begin{pmatrix}
        \dyad{\varphi_t} & 0\\ 0 & 0
    \end{pmatrix}
\end{align}
is block diagonal at all times, the long-time average cannot possibly converge to a full rank stationary state in $\calq$ and thus $\calq$ cannot be minimal, violating the assumption.
This implies that the subspaces $\calq$ and $\calp$ must have the same dimension, i.e. $d_\calq = d_\calp$ and it actually holds that
\begin{align}
  \dyad{\psi_t} = u\dyad{\varphi_t}u^\dag.
\end{align}

Building on this result, unitary equivalence of evolutions implies that both subspaces $\calq$ and $\calp$ have the same dimensionality and there exists a unitary, $U$, acting on the whole Hilbert space spanned by $\calq \oplus \calp$ such that
\begin{align}
    U = 
    \begin{pmatrix}
        0 & u\\ u^\dagger & 0
    \end{pmatrix},
    \qquad U^2 = \mathds{1},
    \label{eq:unitary-diff}
\end{align}
with
\begin{align}
    U \rho_{\xi} U^\dagger
    = 
    \begin{pmatrix}
        0 & u\\ u^\dagger & 0
    \end{pmatrix}
    \begin{pmatrix}
        \tilde \rho_{\xi,\calq} & 0\\ 0 & \tilde \rho_{\xi,\calp}
    \end{pmatrix}
    \begin{pmatrix}
        0 & u \\ u^\dagger & 0
    \end{pmatrix}
    = \rho_{\xi}.
\end{align}
The unitary $U$ is then a dynamical symmetry of the quantum trajectory
\begin{align}
    U\sft^t_\xi(\rho_\xi)U^\dagger
    = \sft^t_\xi(U\rho_\xi U^\dagger).
    \label{appendixeq:dynamical-symmetry}
\end{align}
Since the stochastic terms in \cref{appendixeq:dynamical-symmetry} are mutually independent, it follows for every $k$
\begin{align}
    L_{k,\calq} \tilde \rho_{\xi,\calq} + \tilde \rho_{\xi,\calq} L_{k,\calq}^\dagger
    = u L_{k,\calp} u^\dagger \tilde \rho_{\xi,\calq} 
    + \tilde \rho_{\xi,\calq} u L_{k,\calp}^\dagger u^\dagger.
\end{align}
Denote by $A_k \equiv L_{k,\calq} \tilde \rho_{\xi,\calq}$ and $B_k \equiv u L_{k,\calp} u^\dagger \tilde \rho_{\xi,\calq}$, then the above relation may be rewritten as
\begin{align}
    A_k+A^\dagger_k = B_k+B^\dagger_k.
    \label{eq:AB-master-relation}
\end{align}
This yields $A_k = B_k + \I R_k$, where, at this point, $R^\dagger_k = R_k$ is an arbitrary Hermitian matrix.
Since, by assumption, $\tilde \rho_{\xi,\calq} = \dyad{\psi_t}$ is a pure state, it follows that
\begin{align}
    L_{k,\calq} \dyad{\psi_t}
    = u L_{k,\calp} u^\dagger \dyad{\psi_t} + \I R_k.
    \label{eq:Hermitian}
\end{align}
Now, at any given time $t$, construct an orthogonal basis of the entire subspace $\calq$, where $\ket{\psi_t}$ is considered the first basis vector.
Denote by $\{\ket{\psi^\perp_\alpha}\}$ the remaining basis states that span the reduced subspace $\calq \setminus \{\ket{\psi_t}\}$ with $\braket{\psi_t}{\psi^\perp_\alpha} = 0$ and $\braket{\psi^\perp_\alpha}{\psi^\perp_\beta} = \delta_{\alpha,\beta}$.
In \cref{eq:Hermitian}, multiplying from the right by $\ket{\psi^\perp_\alpha}$ thus yields
\begin{align}
  R_k\ket{\psi^\perp_\alpha} = 0, \ \forall \alpha.
\end{align}
The operator $R_k$ can thus not have support on the reduced subspace $\calq\setminus \{\ket{\psi_t}\}$, and, due to Hermiticity, must hence be of the form
\begin{align}
  R_k(t) = r_k(t) \dyad{\psi_t},
\end{align}
where $r_k(t) \in \mathbb{R}$ is an arbitrary time-dependent real number.
It follows 
\begin{align}
  L_{k,\calq} \dyad{\psi_t}
  = (u L_{k,\calp} u^\dagger + \I r_k(t))\dyad{\psi_t}.
  \label{eq:r(t)}
\end{align}
Define $F_k \equiv L_{k,\calq} - u L_{k,\calp} u^\dagger$, then the above relation may be equivalently expressed as
\begin{align}
  F_k \ket{\psi_t} = \I r_k(t)\ket{\psi_t}
  \label{eq:A-r}
\end{align}
Since diffusive quantum trajectories are continuous \cite{Barchielli2009}, the above relation must hold for a continuum of states and, due to the pathwise ergodic theorem, for a complete basis of the state space.
In particular, in the diffusive unraveling it is impossible to have an orthogonal decomposition of the subspace without violating the assumption of minimality (cf. \cref{sec:absence-of-purification-in-a-minimal-subspace}).
This means the set $\{\ket{\psi_t}\}$ spans $\calq$, and there is no nontrivial orthogonal decomposition $\calq=\bigoplus_i \calq_i$ such that each $\ket{\psi_t}$ lies entirely in a single subspace $\calq_i$.
Then, together with continuity \eqref{eq:A-r} implies that
\begin{align}
  r_k = r_k(t), \ \forall t
\end{align}
is actually time-independent.
From \cref{eq:r(t)} it follows that
\begin{align}
L_{k,\calq} = u^\dagger L_{k,\calp} u + \I r_k\mathds{1}.
\end{align}
Using the dynamical symmetry of the Lindblad equation $U\sfl(\rho_\xi)U^\dagger = \sfl(U\rho_\xi U^\dagger)$ (see \cref{appendixeq:dynamical-symmetry}) we then arrive at 
\begin{align}
  \I[H_\calq,\dyad{\psi_t}] = \I[\tilde H_\calp,\dyad{\psi_t}],
\end{align}
with $\tilde H_\calp \equiv uH_\calp u^\dagger - \frac{1}{2} \sum_k r_k u \left(L_{k,\calp}+L^\dagger_{k,\calp}\right) u^\dagger$.
This relation can again be cast in the form
\begin{align}
  A+A^\dagger = B+B^\dagger,
\end{align}
where $A \equiv iH_\calq \dyad{\psi_t}$ und $B \equiv i\tilde H_\calp \dyad{\psi_t}$.
By applying the same arguments as before, it thus must hold that 
\begin{align}
  H_\calq = \tilde H_\calp + b\mathds{1},
\end{align}
where $b \in \mathbb{R}$.
Finally, this amounts to the identification
\begin{align}
  L_{k,\calq} &= u L_{k,\calp} u^\dagger + \I r_k \mathds{1},\\
  H_\calq &= uH_\calp u^\dagger - \frac{1}{2}
  \sum_k r_k u \left(L_{k,\calp}+L^\dagger_{k,\calp}\right) u^\dagger + b\mathds{1}.
  \label{appendixeq:relation}
\end{align}
Since the inhomogenous transformation 
\begin{align}
  &L_k \to L_k + \alpha_k \mathds{1}, \notag\\
  &H \to H + \frac{1}{2\I}\sum_k (\alpha^\ast_k L_k - \alpha_k L^\dagger_k) + \beta \mathds{1},
\end{align}
with $\alpha_k \in \mathbb{C}$, and $\beta \in \mathbb{R}$ leaves the diffusive quantum trajectory invariant \cite{Wiseman2001} (see \cref{eq:local-oscillator}), we may, without loss of generality, express relation \eqref{appendixeq:relation} as
\begin{align}
  H_\calq = uH_\calp u^\dagger, \qquad L_{k,\calq} = uL_{k,\calp}u^\dagger.
\end{align}
This is case (ii) of \cref{th:incomplete-diff}.
It implies
\begin{align}
    H_{\calq \oplus \calp} &= UH_{\calq \oplus \calp}U^\dagger,\\
    L_{k,\calq \oplus \calp} &= \exp(\I\phi_k) UL_{k,\calq \oplus \calp}U^\dagger,
\end{align}
The unitary $U$ is an intertwiner as in Ref.~\cite[Proposition 16]{Baumgartner2008_2} that transforms stationary states and their support, the orthogonal subspaces, into each other
\begin{align}
    \rho_\calq^\m{s} &= U\rho^\m{s}_\calp U^\dagger, \quad
    P_\calq = UP_\calp U^\dagger, \\
    [H_{\calq \oplus \calp},U] &= [L_{\calq \oplus \calp},U] = 0, \ \forall k
\end{align}
while commuting with the Hamiltonian and the Lindblad jump operators.
This is equivalent to stationary phase relations or coherence blocks in the steady state density matrix of the Lindblad equation $\rho^\m{s}$.
The state space structure is then a noiseless subsystem with 
\begin{align}
    \calq \oplus \calp 
    \simeq \mathds{1}_2 \otimes \calq,
\end{align}
the orthogonal subspaces are degenerate in this sense and the von Neumann algebra generated by the commutant $\{H_{\calq \oplus \calp},L_{k,\calq \oplus \calp},L_{k,\calq \oplus \calp}^\dagger\}^\prime \simeq \mathbb{C}^2\otimes P_\calq$ is generally not abelian, see Ref.~\cite[Proposition 16]{Baumgartner2008_2} (cf. \cref{sec:symmetries}).

\subsubsection{Absence of purification in both minimal subspaces}
\label{sec:absence-of-purification-in-both-subspaces-diff}
We have shown that the absence of purification in a minimal subspace implies that the effective measurement operator 
\begin{align}
  (L_k+L^\dagger_k)_\calq = z_k \mathds{1}_{d_\calq}
\end{align}
must be proportional to the identity (cf. \cref{sec:absence-of-purification-in-a-minimal-subspace}).
We here treat the case where asymptotic purification does not occur in neither $\calq$ nor $\calp$ and thus relations \eqref{eq:Barchielli} and \eqref{eq:Maassen} hold.
According to \cref{eq:valid-trajectories-diff}, the states $\tilde \rho_{\xi,\calq}$ and $\tilde \rho_{\xi,\calp}$ are themselves valid quantum trajectories and we can thus apply the purification theorems, \cref{th:pur-diff,th:pur-jumps}, to each of them individually.
Now, let purification be absent in both $\tilde \rho_{\xi,\calq}$ and $\tilde \rho_{\xi,\calp}$.
By assumption of minimality, we have $(L_k +L^\dag_k)_\calq = z_{k,\calq}\mathds{1}_{d_\calq}$, and $(L_k +L^\dag_k)_\calp = z_{k,\calp}\mathds{1}_{d_\calp}$, yielding
\begin{align}
  z_{k,\calq}\tr[\tilde\rho_{\xi,\calq}]
  = z_{k,\calp}\tr[\tilde \rho_{\xi,\calp}],
\end{align}
which immediately implies $z_k \equiv z_{k,\calq} = z_{k,\calp}$.
This is case (i) of \cref{th:mincomplete-diff}.

\subsubsection{Purification in one minimal subspace}
\label{sec:purification-in-one-subspace-diff}
Purification might happen in only one subspace but not in the other.
It turns out that incomplete localization cannot happen between a minimal purifying subspace and a minimal purity preserving subspace.
Starting from an arbitrarily mixed initial state in the purity preserving subspace will result in state-independent classical noise (cf. \cref{sec:classical-noise}).
However, any initially mixed state in the purifying subspace will transition from mixed to pure under state dependent measurement backaction which can thus not possibly remain constant at all times.
Below, we show this incompatibility explicitly.

Without loss of generality, assume that purification happens only inside of $\calq$ but not in $\calp$ such that $\tilde \rho_{\xi,\calq} = \dyad{\psi_t}$.
In $\calp$ it thus holds that $(L_k +L^\dag_k)_\calp = z_{k,\calp}\mathds{1}_{d_\calp}$ (cf. \cref{sec:absence-of-purification-in-a-minimal-subspace}) and \cref{eq:first-order-diff} then gives
\begin{align}
  \tr[\dyad{\psi_t}(L_k +L^\dag_k)_\calq] = 
  z_{k,\calp}\tr[\tilde \rho_{\xi,\calp}]
  = z_{k,\calp}.
  \label{eq:pur-mixed-diff}
\end{align}
In both subspaces $\calq$ and $\calp$ we perform the independent transformations
\begin{align}
  &\tilde L_{k,\cala} = L_{k,\cala} + a_{k,\cala} \mathds{1}_{d_\cala}, \notag\\
  &\tilde H_\cala = H_\cala + \frac{1}{2\I}\sum_k (a^\ast_{k,\cala} L_k - a_{k,\cala} L^\dagger_k),
\end{align}
where $\cala = \calq,\calp$ (cf. \cref{eq:local-oscillator}) and the complex numbers $a_{k,\cala}$ are chosen such that $(L_k+L^\dag_k)_\calq \to (\tilde L_k+\tilde L^\dag_k)_\calq$ becomes a positive operator and $(L_k+L^\dag_k)_\calp \to 0$.
We obtain
\begin{align}
  \tr[\dyad{\psi_t}(\tilde L_k+\tilde L^\dag_k)_\calq] = 0.
\end{align}
Taking the long-time average and using the pathwise ergodic theorem (cf. \cref{sec:erg-theorem}) then gives
\begin{align}
  \tr[\tilde \rho^\m{s}_\calq (\tilde L_k+\tilde L^\dag_k)_\calq]
  = \sum_j p_j \bra{\Psi_j}(\tilde L_k+\tilde L^\dag_k)_\calq\ket{\Psi_j} = 0.
\end{align}
Since $p_j >0,\ \forall j$, it follows that $(\tilde L_k+\tilde L^\dag_k)_\calq = 0$ and thus $(L_k+L^\dag_k)_\calq = z_{k,\calq}\mathds{1}_{d_\calq}$ and the state $\ket{\psi_t}$ must have been pure from the beginning.
Revisiting again \cref{eq:pur-mixed-diff} further specifies $z_{k,\calq} = z_{k,\calp} \equiv z_k$, which is case (i) of \cref{th:mincomplete-diff}.

\subsection{Incomplete localization of jump trajectories between two minimal subspaces}
\label{sec:proof-no-selection-jumps}
We prove the following theorem.
\begin{theorem}[Conditional on \rconj{appendix-pur}]
\label{th:mincomplete-jump}
  Let the quantum jump trajectory \cref{eq:jump-sme} have support exclusively on two minimal subspaces $\calq \oplus \calp$.
  Then, provided \rconj{appendix-pur} holds, there is no subspace selection on the trajectory level between $\calq$ and $\calp$ if and only if at least one of the following holds $\forall k$.\begin{enumerate}[(i)]
      \item Independent trajectories:
      \begin{align} 
          (L^\dagger_k L_k)_{\calq} = z_k \mathds{1}_{d_\calq}, \quad 
          (L^\dagger_k L_k)_{\calp} = z_k \mathds{1}_{d_\calp}.
          \label{appendixeq:pur-jumps}
      \end{align}
      \item[(iia)] Unitarily equivalent trajectories: 
      \begin{align} 
          H_\calq = uH_\calp u^\dagger, \qquad 
          L_{k,\calq} = \exp(\I\phi_k)uL_{k,\calp}u^\dagger.
          \label{appendixeq:u-jumps}
      \end{align}
      \item[(iib)] Unitarily equivalent trajectories:
      \begin{align}
        &H_\calq = \bigoplus_{\alpha=1} h_{\calq_\alpha}, \quad 
        u H_\calp u^\dagger = \bigoplus_{\alpha=1} \left(h_{\calq_\alpha} + r_\alpha\mathds{1}_{d_\alpha}\right),
        \label{appendixeq:block-H}\\
        &L_{k,\calq} = \exp(\I\phi_k)uL_{k,\calp}u^\dagger, \\
        &P_\beta \left(\sum_k (L^\dag_kL_k)_\calq\right)P_\gamma = 0, \ \forall \beta\neq \gamma\\
        &P_\beta  L_{k,\calq} P_\alpha = 0 \lor P_\gamma L_{k,\calq} P_\alpha = 0, \ \forall \alpha,\beta \neq \gamma
        \label{appendixeq:sim-L},
      \end{align}
  \end{enumerate}
  with $z_k \ge 0$, $\phi_k,r_\alpha \in \mathbb{R}$ and orthogonal projectors $P_\alpha$.
\end{theorem}
Consider an arbitrary quantum jump trajectory $\rho_J(t)$ in finite dimensions and denote by 
\begin{align}
  \overline{\rho}_J = \frac{1}{t} \int_0^t \dd{s} \rho_J(s),
\end{align}
the infinite time averaged state.
From the pathwise ergodic theorem (Section VIII) it follows that this state is a stationary state of the Lindblad equation, i.e. $\sfl(\overline{\rho}_J) = 0$.
Decompose the support of $\overline{\rho}_J$ into two minimal orthogonal subspaces $\calq \oplus \calp$, then $\overline{\rho}_J$ has full rank on $\calq \oplus \calp$.
Asymptotic dynamics can thus take place only on the combined subspace $\calq \oplus \calp$. 
Localization does not occur between $Q$ and $P$ if and only if the following holds
\begin{align}
    |\calp(t)|^2\tr[\rho_J(t)(L_k^\dagger L_k)P_\calq]
    = |\calq(t)|^2\tr[\rho_J(t)(L_k^\dagger L_k)P_\calp],
    \label{eq:jump-freeze}
\end{align}
for all times $t > T$, where $T$ is a random variable that is generally different for every realization.
The trajectory is thus unable to decide between the two subspaces and gets stuck in between.
If this happens then $\forall t > T$ the probabilities to be in the respective subspace become stationary, $0< |\calq(t)|^2 = \m{const.} <1$ and $0<|\calp(t)|^2 = \m{const.}<1$, and it still holds that $|\calq(t)|^2+|\calp(t)|^2=1$ (the subspaces span the whole support of $\rho_J$).
We define effective states in each subspace
\begin{align}
    \rho_{J,\cala} &\equiv P_\cala \rho_J(t) P_\cala,\\
    \tilde \rho_{J,\cala} &\equiv \frac{\rho_{J,\cala}}{\tr[\rho_{J,\cala}]}
    = \frac{\rho_{J,\cala}}{|\cala|^2_J},
\end{align}
with $\cala = \calq,\calp$ and, in accordance with the definitions in \cref{sec:erg-theorem,sec:Born}, we denote by $|\cala|^2_J \equiv |\cala(t)|^2, \ t>T$, the random variable, constant for every realization.
\Cref{eq:jump-freeze} can then be written as 
\begin{align}
  \label{eq:first-order-jump}
  \tr[\tilde \rho_{J,\calq}(L_k^\dagger L_k)_\calq] = 
    \tr[\tilde \rho_{J,\calp}(L_k^\dagger L_k)_\calp]
\end{align}
and the evolution of $\tilde \rho_{J,\calq}$ and $\tilde \rho_{J,\calp}$ in each subspace may thus be considered an independent trajectory
\begin{align}
  \label{eq:valid-trajectories-jump}
  \dd{\tilde \rho}_{J,\calq}
  =& -\I[H_\calq,\tilde \rho_{J,\calq}] \dd{t} \notag \\
  &+ \sum_k\left(\langle L_{k,\calq}^\dagger L_{k,\calq}\rangle_\calq \tilde \rho_{J,\calq} - \frac{1}{2} \{L_{k,\calq}^\dagger L_{k,\calq},\tilde \rho_{J,\calq}\}\right) \dd{t} \notag \\
  &+ \left[\frac{L_{k,\calq}\tilde \rho_{J,\calq} L_{k,\calq}^\dagger}{\langle L_{k,\calq}^\dagger L_{k,\calq}\rangle_\calq}-\tilde \rho_{J,\calq}\right] \dd{N}_k,
\end{align}
since \cref{eq:jump-freeze} is equivalent to
\begin{align}
  \tr[\rho_J (L_k^\dagger L_k) P_\calq] = |\calq(t)|^2 \langle L_k^\dagger L_k\rangle,
\end{align}
and therefore
\begin{align}
\langle L^\dagger_k L_k\rangle 
= \tr[\tilde \rho_{J,\calq}(L_k^\dagger L_k)_\calq]
= \langle L^\dagger_{k,\calq} L_{k,\calq}\rangle_\calq
\label{eq:meas-record-jump}
\end{align}
and the same for $\dd{\tilde \rho}_{J,\calp}$.

\subsection{Unitary equivalence---case (iia)}
\label{sec:unitary-equivalence-jump}
Before proceeding, we need to distinguish two cases.
Either, as we will now assume, the trajectories $\tilde \rho_{J,\calq}$ and $\tilde \rho_{J,\calp}$ asymptotically purify individually or at least one of them does not.
The latter case is deferred to \cref{sec:absence-of-purification-in-both-subspaces-jumps,sec:purification-in-one-subspace-jumps}.

If the asymptotic purification theorem holds in both subspaces, we can assume that 
\begin{align}
    \tilde \rho_{J,\calq} = \dyad{\psi_t}, \qquad 
    \tilde \rho_{J,\calp} = \dyad{\varphi_t}
\end{align}
are pure states for all times $t > T$.
In general the minimal subspaces $\calq$ and $\calp$ might have different dimensions.
Without loss of generality assume that $\calq$ is larger, i.e, $d_\calq \ge d_\calp$.
Now, introduce the inflated pure state projector
\begin{align}
  \dyad{\tilde \varphi_t} = 
  \begin{pmatrix}
    \dyad{\varphi_t} & 0 \\ 0 & 0
  \end{pmatrix} \in \mathbb{C}^{d_\calq \times d_\calq},
\end{align}
such that both $\dyad{\psi_t}$ and $\ket{\varphi_t}$ live on a $d_\calq$-dimensional Hilbert space.
At each instant in time there thus exist a unitary matrix $u_t$ that transforms the states into each other, viz.
\begin{align}
  \dyad{\psi_t} = u_t \dyad{\tilde \varphi_t} u^\dag_t.
\end{align}
This needs to hold for all asymptotic times and in particular taking the stochastic differential on both sides yields
\begin{align}
  \dd{(\dyad{\psi_t})}
  =& (\dd{u_t})\dyad{\tilde \varphi_t}u^\dag_t + u_t\dd{(\dyad{\tilde \varphi_t})}u^\dag_t \notag\\
  &+ u_t\dyad{\tilde \varphi_t}(\dd{u^\dag_t}) + (\dd{u_t})\dd{(\dyad{\tilde \varphi_t})}u^\dag_t \notag\\
  &+ u_t\dd{(\dyad{\tilde \varphi_t})}(\dd{u^\dag_t}) + (\dd{u_t})\dyad{\tilde \varphi_t}(\dd{u^\dag_t}),
  \label{eq:ito-unitary-jump}
\end{align} 
Generally, the unitary matrix may depend on the noise channels according to
\begin{align}
  \dd{u_t} = (A_t \dd{t} + \sum_k B_{t,k} \dd{N_k})u_t,
\end{align}
where, in order to preserve unitarity, the matrices need to satisfy $A_t = -A_t^\dag$ and $B_{t,k}+B^\dag_{k,t} + B_{t,k}B^\dag_{k,t} = 0$.
\Cref{eq:ito-unitary-jump} is a valid quantum trajectory. In order for the right hand side to be consistent, the term $(\dd{u_t})\dd{(\dyad{\tilde \varphi_t})}u^\dag_t + u_t\dd{(\dyad{\tilde \varphi_t})}(\dd{u^\dag_t}) + (\dd{u_t})\dyad{\tilde \varphi_t}(\dd{u^\dag_t})$ needs to vanish identically, or equivalently the unitary does not depend on the noise and $B_{t,k} = 0, \ \forall k$.
Using \cref{eq:unitary-relation} to replace $\dyad{\tilde \varphi_t}$ and the fact that both trajectories have identical noise histories and measurement records (cf. \cref{eq:first-order-jump,eq:meas-record-jump}), we identify an effective Hamiltonian $\tilde H_\calp$ and jump operators $\tilde L_{k,\calp}$ such that relation \eqref{eq:ito-unitary-jump} may be written as
\begin{align}
  \dd{\rho_\calq} 
  = \sfl_{J,\calq}(\rho_\calq) 
  = \tilde \sfl_{J,\calp}(\rho_\calq),
  \label{eq:cast-jumps}
\end{align}
with
\begin{align}
  &= -\I [\tilde H_\calp,\rho_\calq] \dd{t}\notag\\
  &\quad + \sum_k\left(\langle \tilde L_{k,\calp}^\dag \tilde L_{k,\calp}\rangle_\calq \rho_\calq
  - \frac{1}{2} \{(\tilde L^\dag_k\tilde L_k)_\calp,\rho_\calq\}\right)\dd{t} \notag\\
  &\quad +\left[\frac{\tilde L_{k,\calp} \rho_\calq \tilde L_{k,\calp}^\dag}{\langle \tilde L_{k,\calp}^\dag \tilde L_{k,\calp}\rangle_\calq} - \rho_\calq\right] \dd{N_k},
\end{align}
where for the sake of brevity we have used the shorthand $\rho_\calq = \dyad{\psi_t}$.
The effective operators read
\begin{align}
  \tilde H_\calp = H_\calp +\I (\dd{u_t})u^\dag_t \oplus \vb{0}, \quad \tilde L_{k,\calp} = u_tL_{k,\calp}u^\dag_t \oplus \vb{0}.
\end{align}
$\dyad{\psi_t}$ is the uniqe solution of the quantum stochastic differential equations $\sfl_{\xi,\calq}$ and $\tilde \sfl_{\xi,\calp}$.
Since they produce the same unique solution, the tuples $(\tilde H_\calp,\tilde L_{k,\calp})$ and $(H_\calq, L_{k,\calq})$ must be related by a gauge transformation that transforms the operators but leaves the equation invariant.
Recently, the gauge invariance of quantum jump trajectories has been completely classified by Brown, Macieszczak and Jack in Ref.~\cite{Brown2025}.
Crucially, these transformations take time-independent operators to time-independent operators and therefore $u_t$ must not be time-dependent.
Since \cref{eq:cast-jumps} does not uniquely determine $(\tilde H_\calp,\tilde L_{k,\calp})$, all that can be said at this point is that the trajectory in $\calq$ and the inflated state in $\calp$ must thus be related by a time-independent unitary according to 
\begin{align}
  \dyad{\psi_t} = u\dyad{\tilde \varphi_t}u^\dag.
\end{align}
But since 
\begin{align}
    \dyad{\psi_t} = u\dyad{\tilde \varphi_t}u^\dag = 
    \begin{pmatrix}
        \dyad{\varphi_t} & 0\\ 0 & 0
    \end{pmatrix}
\end{align}
is block diagonal at all times, the long-time average cannot possibly converge to a full rank stationary state in $\calq$ and thus $\calq$ cannot be minimal.
This implies that the subspaces $\calq$ and $\calp$ must have the same dimension, i.e. $d_\calq = d_\calp$ and it actually holds that
\begin{align}
  \dyad{\psi_t} = u\dyad{\varphi_t}u^\dag.
\end{align}

Unitary equivalence of evolutions implies that there exists a unitary, $U$, acting on the whole Hilbert space spanned by $\calq \oplus \calp$ such that
\begin{align}
    U = 
    \begin{pmatrix}
        0 & u\\ u^\dagger & 0
    \end{pmatrix},
    \qquad U^2 = \mathds{1},
    \label{eq:unitary-jump}
\end{align}
with
\begin{align}
    U \rho_J U^\dagger
    = 
    \begin{pmatrix}
        0 & u\\ u^\dagger & 0
    \end{pmatrix}
    \begin{pmatrix}
        \tilde \rho_{J,\calq} & 0\\ 0 & \tilde \rho_{J,\calp}
    \end{pmatrix}
    \begin{pmatrix}
        0 & u \\ u^\dagger & 0
    \end{pmatrix}
    = \rho_J,
\end{align}
and
\begin{align}
    U\sft_J^t(\rho_J)U^\dagger
    = \sft_J^t(U\rho_J U^\dagger).
    \label{eq:dynamical-symmetry-jump}
\end{align}
Since the stochastic update terms are mutually independent, we obtain
\begin{align}
    L_{k,\calq}\tilde \rho_{J,\calq} L_{k,\calq}^\dagger = 
    u L_{k,\calp}u^\dagger \tilde \rho_{J,\calq}u L_{k,\calp}^\dagger u^\dagger.
    \label{eq:update-terms}
\end{align}
Let $A^\dagger_k \equiv (L_{k,\calq}\tilde{\rho}_{J,\calq})$ and $B^\dagger_k \equiv (u L_{k,\calp} u^\dagger\tilde{\rho}_{J,\calq})$.
\Cref{eq:update-terms} can then be equivalently expressed as
\begin{align}
    A^\dagger_k A_k = B^\dagger_k B_k.
\end{align}
According to Douglas's lemma, there exists a partial isometry $C_k$ such that $A_k = C_kB_k$ \cite{Douglas1966}, or $A^\dagger_k = B^\dagger_k C^\dagger_k$.
Since, by assumption, $\tilde \rho_{J,\calq} = \dyad{\psi_t}$ is pure, we have
\begin{align}
  L_{k,\calq}\tilde \rho_{J,\calq} = u L_{k,\calp} u^\dagger \tilde \rho_{J,\calq} C^\dagger_k.
  \label{eq:rhotilde}
\end{align}
Consider the subspace that belongs to the image, $\Im(L_{k,\calq})$, of $L_{k,\calq}$, i.e. states $\ket{\psi_t} \in \calq$ for which $L_{k,\calq}\ket{\psi_t} \neq 0$.
Now, at any given time $t$, construct an orthogonal basis of the entire subspace $\Im(L_{k,\calq})$, where $\ket{\psi_t}$ is considered the first basis vector.
Denote by $\{\ket{\psi^\perp_\alpha}\}$ the remaining basis states that span the reduced subspace $\Im(L_{k,\calq}) \setminus \{\ket{\psi_t}\}$ with $\braket{\psi}{\psi^\perp_\alpha} = 0$ and $\braket{\psi^\perp_\alpha}{\psi^\perp_\beta} = \delta_{\alpha,\beta}$.
Multiplying from the right by $\ket{\psi^\perp_\alpha}$ and taking the adjoint yields for all states $\ket{\psi_t}$ not in the kernel of $L_{k,\calq}$
\begin{align}
  \bra{\psi^\perp_\alpha}C_k\ket{\psi_t} = 0, \ \forall \alpha.
\end{align}
Since $C_k$ is a partial isometry, this results in the block form $C_k = \exp(-i\phi_k)\oplus c_k$, where $\phi_k$ is a phase and $c_k$ is a partial isometry on the reduced subspace $\Im(L_{k,\calq})\setminus \{\ket{\psi_t}\}$.
It immediately implies
\begin{align}
  C_k(t)\ket{\psi_t} = e^{-i\phi_k(t)}\ket{\psi_t},
\end{align}
where we have made explicit a potential time dependence of $C_k(t)$ and $\exp(\I\phi_k(t))$ that can, at this point, not be excluded.
If we plug this back in into \cref{eq:rhotilde} we further obtain
\begin{align}
  L_{k,\calq} \dyad{\psi_t} = e^{\I\phi_k(t)}u L_{k,\calp}u^\dagger\dyad{\psi_t}.
  \label{eq:ket-phase}
\end{align}

The pathwise ergodic theorem states that any quantum trajectory converges to a full rank stationary state (see \cref{sec:erg-theorem}).
It is hence always possible to select a sequence of states at different times such that the set $\{\ket{\psi_{t_1}},\ket{\psi_{t_2}},\ldots,\ket{\psi_{t_n}}\}$ with $\ket{\psi_{t_i}} \in \{\ket{\psi_t}\}$ provides a complete basis of the reduced subspace $\Im(L_{k,\calq})$, which might however not be orthogonal.
Using the shorthand notation $\ket{\psi_i} \equiv \ket{\psi_{t_i}}$,
we may rewrite \cref{eq:ket-phase} as
\begin{align}
  L_{k,\calq} \dyad{\psi_i} = e^{\I\phi_{i,k}}u L_{k,\calp}u^\dagger\dyad{\psi_i}.
  \label{eq:i-phase}
\end{align}
In order to deduce statements about the operators $L_{k,\calq}$ and $L_{k,\calp}$ we transform to an orthogonal basis.
We do so by means of symmetric orthogonalization, employing the Gram matrix \cite{Horn2012}.
Given the non-orthogonal basis $\{\ket{\psi_i}\}$, denote by $A = \begin{pmatrix}\ket{\psi_1}&\ket{\psi_2}&\ldots&\ket{\psi_n}\end{pmatrix}$ the matrix with the vectors $\ket{\psi_i}$ as its columns.
From it, construct the positive semidefinite Hermitian matrix $A^\dagger A$, which can be unitarily diagonalized according to $A^\dagger A = V^\dagger D V$.
Defining further the matrix $H = V^\dagger D^{-1/2}V$, there is a new matrix $\tilde A = AH$ whose column vectors form an orthonormal basis with $\tilde A \tilde A^\dagger = \mathds{1}$.
After application of $\ket{\psi_i}$ from the right, we may hence write \cref{eq:i-phase} as
\begin{align}
  L_{k,\calq} A = u L_{k,\calp} u^\dagger A \Phi_k,
\end{align}
where $\Phi_k = \m{diag}(e^{\I\phi_{1,k}},e^{\I\phi_{2,k}},\ldots,e^{\I\phi_{n,k}})$ is a matrix diagonal in the canonical basis accounting for the phase factors in \cref{eq:i-phase}.
Multiplying with $H$ from the right then yields
\begin{align}
  L_{k,\calq} \tilde A = u L_{k,\calp} u^\dagger \tilde A H^{-1} \Phi_k H.
\end{align}
This relation is time-independent by construction and holds for a complete orthonormal basis. 
Defining $W_k \equiv \tilde A H^{-1}\Phi_k H \tilde A^\dagger$, it follows that
\begin{align}
  L_{k,\calq} = u L_{k,\calp}u^\dagger W_k.
  \label{eq:W}
\end{align}
The matrix $\Phi_k$ is unitary and diagonal and we can identify the similarity transform $W_k = P^{-1}\Phi_k P$, where $P = \tilde AH^{-1}$ is the transformation matrix.
The matrix $W_k$ is thus similar to a unitary.
Using \cref{eq:W} and transforming the relation \cref{eq:ket-phase} into the eigenbasis of $\Phi_k$ gives
\begin{align}
  &PL_{k,\calq}P^{-1} P\dyad{\psi_t}P^\dagger =\\
  &P L_{k,\calq} P^{-1}PW^{-1}e^{\I\phi_k(t)}P^{-1}P\dyad{\psi_t}P^\dagger.
\end{align}
Since $L_{k,\calq}$ is invertible on its support, this is equivalent to 
\begin{align}
  P\dyad{\psi_t}P^\dagger = \Phi_k^\dagger e^{\I\phi_k(t)}P\dyad{\psi_t}P^\dagger.
\end{align}
Defining the new states $\ket{\gamma_t} \equiv P \ket{\psi_t}$, we then obtain
\begin{align}
  \Phi_k \dyad{\gamma_t} = e^{\I\phi_k(t)} \dyad{\gamma_t},
  \label{eq:unitary-phase}
\end{align}
where the states $\ket{\gamma_t}$ explore at least a complete basis.
We can thus conclude that 
\begin{align}
  \Phi_k \ket{\gamma_t} = e^{\I\phi_k(t)} \ket{\gamma_t}
\end{align}
holds for all $t$.
The stochastic process $\ket{\psi_t}$ is not necessarily continuous (see \cref{eq:jump-sme}) and the set $\{\ket{\psi_t}\}$ might contain only a finite number of different states, which precludes the application of the same arguments as below \cref{eq:A-r} for the diffusive case.
Instead, to extract more information about the phase factor, we take the derivative with respect to time on both sides to obtain
\begin{align}
  \Phi_k \ket{\dot{\gamma}_t} = e^{\I\phi_k(t)}\ket{\dot{\gamma}_t} + i\dot{\phi}_k(t) \ket{\gamma_t}.
\end{align}
Applying the adjoint of \cref{eq:unitary-phase}, $\bra{\gamma_t}\Phi^\dagger_k = \exp(-i\phi_k(t)) \bra{\gamma_t}$, from the left amounts to
\begin{align}
  \braket{\gamma_t}{\dot{\gamma}_t} = \braket{\gamma_t}{\dot{\gamma}_t} + i \dot{\phi}_k(t) \braket{\gamma_t}{\gamma_t},
\end{align}
and it follows that the phase cannot be time-dependent, $\dot{\phi}_k(t) = 0$.
Equipped with this newly acquired information, we revisit again \cref{eq:unitary-phase}
\begin{align}
  \Phi_k \ket{\gamma_t} = e^{\I\phi_k} \ket{\gamma_t},
\end{align}
which has to hold for a complete basis.
Since both $\Phi_k$ and $\exp(\I\phi_k)$ are time-independent, we can finally determine that
\begin{align}
  \Phi_k = e^{\I\phi_k}\mathds{1}.
\end{align}
In general, we thus obtain from \cref{eq:W}
\begin{align}
    L_{k,\calq} = \exp(\I\phi_k) u L_{k,\calp} u^\dagger.
    \label{eq:Lrelation}
\end{align}
This determines the relation between the jump operators.
Note that this would follow much more straightforwardly if we were to assume a complete and orthonormal basis which is, however, not immediately implied by pathwise ergodicity.

To obtain the relation of the Hamiltonians $H_\calq$ and $H_\calp$, we use \cref{eq:Lrelation} and the dynamical symmetry of the Lindblad equation $U\sfl(\rho_J)U^\dagger = \sfl(U\rho_JU^\dagger)$ (see \cref{eq:dynamical-symmetry-jump}), resulting in 
\begin{align}
  \I[H_\calq,\dyad{\psi_t}] = \I[u H_\calp u^\dagger,\dyad{\psi_t}].
\end{align}
Defining $A \equiv iH_\calq \dyad{\psi_t}$ and $B \equiv iu H_\calp u^\dagger \dyad{\psi_t}$ gives 
\begin{align}
  A+A^\dagger = B+B^\dagger.
\end{align}
from which it follows that $A = B + \I R$, which translates to
\begin{align}
  H_\calq \dyad{\psi_t} = u H_\calp u^\dagger\dyad{\psi_t} + R(t),
  \label{eq:HqHp}
\end{align}
where $R(t)$ is a Hermitian matrix.
Now, at any given time, construct an orthogonal basis of the entire subspace $\calq$, where $\ket{\psi}$ is considered the first basis vector.
Denote by $\{\ket{\psi^\perp_\alpha}\}$ the remaining basis states that span the reduced subspace $\calq \setminus \{\ket{\psi}\}$ with $\braket{\psi}{\psi^\perp_\alpha} = 0$ and $\braket{\psi^\perp_\alpha}{\psi^\perp_\beta} = \delta_{\alpha,\beta}$.
Multiplying from the right by $\ket{\psi^\perp_\alpha}$ thus yields
\begin{align}
  R(t)\ket{\psi^\perp_\alpha} = 0, \ \forall \alpha.
\end{align}
The operator $R(t)$ can thus not have support on the reduced subspace $\calq\setminus \{\ket{\psi}\}$, and, due to Hermiticity, must hence be of the form
\begin{align}
  R(t) = r(t) \dyad{\psi_t},
\end{align}
where $r(t) \in \mathbb{R}$.
Employing the form of $R(t)$ in \cref{eq:HqHp} results in
\begin{align}
  H_\calq \dyad{\psi_t} = (u H_\calp u^\dagger + r(t)\mathds{1})\dyad{\psi_t},
\end{align}
or, equivalently
\begin{align}
  X \ket{\psi_t} = r(t)\ket{\psi_t},
  \label{eq:X}
\end{align}
where $r(t) \in \mathbb{R}$ and we have defined the Hermitian operator $X \equiv H_\calq- u H_\calp u^\dagger$.

At this point, two cases need to be distinguished.
We have a complete possibly non-orthogonal basis of states $\{\ket{\psi_i}\}$ visited by the trajectory in the course of time.
First assume that the set $\{\ket{\psi_t}\}$ irreducible spans the subspace $\calq$, i.e., there exists no orthogonal decomposition $\calq=\calq_1 \oplus \calq_2$ such that each $\ket{\psi_t}$ lies entirely within either $\calq_1$ or $\calq_2$.
\begin{align}
  \nexists \calq_1 \oplus \calq_2 = \calq \text{ s.t. } \forall t: \ket{\psi_t} \in \calq_1 \text{ or } \ket{\psi_t} \in \calq_2,
  \label{eq:non-ortho-assumption}
\end{align}
but not both.
In particular, this implies 
\begin{align}
  X \ket{\psi_t} &= r(t)\ket{\psi_t},\\
  \bra{\psi_{t^\prime}} X &= \bra{\psi_{t^\prime}} r({t^\prime}),
\end{align}
with 
\begin{align}
  \bra{\psi_{t^\prime}} X \ket{\psi_t} 
  = r(t) \braket{\psi_{t^\prime}}{\psi_t}
  = r(t^\prime) \braket{\psi_{t^\prime}}{\psi_t}
\end{align}
Since $X$ is Hermitian and $r(t)$ is real, it follows that $r(t) = r(t^\prime)$.
Moreover, \cref{eq:X} holds true for a complete basis and it therefore must hold that 
\begin{align}
  X = r\mathds{1},
\end{align}
with $r$ independent of time.
Since quantum evolution is invariant under the transformation $H \to H + r\mathds{1}$, without loss of generality, we finally arrive at the identification
\begin{align}
  L_{k,\calq} = \exp(\I\phi_k)u L_{k,\calp}u^\dagger,\quad  H_\calq = u H_\calp u^\dagger.
\end{align}
This is case (ii) of \cref{th:incomplete-jump}.
It implies
\begin{align}
    H_{\calq \oplus \calp} &= UH_{\calq \oplus \calp}U^\dagger,\\
    L_{k,\calq \oplus \calp} &= \exp(\I\phi_k) UL_{k,\calq \oplus \calp}U^\dagger,
\end{align}
where $U$ intertwines between enclosures and maps stationary states of the Lindblad equation into each other
\begin{align}
    \rho_\calq^\m{s} = U\rho^\m{s}_\calp U^\dagger, \qquad
    P_\calq = UP_\calp U^\dagger.
\end{align}
Crucially however, the unitary $U$ does not commute with the Lindblad jump operators in general $[L_{k,\calq\oplus \calp},U] \neq 0$ and Ref.~\cite[Proposition 16]{Baumgartner2008_2} is not satisfied.
If $\exp(\I\phi_k) \neq 1$ for at least one $k$, there will not be stationary phase relations or coherence blocks in the steady state density matrix $\rho^\m{s}$ and the structure of the state space does not form a noiseless subsystem (cf. \cref{sec:symmetries}).

\begin{figure}[t]
  \centering
  \includegraphics{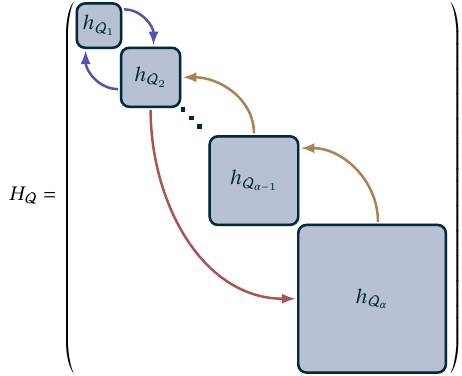}
  \caption{Schematic representation of the structure of the Hamiltonian $H_\calq$ (cf. \cref{eq:special-H}) to illustrate the dynamics in the unitary equivalence case (iib) of \cref{th:incomplete-jump}. 
  The Hamiltonian has a block diagonal form with individual Hermitian blocks $h_{\calq_\alpha}$ of dimension $d_{\calq_\alpha}$. 
  According to \cref{eq:ortho-assumption}, the evolution cannot create coherences between blocks and the pure state $\tilde \rho_{J,\calq}$ must thus be located entirely inside only one of these blocks at each point in time.
  On this given block structure, arrows indicate admissible jumps between blocks induced by the Lindblad jump operators $L_{k,\calq}$.
  Different colors correspond to different jump operators and can thus not occur simultaneously.
  Only complete jumps are allowed that transfer the entire state.
  The system may exchange probability between two blocks (blue) or transfer probability unidirectionally from one block to another (red and orange).
  Jumps can also happen independently inside of each block (not shown).
  Minimality of the subspace $\calq$ implies that the entire flow on the graph (where, arrows $\hat{=}$ edges, blocks $\hat{=}$ nodes) must be irreducible such that every subspace $\calq_\alpha$ can be reached from any other subspace $\calq_\beta$.}
  \label{fig:iib}
\end{figure}

\subsection{Unitary equivalence---case (iib)}
\label{sec:proof-iib}
Suppose that assumption \cref{eq:non-ortho-assumption} does not hold and there exists a collection of mutually orthogonal subspaces $\calq_\alpha$ with corresponding orthogonal projectors $P_\alpha$, i.e.
\begin{align}
  \calq = \bigoplus_\alpha \calq_\alpha,
  \label{eq:ortho-assumption}
\end{align}
and, at each time, the state is located entirely inside only one of these subspaces, i.e
\begin{align}
  \forall t> T, \ \exists P_\alpha \ \m{s.t.} \ P_\alpha\dyad{\psi_t}P_\alpha = \dyad{\psi_t}.
  \label{eq:state-in-block form}
\end{align}
Since $\calq$ itself is minimal by assumption, the state must still transition between the subspaces $\calq_\alpha$.
To proceed we must supply that the decomposition \eqref{eq:ortho-assumption} must be a structural property and thus holds for all states $\rho$.
Let the trajectory be located inside of $\calq_\alpha$, it hence must hold that (transitions take place)
\begin{align}
  P_\alpha\rho P_\alpha: P_\alpha^\perp \dot{\rho} P_\alpha^\perp \neq 0 \Leftrightarrow \exists k: P^\perp_\alpha L_{k,\calq}P_\alpha \neq 0,
  \label{eq:off-diag}
\end{align}
and (state adheres to the block form \cref{eq:state-in-block form})
\begin{align} 
  \forall \rho = P_\alpha\rho P_\alpha: \ P_\beta\rho P_\gamma = 0, \ P_\beta\dot{\rho} P_\gamma = 0, \ \forall \beta \neq \gamma,
\end{align}
from which it follows for the deterministic and stochastic contributions respectively (see also \cite[Lemma 11]{Baumgartner2008_2})
$\forall \rho = P_\alpha \rho P_\alpha$ and $\forall \beta\neq \gamma$
\begin{align}
  P_\beta \rho P_\beta\left(\I H_\calq - \frac{1}{2} \sum_k L^\dag_{k,\calq}L_{k,\calq}\right) P_\gamma &= 0,\\
  P_\beta \left(-\I H_\calq - \frac{1}{2} \sum_k L^\dag_{k,\calq}L_{k,\calq}\right)P_\gamma \rho P_\gamma &= 0,\\
  P_\beta  L_{k,\calq}\left(P_\alpha\rho P_\alpha\right) L^\dag_{k,\calq} P_\gamma &= 0,\label{eq:entire-jumps}
\end{align}
The first two relations imply
\begin{align}
  \I P_\beta H_\calq P_\gamma &= P_\beta \left(\frac{1}{2} \sum_k L^\dag_{k,\calq}L_{k,\calq}\right)P_\gamma,\\
  \I P_\beta H_\calq P_\gamma &= -P_\beta \left(\frac{1}{2} \sum_k L^\dag_{k,\calq}L_{k,\calq}\right)P_\gamma.
\end{align}
It therefore follows that the Hamiltonian and and the sum of the jump operators have a common block diagonal decomposition
\begin{align}
  P_\beta H_\calq P_\gamma = 0, \quad P_\beta \left(\sum_k L^\dag_{k,\calq}L_{k,\calq}\right)P_\gamma = 0, \ \forall \beta\neq \gamma,
  \label{eq:Hij}
\end{align}
with $H_\calq = \bigoplus_\alpha h_{\calq_\alpha}$ and $\sum_k L^\dag_{k,\calq}L_{k,\calq} = \bigoplus_\alpha (l^\dag_{k,\calq_\alpha}l_{k,\calq_\alpha})$
Since the subspace $\calq$ is minimal $H_\calq$ and $L_{k,\calq}$ cannot be simultaneously block diagonal which is in fact guaranteed by \cref{eq:off-diag}.
The third relation \cref{eq:entire-jumps} implies $\forall \alpha,\beta \neq \gamma$
\begin{align}
  P_\beta L_{k,\calq}P_\alpha = 0 \lor P_\gamma L_{k,\calq}P_\alpha = 0,
  \label{eq:entire-jumps-block form}
\end{align}
which means that in each column with index $\beta$ there can thus be at most a single block matrix $P_\beta L_{k,\calq}P_\alpha$ different from zero.
If we require that any arbitrary initial state eventually adheres to the block form (cf. \cref{eq:state-in-block form}) we need corresponding jump operators that can kill coherences between any two blocks. A sufficient condition to guarantee this is if every $L_{k,\calq}$ contains only one block matrix different from zero (cf. red arrow in \cref{fig:iib}).
In \cref{fig:iib} we schematically demonstrate the common block form of $H_\calq$ and $\dyad{\psi_t}$ along with possible transitions induced by the jump operators $L_{k,\calq}$.

Invoking again \cref{eq:X}, it still must hold for all states that
\begin{align}
  X \ket{\psi_t} = r(t) \ket{\psi_t}.
\end{align}
This means $X$ is of block diagonal form with $X = \bigoplus_\alpha x_\alpha$, where $x_\alpha$ are Hermitian matrices.
By assumption \cref{eq:non-ortho-assumption} holds again in each subspace $\calq_\alpha$, thus yielding
\begin{align}
  x_\alpha = r_\alpha\mathds{1}_{d_\alpha},
\end{align}
where $d_\alpha = \m{dim}(\calq_\alpha)$ is the corresponding subspace dimension.
Eventually, this further constrains the Hamiltonians to admit a special block structure
\begin{align}
  H_\calq = \bigoplus_\alpha h_{\calq_\alpha}, \quad u H_\calp u^\dag = \bigoplus_\alpha \left(h_{\calq_\alpha} + r_j \mathds{1}_{d_\alpha}\right).
  \label{eq:special-H}
\end{align}
The jump operators are still related by a unitary transformation and a phase factor (see \cref{eq:Lrelation}), but they must obey the additional constraints in \cref{eq:Hij,eq:entire-jumps-block form}.

Denote by $V$ the class of operators that satisfy \eqref{eq:entire-jumps-block form}, i.e. $L_{k,\calq} \in V$.
Then the commutator with the Hamiltonian must again belong to the same class since
\begin{align}
  P_\beta[H_\calq,L_{k,\calq}]P_\alpha &= P_\beta H_\calq P_\beta L_{k,\calq}P_\alpha - P_\beta L_{k,\calq} P_\alpha H_\calq P_\alpha\\
  P_\gamma[H_\calq,L_{k,\calq}]P_\alpha &= P_\gamma H_\calq P_\gamma L_{k,\calq}P_\alpha - P_\gamma L_{k,\calq} P_\alpha H_\calq P_\alpha
\end{align}
are respectively zero if either $P_\beta L_{k,\calq}P_\alpha = 0$ or $P_\gamma L_{k,\calq}P_\alpha = 0$.
Zero block matrices in the jump operators thus transitively translate to their commutator with the Hamiltonian according to
\begin{align}
  P_\beta L_{k,\calq} P_\alpha = 0 \Rightarrow P_\beta([H_\calq,L_{k,\calq}]) P_\alpha = 0.
  \label{eq:eigenoperator}
\end{align}
\Cref{eq:Hij,eq:entire-jumps-block form,eq:eigenoperator} contain as a special case systems where the jump operators $L_{k,\calq}$ act as ladder operators on the eigenstates of the Hamiltonian.
This includes for instance a two-level system coupled to a thermal bath of harmonic oscillators with Hamiltonian $H_\calq = \omega \sigma^z$ and jump operators $L_{1,\calq} = \sqrt{\Gamma n_\m{th}} \sigma^+$ and $L_{2,\calq} = \sqrt{\Gamma (1+n_\m{th})} \sigma^-$, satisfying
\begin{align}
  [H_\calq,L_{k,\calq}] \propto L_{k,\calq}.
\end{align}
Here $\Gamma$ is an arbitrary real parameter controlling the coupling strength and $n_\m{th} = [\exp(\beta\omega)-1]^{-1}$ is the Bose--Einstein distribution with inverse temperature $\beta$.
The quantum jump unraveling of this system gives rise to discrete jumps between the eigenstates of $\sigma^z$.
Consequently, the asymptotic evolution has nonzero overlap only with these two states, viz. $\{\ket{\psi_t}\} = \{\ket{0},\ket{1}\}$.

Here too, the unitary $U$ does not commute with the Lindblad jump operators in general $[L_{k,\calq\oplus \calp},U] \neq 0$ and Ref.~\cite[Proposition 16]{Baumgartner2008_2} is not satisfied.
Since the Hamiltonians restricted to the subspaces $\calq$ and $\calp$ are not necessarily unitarily related either, the composite Hamiltonian does not generally commute with $U$, i.e. $[H_{\calq\oplus \calp},U] \neq 0$, again preventing stationary phase relations or coherence blocks in the steady state density matrix $\rho^\m{s}$. The structure of the state space does not form a noiseless subsystem (cf. \cref{sec:symmetries}).

\subsubsection{Absence of purification in both minimal subspaces}
\label{sec:absence-of-purification-in-both-subspaces-jumps}
We have conjectured that the absence of purification in a minimal subspace implies that the effective measurement operator 
\begin{align}
  (L^\dagger_kL_k)_\calq = \mathds{1}_{d_\calq}
\end{align}
must be proportional to the identity (cf. \cref{sec:absence-of-purification-in-a-minimal-subspace}).
We treat the case where asymptotic purification does not occur in neither subspace and \cref{eq:Barchielli,eq:Maassen} does hold.
According to \cref{eq:valid-trajectories-diff}, the states $\tilde \rho_{J,\calq}$ and $\tilde \rho_{J,\calp}$ are themselves valid quantum trajectories and we can thus apply \cref{th:pur-diff,th:pur-jumps} to each of them individually.
Now let purification be absent in both $\tilde \rho_{J,\calq}$ and $\tilde \rho_{J,\calp}$ such that \cref{eq:Barchielli,eq:Maassen} holds in both subspaces.
Both trajectories $\tilde \rho_{J,\calq}$ and $\tilde \rho_{J,\calp}$ will explore their whole subspace respectively.
By assumption of minimality, we have $(L^\dagger_kL_k)_\calq = z_{k,\calq}\mathds{1}_{d_\calq}$, and $(L^\dagger_kL_k)_\calp = z_{k,\calp}\mathds{1}_{d_\calp}$, yielding
\begin{align}
  z_{k,\calq}\tr[\tilde\rho_{J,\calq}]
  = z_{k,\calp}\tr[\tilde \rho_{J,\calp}],
\end{align}
which implies $z_k \equiv z_{k,\calq} = z_{k,\calp}$.
This is case (i) of \cref{th:mincomplete-diff,th:mincomplete-jump}.

\subsubsection{Purification in one minimal subspace}
\label{sec:purification-in-one-subspace-jumps}
Purification might happen in only one subspace but not in the other.
As outlined in \cref{sec:purification-in-one-subspace-diff}, incomplete localization does not allow for a purifying minimal subspace to coexist with a purity preserving subspace.
Below, we show this incompatibility explicitly.

Without loss of generality, assume that purification happens only inside of $\calq$ but not in $\calp$ such that $\tilde \rho_{J,\calq} = \dyad{\psi_t}$.
In $\calp$ it thus holds that $(L^\dag_kL_k)_\calp = z_{k,\calp}\mathds{1}_{d_\calp}$ and \cref{eq:first-order-jump} then gives
\begin{align}
  \tr[\dyad{\psi_t}(L^\dag_kL_k)_\calq] = 
  z_{k,\calp}\tr[\tilde \rho_{\xi,\calp}]
  = z_{k,\calp}.
  \label{eq:pur-mixed-jumps}
\end{align}
For a single trajectory that traverses a complete but non-orthogonal basis, this condition does not uniquely determine the structure of the operator $(L^\dag_kL_k)_\calq$.
Instead, we need to invoke all possible trajectories and require that \eqref{eq:pur-mixed-jumps} should hold for any of them.
In particular it needs to hold for any initial state $\ket{\psi_0}$ in $\calq$, which immediately implies that
\begin{align}
  (L^\dag_kL_k)_\calq = z_k \mathds{1}_{d_\calq},
\end{align}
where we have defined $z_k \equiv z_{k,\calp}$.
This is case (i) of \cref{th:mincomplete-jump}.

\section{Incomplete localization between arbitrary orthogonal subspaces}
\label{sec:incomplete-composite}

We first establish that any trajectory gets stuck between two orthogonal subspaces $\calq$ and $\calp$ that span the support of $\rho_\m{c}$ if and only if they belong to the decomposition of $\calr$ induced by the Lindblad equation (cf. \cref{eq:decomposition}).
Consider the change in time of the overlap with the subspace $\calq$, for the diffusive unraveling
\begin{align}
    \dd{(|\calq(t)|^2)} 
    =& \tr[\sfl(\rho_\xi)P_\calq] \notag\\
    &+ \sum_k \bigg(\tr[\rho_\xi (L_k P_\calq + P_\calq L^\dag_k)] \notag\\
    &- |\calq(t)|^2\langle L_k+L_k^\dagger\rangle\bigg) \dd{W}_k
    \overset{!}{=} 0,
\end{align}
and the quantum jump unraveling
\begin{align}
    &\dd{(|\calq(t)|^2)} 
    = \tr[\sfl(\rho_J)P_\calq] \notag\\
    &+ \sum_k \left(\frac{\tr[\rho_J(L_k^\dag P_\calq L_k)]}{\langle L^\dag_k L_k\rangle}-|\calq(t)|^2\right) \dd{\tilde N}_k
    \overset{!}{=} 0.
\end{align}
The stochastic and deterministic are independent, so they need to vanish separately.
In particular
\begin{align}
    \tr[\sfl(\rho_\m{c})P_\calq] = \tr[\rho_\m{c} \sfl^\dagger(P_\calq)] = 0.
\end{align}
This can hold for all $\rho_\m{c}$ if and only if the projector on the orthogonal subspace $\calq$ is a conserved quantity of the Lindblad equation, i.e. $\dot{P}_\calq = \sfl^\dagger(P_\calq) = 0$.
A projector is conserved if and only if it is an element of the commutant $P_\calq \in \{H,L_k,L_k^\dag\}^\prime$ \cite[Lemma 7]{Baumgartner2008_2}.
The orthogonal subspace $\calq$ and, by extension, the subspace $\calp$ thus belong to the decomposition of the Hilbert space into orthogonal subspaces imposed by the Lindblad equation (\cref{sec:space-structure}) and localization can thus generally be incomplete only between such structures.

Now continue with orthogonal subspaces that can be further decomposed into minimal ones according to
\begin{align}
    \calq = \bigoplus_j \calq_j, \qquad \calp = \bigoplus_l \calp_l.
\end{align}
Denote by 
\begin{align}
      M_{k,\calq} = 
    \begin{dcases}
        (L_k+L^\dagger_k)_\calq, &\text{ diffusion}\\
        (L^\dagger_kL_k)_\calq, &\text{ jumps}
    \end{dcases},
\end{align}
the Hermitian measurement operators of both unravelings.
A quantum trajectory is unable to localize between $\calq$ and $\calp$ if
\begin{align}
  |\calp(t)|^2\tr[\rho_\m{c}(t)M_{k,\calq}P_\calq]
  = |\calq(t)|^2 \tr[\rho_\m{c}(t)M_{k,\calp}P_\calp],
\end{align}
from which it again follows that (cf. \cref{eq:first-order-diff,eq:first-order-jump})
\begin{align} 
    \tr[\tilde\rho_{\m{c},\calq}M_{k,\calq}]
   &= \tr[\tilde\rho_{\m{c},\calp}M_{k,\calp}].
\end{align}

\subsection{Absence of purification in an arbitrary subspace}
We are now in position to consider the generalization of \cref{sec:absence-of-purification-in-a-minimal-subspace} to an arbitrary subspace, not necessarily minimal.
Before we can compare two arbitrary orthogonal subspaces, we need to investigate what happens in an individual subspace when purification does not take place.
The following observation is crucial.
\begin{observation}
  \label{obs:purification-minimality}
  Asymptotic purification in a subspace implies minimality of that subspace.
\end{observation}
Suppose the trajectory is supported on $\calq$ which is decomposable into minimal subspaces $\calq = \bigoplus_j \calq_j$, then the trajectory can either get stuck between at least two orthogonal subspaces or it localizes completely.
If localization is complete then the asymptotic support is a minimal subspace.
If localization is incomplete, then there are valid trajectories in at least two subspaces and purification can at most continue inside them but not globally.
For this reason, the case where purification occurs in both $\calq$ and $\calp$ is the same as the unitary equivalence case (\cref{sec:unitary-equivalence-diff,sec:unitary-equivalence-jump}) and does not need to be considered here again.

Assume a quantum trajectory $\tilde \rho_\m{c}(t)$ is supported on an arbitrary orthogonal subspace $\calq$ and purification is absent.
We assume that there is no further localization in $\calq$ such that the state must retain its support and the long-time average converges to a full rank stationary state.
The overlaps with the individual minimal subspaces $|\calq_j(t)|^2$ are bounded martingales and therefore converge to a constant, random variable in the interval $(0,1)$.
Probability can thus not leak out of any one of the subspaces $\calq_j$ anymore and the trajectory is hence stuck between all minimal subspaces $\calq_j$.
Since 
\begin{align}
  \dd{(|\calq_j(t)|^2)} = 0, \ \forall j,
\end{align}
we may arbitrarily bipartition the support of $\tilde \rho_{\m{c},\calq}$ into $\calq = \calq_j \oplus \bigoplus_{l\neq j} \calq_l$ and write (cf. \cref{eq:calq,eq:calq_jump})
\begin{align}
  (1-|\calq_j(t)|^2)& \tr[\tilde \rho_{\m{c},j}M_{k,\calq_j}] = \notag
  \\ &|\calq_j(t)|^2 \sum_{l\neq j} |\calq_l(t)|^2 \tr[\tilde \rho_{\m{c},l}M_{k,\calq_l}],
\end{align}
which results in
\begin{align}
  \tr[\tilde \rho_{\m{c},j}M_{k,\calq_j}] = 
  \sum_l p_l \tr[\tilde \rho_{\m{c},l}M_{k,\calq_l}], \ \forall j.
\end{align}
This in turn implies
\begin{align}
    \tr[\tilde \rho_{\m{c},1}M_{k,\calq_1}] = \notag
    \tr[\tilde \rho_{\m{c},2}M_{k,\calq_2}] &= \ldots\\
    &= \tr[\tilde \rho_{\m{c},n}M_{k,\calq_n}].
\end{align}
Every pair can be considered separately and reduces to one of the cases in \cref{sec:unitary-equivalence-diff,sec:unitary-equivalence-jump}, \cref{sec:absence-of-purification-in-both-subspaces-diff,sec:absence-of-purification-in-both-subspaces-jumps} or \cref{sec:purification-in-one-subspace-diff,sec:purification-in-one-subspace-jumps}.
The relation of a single pair thus becomes transitive and extends to every pair such that (i), provided \rconj{appendix-pur} is true, in every subspace $\calq_j$ the operators $M_{k,\calq_j} = z_k \mathds{1}_{d_{\calq_j}}$ are all proportional to the identity with the same constant $z_k$, or (ii), all states in minimal subspaces are unitarily equivalent according to
\begin{align}
    \dyad{\psi_1} 
    &= u_2\dyad{\psi_2}u_2^\dagger
    = \ldots
    = u_n\dyad{\psi_n}u_n^\dagger,
    \label{eq:all-equivalent}\\
    M_{k,\calq_1}
    &= u_2M_{k,\calq_2}u_2^\dagger
    = \ldots = u_nM_{k,\calq_n}u_n^\dagger,
    \label{eq:u-measurement}
\end{align}
where $u_j$ are arbitrary unitary matrices.
Here, \cref{sec:proof-no-selection-diff,sec:proof-no-selection-jumps} apply and guarantee that the type of unitary equivalence evolution must be the same between any pair of subspaces.

\begin{figure}[t] 
  \centering
  \includegraphics{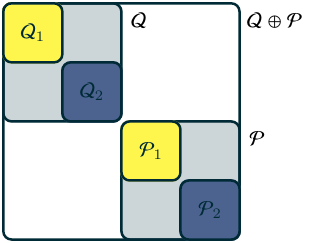}
  \caption{Decomposition of the space of bounded linear operators $\mathfrak{B}(\calh)$. We consider times $t>T$, where the decaying subspace has already been completely emptied and the trajectory has partially localized in $\calr$ with full support on a subspace $\calq \oplus \calp \subseteq \calr$ spanned by two orthogonal subspaces $\calq$ and $\calp$ that are not minimal. Evolution inside blocks of the same color is unitarily equivalent. Although the evolutions in $\calq$ and $\calp$ are unitarily equivalent, the trajectory can still decide between the minimal subspaces in blue ($\calq_1,\calp_1$) and yellow ($\calq_2,\calp_2$).
  This localization transition is however not observable on the coarse-grained level of the larger subspaces $\calq$ and $\calp$.}
  \label{fig:unitary-equivalence}
\end{figure}

\subsection{Incomplete localization}
The same considerations apply of course to the subspace $\calp$, where it either holds that $M_{k,\calp} = z_k\mathds{1}_{d_\calp}$ or \cref{eq:all-equivalent,eq:u-measurement} hold for $\calp$.
Going back again to the original relation between the composite subspaces $\calq$ and $\calp$, we still have
\begin{align}
    \tr[\tilde \rho_{\m{c},\calq}(t)M_{k,\calq}]
    = \tr[\tilde \rho_{\m{c},\calp}(t)M_{k,\calp}],
\end{align}
which, using relation \eqref{eq:first-order-diff} and \eqref{eq:first-order-jump} respectively, results in either one of the following relations $\forall k$
\begin{align}
    \tr[\dyad{\Psi_{\calq_j}(t)}M_{k,\calq_j}] &= \\
    \tr[\dyad{\Psi_{\calp_l}(t)}M_{k,\calp_l}], \ \forall j,l \\
    \tr[\dyad{\Psi_{\calq_j}(t)}M_{k,\calq_j}] &= z_{k,\calp}, \ \forall j, \\
    z_{k,\calq} &= z_{k,\calp}.
\end{align}
Note that the first two relations hold between two minimal and thus indecomposable subspaces $\calq_j$ and $\calp_l$.
The properties of the individual minimal subspaces thus extends to the coarse-grained level of $\calq$ and $\calp$

\section{On the unitary equivalence of composite subspaces}
\label{sec:minimal-equivalence}
In \cref{sec:incomplete-composite} we have proven that a quantum trajectory is unable to decide between two general orthogonal subspaces $\calq$ and $\calp$ if the respective evolutions are unitarily equivalent.
This has to hold in particular for minimal orthogonal subspaces.
Assume that $\calq$ and $\calp$ were to be further decomposable into minimal orthogonal subspaces that do not give rise to unitary equivalent evolutions between themselves, then any quantum trajectory could still decide between them and localize further.
This would however not be noticeable on the coarse-grained level of the larger subspaces $\calq$ and $\calp$.
In order to correctly identify all unitarily equivalent cases one always has to consider the lowest level of the decomposition, namely the minimal subspaces.
This situation is illustrated in \cref{fig:unitary-equivalence}.

\bibliography{aqt}
\end{document}